\documentclass[aps, prd, 10pt, twocolumn, superscriptaddress,noshowpacs, preprintnumbers,
longbibliography,nofootinbib,bibnotes,hyperref,floatfix]{revtex4-2}

\usepackage[dvipsnames]{xcolor}
\usepackage[colorlinks=true,breaklinks=true]{hyperref}
\hypersetup{allcolors=[rgb]{0.0 0.0 0.70},linkcolor=[rgb]{0.75 0.05 0.05}}
\usepackage{orcidlink}
\usepackage{microtype}

\usepackage{amsmath}
\usepackage{amsfonts}
\usepackage{amssymb}
\usepackage{mathtools}
\usepackage{multirow}
\usepackage{bm}
\usepackage[utf8]{inputenc}

\usepackage{siunitx}
\DeclareSIUnit\erg{erg}

\usepackage{pifont}

\newcommand{\optbar}[1]{\vphantom{#1}\smash{\overset{\scalebox{0.7}{$\scriptscriptstyle(-)$}}{#1}}}

\begin{document}

\title{Neutrino flavor conversion alters the lepton-emission self-sustained asymmetry\\ in core-collapse supernovae
}

\author{Noah Roux \orcidlink{0009-0005-1438-0432}}
\affiliation{Niels Bohr International Academy and DARK, Niels Bohr Institute, University of Copenhagen,\\
Blegdamsvej 17, 2100, Copenhagen, Denmark}

\author{Irene Tamborra \orcidlink{0000-0001-7449-104X}}
\affiliation{Niels Bohr International Academy and DARK, Niels Bohr Institute, University of Copenhagen,\\
Blegdamsvej 17, 2100, Copenhagen, Denmark}

\author{Maryna Mesiura \orcidlink{0009-0002-9362-5108}}
\affiliation{Niels Bohr International Academy and DARK, Niels Bohr Institute, University of Copenhagen,\\
Blegdamsvej 17, 2100, Copenhagen, Denmark}

\begin{abstract}
We solve the neutrino equations of motion in a half annulus, accounting for two spatial dimensions. At any location in the annulus, the neutrino momentum is characterized by a polar-angle distribution, two effective azimuthal angles, and an energy distribution.
We explore how neutrinos change their flavor while decoupling from matter in the presence of the neutrino-driven hydrodynamical instability LESA (i.e., the Lepton-number Emission Self-sustained Asymmetry)  during the stalled-shock phase of a core-collapse supernova. In our setup, LESA is responsible for an excess of $\nu_e$'s ($\bar\nu_e$'s) in the Northern (Southern) region of the annulus.
Crossings in the $\nu_e$--$\bar\nu_e$ angular distributions appear in the ELN-depleted Southern region of the annulus,
while crossings in the energy distributions of $\nu_e$--$\nu_x$ and $\bar\nu_e$--$\bar\nu_x$ are present throughout the annulus. We find that collisional  instabilities drive flavor conversion just before neutrino decoupling, fast instabilities are prominent in the Southern region of the annulus, and slow instabilities dominate otherwise. Flavor conversion is most efficient in the Southern region  and 
is responsible for a $5$--$10\%$ enhancement of the LESA dipole, independent of the neutrino-mass ordering. Our findings suggest that  direction-dependent flavor conversion  can have implications for the development of global neutrino asymmetries and hydrodynamical instabilities in core-collapse supernovae.
\end{abstract}

\maketitle

\section{Introduction}

Neutrinos are key particles in driving the core-collapse supernova (CCSN) explosion, as well as affecting its multi-messenger emission and the nucleosynthesis outcome~\cite{Mirizzi:2015eza,Raffelt:2025wty,Tamborra:2024fcd,Janka:2025tvf,Burrows:2020qrp}. State-of-the-art hydrodynamical CCSN simulations do not account for the fact that neutrinos change their flavor in the CCSN core~\cite{Mezzacappa:2020oyq}. However, the neutrino density in the CCSN core is so large that the forward scattering of neutrinos on each other  is responsible for triggering dramatic changes in the (anti)neutrino flavor  distribution~\cite{Mirizzi:2015eza,Tamborra:2020cul,Volpe:2023met,Johns:2025mlm,Duan:2010bg,Fiorillo:2026gzy}. 
 As neutrinos decouple from dense matter in the CCSN core, they may undergo collisional, fast, and slow flavor instabilities~\cite{Johns:2021qby,Johns:2022yqy,Chakraborty:2016lct,Chakraborty:2016yeg,Izaguirre:2016gsx,Fiorillo:2025gkw,Fiorillo:2025zio,Fiorillo:2024uki}. Collisional instabilities stem from asymmetries between the neutrino and antineutrino interaction rates~\cite{Johns:2021qby,Padilla-Gay:2022wck,Xiong:2022zqz},  fast flavor instabilities depend on the presence of crossings in the  lepton number  angular distribution of electron and heavy-flavor neutrinos (ELN-XLN distribution, with $\nu_x$ being $\nu_\mu$ or $\nu_\tau$)~\cite{Izaguirre:2016gsx,Morinaga:2021vmc,Fiorillo:2024bzm}, whereas slow instabilities develop thanks to the neutrino-mass splittings and crossings in the $e$--$x$ energy distributions of neutrinos and antineutrinos~\cite{Dasgupta:2021gfs,Raffelt:2007cb,Fiorillo:2024pns,Fiorillo:2025ank}. 
Flavor conversion due to slow collective instabilities has been mainly investigated in the context of the neutrino-bulb model, cf.~Refs.~\cite{Mirizzi:2015eza,Duan:2010bg} for reviews on the topic. Recent work, relaxing the bulb model approximation, suggests that slow flavor conversion may operate efficiently in the CCSN core~\cite{Fiorillo:2025gkw,Shalgar:2024gjt,Shalgar:2025oht,Padilla-Gay:2025tko}. Neutrino propagation and non-forward collisions with the background medium  are expected to further contribute to shaping the quasi-steady state flavor configuration~\cite{Shalgar:2019qwg,Padilla-Gay:2020uxa,Martin:2021xyl,Shalgar:2020wcx,Sigl:2021tmj,Hansen:2020vgm,Sasaki:2021zld,Shalgar:2022rjj,Shalgar:2022lvv}.

The solution of  the neutrino kinetic equations in their full dimensionality (i.e., three dimensions for the spatial coordinates, time, and three dimensions for the neutrino momentum coordinates)  is beyond state-of-the-art computational capabilities. Therefore, several symmetry assumptions are imposed to  make the problem tractable and facilitate its numerical solution. 
In parallel, recent efforts aim to provide semi-analytical criteria that can model the quasi-steady state neutrino flavor configuration reliably~\cite{Nagakura:2022xwe,Xiong:2023vcm,Xiong:2024pue,Zaizen:2023ihz,Zaizen:2022cik,Bhattacharyya:2022eed,Johns:2023jjt,Nagakura:2023jfi,Johns:2024dbe,Goimil-Garcia:2026flm,Goimil-Garcia:2025ozm,Chen:2026oym}. This vigorous activity is motivated by the fact that preliminary attempts to embed flavor conversion in a  parameterized fashion in hydrodynamic simulations suggest that it can favor or hinder the CCSN explosion as well as have an impact on the properties of the compact remnant~\cite{Ehring:2023abs,Ehring:2023lcd,Akaho:2026kff,Gogilashvili:2026kef,Gogilashvili:2026epg,Nagakura:2023mhr,Mori:2025cke}. 
 
Our understanding of the quasi-steady state flavor configuration is further complicated by the fact that the latter  depends on the choice of the boundary conditions assumed in the  solution of the kinetic equations~\cite{Cornelius:2023eop}. Moreover, flavor coherence dynamically  spreads across multiple spatial and angular domains~\cite{Cornelius:2024zsb,Richers:2021xtf,Shalgar:2019qwg,Padilla-Gay:2020uxa,Shalgar:2021oko}. These poorly explored aspects urgently call for the exploration  of  neutrino kinetics in multiple dimensions, and within realistic settings, to deepen our conceptual understanding of the problem and ensure that any prescription of the flavor configuration properly describes the flavor conversion phenomenology.
 
Three dimensional hydrodynamic simulations of CCSNe highlight the development of large-scale asymmetries characterizing  the neutrino emission properties--cf.~Refs.~\cite{Janka:2016fox,Burrows:2006ci} for an overview. 
In particular, Ref.~\cite{Tamborra:2014aua} discovered that, during the stalled-shock phase, the ELN flux exhibits a pronounced, large-scale, dipolar pattern. This phenomenon was dubbed ``Lepton-number Emission Self-sustained Asymmetry'' (LESA). LESA is expected to have an impact on  the neutron-to-proton ratio in the CCSN ejecta, neutron-star kicks~\cite{Tamborra:2014aua,Janka:2016nak,Glas:2018vcs,Janka:2024xbp,Stockinger:2020hse}, and presumably flavor conversion. 

The existence of LESA implies that flavor conversion may lead to direction-dependent flavor configurations, with a potentially rich flavor phenomenology that depends on the  emission direction. The impact of LESA on self-induced flavor conversion was first explored in Ref.~\cite{Chakraborty:2014lsa}, relying on the bulb model approximation:
it was found the large matter density characterizing the region below the shock suppresses the development of any flavor instability. 
However, once the bulb model approximation is relaxed, matter effects on flavor evolution appear to be qualitatively different, especially in the instance of inhomogeneous and non-stationary systems~\cite{Abbar:2015fwa,Dasgupta:2015iia,Dasgupta:2016dbv,Abbar:2017pkh,Shalgar:2025oht}.

In the light of these recent developments, for the first time, we explore neutrino quantum kinetics in a multi-dimensional setup in the presence of LESA. Employing the GPU-accelerated software {\tt FOCACCIA} (the Flavor Oscillation Code with Advection and Collisions for Compact-objects In Astrophysics), we solve the quantum kinetic equations in a half annulus exhibiting an excess of $\nu_e$'s ($\bar\nu_e$'s) in the Northern (Southern) hemisphere. 
We account for charged- and neutral-current collisions of (anti)neutrinos with the matter background around the decoupling region and track flavor evolution in space and time for multi-energy (anti)neutrino ensembles with a non-isotropic polar distribution and two effective azimuthal bins for neutrino momentum. 
We find that collisional instabilities affect the region  of neutrino decoupling in our half annulus, with fast flavor instabilities being prominent in the Southern part of our annulus and slow instabilities everywhere else. We find that flavor conversion is responsible for a $5\%$ to $10\%$ increase of the electron lepton number dipole.

This paper is organized as follows. The neutrino quantum kinetic equations are introduced in Sec.~\ref{sec:eom}. Our simulation setup and modeling of the LESA dipole are outlined in Sec.~\ref{sec:model}.  Section~\ref{sec:classical} presents the properties of neutrinos in our half annulus, in the presence of LESA and in the absence of flavor conversion. In order to identify the types of flavor instabilities triggering flavor conversion, we carry out a  linear stability analysis (LSA) in Sec.~\ref{sec:LSA}; we also investigate the impact of the attenuation of the self-interaction strength (employed to speed up the numerical simulations) and of the azimuthal angle on the development of  flavor instabilities. We present our findings in Sec.~\ref{sec:quantum}.
Section~\ref{sec:LESA} explores the impact of flavor conversion on LESA. Finally, we discuss our results and conclude in Sec.~\ref{sec:conclusions}. 
In addition, Appendix~\ref{sec:collisional_rates} provides a summary of the modeling of the  reactions entering the collision functionals as well as a comparison between the neutrino emission properties that we obtain by solving the Boltzmann equations for neutrinos and antineutrinos and the output of our benchmark CCSN model. Appendix~\ref{sec:convergence} offers additional details on the numerical convergence of the  simulations. Appendix~\ref{sec:1d} compares the quasi-steady state flavor configuration obtained in our half annulus with the one obtained by solving the equations of motion in one spatial dimension (i.e., along selected $\Theta$'s of the annulus); our goal is to assess the role of tangential advection on the quasi-steady state flavor configuration.

\section{Equations of motion}
\label{sec:eom}

\begin{figure}[b]
    \centering
    \includegraphics[width=\linewidth]{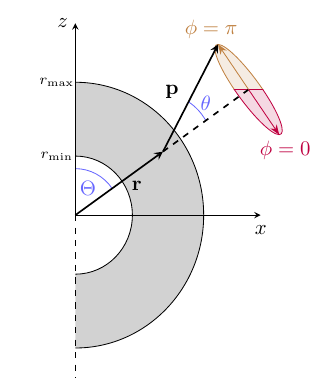}
    \vskip-4pt
    \caption{Schematic representation of the half annulus adopted to solve the neutrino quantum kinetic equations. Each point in the annulus is defined by the radius $r$, varying  between $r_{\mathrm{min}}$ and $r_\mathrm{max}$, and the polar angle  $\Theta \in [0, \pi]$, under the assumption of spatial azimuthal symmetry. The direction of  neutrino momentum ($\mathbf{p}$) is  described by the local polar and azimuthal angles, $\theta$ and  $\phi$, respectively. For simplicity, we only consider two azimuthal bins centered at $\phi=0$ and $\phi=\pi$. 
    }
    \label{fig:fig1}
    \vskip-4pt
\end{figure}

We solve the neutrino kinetic equations in a half annulus. 
As illustrated in Fig.~\ref{fig:fig1}, each point in the annulus is defined by the vector $\mathbf{r}$ which is a function of the radius $r \in [r_\mathrm{min}, r_\mathrm{max}]$ and the polar angle $\Theta$, under the assumption of spatial azimuthal symmetry.
For a fixed point in the annulus, the neutrino momentum $\mathbf{p}=E\mathbf{\hat{\Omega}}(\theta,\phi)$ depends on the neutrino energy ($E\gg m_\nu$) and the polar and azimuthal angles in momentum space ($\theta$ and $\phi$) defined in the local spherical coordinate system. Here, $\mathbf{\hat\Omega}(\theta, \phi)$ is the unit vector in the direction defined by $\theta$ and $\phi$.
For  simplicity, we restrict ourselves to $N_\phi=2$ effective azimuthal angles of neutrino momentum centered at $\phi=0$ and $\phi=\pi$, representing $\cos\phi>0$ and $\cos\phi<0$, respectively.

We consider two neutrino flavors $\nu_e$ and $\nu_x$
and introduce $2\times2$ density matrices,  whose diagonal elements $\varrho_{\alpha\alpha}$ (with $\alpha = e, x$) represent neutrino occupation numbers.
The off-diagonal terms of such matrices instead track coherence between flavors.

The equations of motion for neutrinos and antineutrinos are~\cite{Sigl:1993ctk}:
\begin{eqnarray}\label{eq:qkes}
i \left(\frac{\partial}{\partial t}+ \mathbf{v} \cdot \mathbf{\nabla}_\mathbf{r}\right) \varrho(\mathbf{r},\mathbf{p}) &=&  [\mathcal{H}(\mathbf{r},\mathbf{p}),\varrho(\mathbf{r},\mathbf{p})] + i\mathcal{C}[\varrho]\, ,\\
i \left(\frac{\partial}{\partial t}+ \mathbf{v} \cdot \mathbf{\nabla}_\mathbf{r}\right) \bar\varrho(\mathbf{r},\mathbf{p}) &=&  [\bar{\mathcal{H}}(\mathbf{r},\mathbf{p}),\bar\varrho(\mathbf{r},\mathbf{p})] + i\bar{\mathcal{C}}[\bar\varrho]\, , \label{eq:qkes1}
\end{eqnarray}
where we have kept the time dependence of the density matrices and the Hamiltonians implicit.

In flat spacetime, the advection term  on the left hand side of the equations of motion  is~\cite{Nagakura:2022qko}:
\begin{equation}
\begin{aligned}
    \left(\mathbf{v} \cdot \mathbf{\nabla}_\mathbf{r}\right) \varrho&= \frac{1}{r^2}\frac{\partial}{\partial r}(r^2\cos\theta \varrho) + \frac{1}{r\sin\Theta}\frac{\partial}{\partial \Theta}(\sin\Theta\sin\theta\cos\phi \varrho)\\
    &+ \frac{1}{r\sin\Theta}\frac{\partial}{\partial \Phi}(\sin\theta\sin\phi \varrho) - \frac{1}{r\sin\theta}\frac{\partial}{\partial \theta}(\sin^2\theta \varrho)\\
    &- \frac{1}{r\tan\Theta}\frac{\partial}{\partial \phi}(\sin\theta\sin\phi \varrho)\, .
\end{aligned}
\end{equation}
Under the assumption of spatial azimuthal symmetry and after taking the averages $\varrho_0$ and $\varrho_\pi$ over $\phi$ in the ranges defined by $\cos\phi>0$ and $\cos\phi<0$, respectively, the advection term can be rewritten as
\begin{equation} \label{eq:advection}
  \left(\mathbf{v} \cdot \mathbf{\nabla}_\mathbf{r}\right) \varrho_{0, \pi} =
    \cos\theta\frac{\partial\varrho_{0,\pi}}{\partial r}\pm\frac{2}{\pi}\frac{\sin\theta}{r}\frac{\partial\varrho_{0,\pi}}{\partial\Theta} - \frac{\sin\theta}{r}\frac{\partial\varrho_{0,\pi}}{\partial\theta}\, ,
\end{equation}
where the sign $+$ ($-$) corresponds to $\varrho_0$ ($\varrho_\pi$). In the limit of azimuthal symmetry in momentum space, integrals over $\phi$ are exact because of the $2/\pi$ factor.

The Hamiltonians appearing in the commutator on the right hand side of Eqs.~\eqref{eq:qkes} and \eqref{eq:qkes1} are defined as 
\begin{eqnarray}
\mathcal{H}(\mathbf{r},\mathbf{p}) &=& \mathcal{H}_{\mathrm{vac}}(\mathbf{r},\mathbf{p})+\mathcal{H}_{\nu\nu}(\mathbf{r},\mathbf{p})\, ,\\
\bar{\mathcal{H}}(\mathbf{r},\mathbf{p}) &=& - \mathcal{H}_{\mathrm{vac}}(\mathbf{r},\mathbf{p})+\mathcal{H}_{\nu\nu}(\mathbf{r},\mathbf{p})\, .
\end{eqnarray}
The vacuum term is
\begin{equation}
    \mathcal{H}_{\mathrm{vac}}(\mathbf{r},\mathbf{p}) = \frac{\omega}{2}\begin{pmatrix}
        -\cos2\tilde\vartheta & \sin2\tilde\vartheta\\
        \sin2\tilde\vartheta & \cos2\tilde\vartheta
    \end{pmatrix}\,,
\end{equation}
where $\omega=\pm {\Delta m^2}/{2E}$ is the vacuum  frequency with $\Delta m^2=2.5\times10^{-3}\,\mathrm{eV}^2$ being the neutrino mass squared difference, $\omega >0$ ($\omega <0$) in normal (inverted) mass ordering, 
and $\tilde\vartheta$ is the vacuum mixing angle. Note that,  unless otherwise specified, we adopt natural units and  set $c=\hbar=k_\mathrm B=1$. 
For simplicity, we neglect the matter term in the Hamiltonian and consider an effective mixing angle $\tilde\vartheta=10^{-3}$~\cite{Hannestad:2006nj}; Appendix A of Ref.~\cite{Shalgar:2025oht} shows that the quasi-steady-state flavor configuration is  negligibly affected by such a simplification. 

The self-interaction Hamiltonian is given by 
\begin{equation}
\label{eq:Hnunu}
\begin{aligned}
    \mathcal{H}_{\nu\nu}(\mathbf{r},\mathbf{p}) &= \zeta(r) \sqrt{2}G_\mathrm F\int\frac{\mathrm dE'(E')^2}{(2\pi)^3}\int\mathrm d\Omega'\\
    &\quad\times\left(1-\mathbf{\hat\Omega}\cdot\mathbf{\hat\Omega'}\right)\left[\varrho({\bf{r}}, E', \mathbf{\hat\Omega'}) - \bar\varrho({\bf{r}}, E', \mathbf{\hat\Omega'})\right]\, ,
\end{aligned}
\end{equation}
where $\mathrm d\Omega=\mathrm d\cos\theta\mathrm d\phi$.
Given the computational challenges related to the multi-dimensional solution of the (anti)neutrino equations of motion, we employ the following attenuation function: 
\begin{equation} \label{eq:zeta}
    \zeta(r) =
    \begin{cases}
         \zeta_0 &\mathrm{if~}r\geq r_\mathrm{c}\\
         0 &\mathrm{if~}r<r_\mathrm c
    \end{cases},
\end{equation}
where $\zeta_0 \leq 1$ is a constant factor and $r_\mathrm{c}$ is a cut-off radius.
This choice allows to  turn off $\mathcal{H}_{\nu\nu}$  below $r_\mathrm c$, where  the computational cost is the  highest due to the large (anti)neutrino density but flavor instabilities are not expected.

The collision functionals on the right hand side of Eqs.~\eqref{eq:qkes} and \eqref{eq:qkes1} are modeled following Refs.~\cite{thompson2002,OConnor:2014sgn}.
The collision terms include emission, absorption, and direction-changing reactions:
\begin{eqnarray}
\label{eq:coll_term}
    \mathcal C[\varrho] &=& \mathcal C_\mathrm{em}(\mathbf{r}, E) - \mathcal C_\mathrm{abs}(\mathbf r, E)\odot \varrho({\bf{r}}, E, \mathbf{\hat\Omega}) 
    + \frac{1}{4\pi}\int\mathrm d\Omega' \nonumber \\ &&\mathcal C_\mathrm{dirch}(\mathbf r, E, \mathbf{\hat\Omega},\mathbf{\hat\Omega'})\odot\left[\varrho({\bf{r}}, E, \mathbf{\hat\Omega'}) - \varrho({\bf{r}}, E, \mathbf{\hat\Omega})\right]\ , 
    \nonumber \\
\end{eqnarray}
where $\odot$ designates point-wise matrix multiplication and the collision coefficients are $2\times2$ momentum- and space-dependent matrices in flavor space. The direction-changing reactions, $\mathcal C_\mathrm{dirch} = C_\mathrm{dirch}^{(0)} + C_\mathrm{dirch}^{(1)}(\mathbf{\hat\Omega}\cdot \mathbf{\hat\Omega'})$, depend on  isotropic ($\mathcal C_\mathrm{dirch}^{(0)}$) and anisotropic ($\mathcal C_\mathrm{dirch}^{(1)}$) contributions (see Appendix~\ref{sec:collisional_rates}). 
The emission and absorption rates account for beta reactions for the electron flavors,
as well as pair processes and Bremsstrahlung for all flavors. Such  emission and absorption terms are related via Kirchhoff's law. 
Moreover, we neglect non-elastic neutrino-electron scatterings, due to their subdominant role, especially for electron flavors. 
We refer the interested reader to Appendix~\ref{sec:collisional_rates} for additional information on the modeling of the  reaction channels entering the collision functionals.

We stress that the attenuation of $\mathcal{H}_{\nu\nu}$ (Eq.~\ref{eq:zeta})
comes at the expense of affecting the relative ratios  among the  physical timescales of our problem, namely the self-interaction strength [$\mu^{-1}=(\sqrt2G_\mathrm Fn_\nu)^{-1}$, with $n_\nu$ being the all-flavor neutrino number density], the vacuum oscillation frequency [$\omega^{-1}$], the advection timescale 
[$\tau_\mathrm{adv}\sim(r_\mathrm{max}-r_\mathrm{min})/c$], and the collisional timescale.
For fast flavor conversion,
the attenuation of $H_{\nu\nu}$ has a negligible effect on the  quasi-steady state flavor configuration, as long as $\zeta_0$ remains sufficiently large~\cite{Nagakura:2022kic}.
However, in the instance of slow and collisional flavor instabilities, the attenuation of $\mathcal{H}_{\nu\nu}$ may have a non-trivial impact on the quasi-steady state configuration  and should be used with caution~\cite{Shalgar:2025oht} (see also discussion in Sec.~\ref{sec:quantum}).

\section{Model setup}
\label{sec:model}

In this section, we outline our numerical setup and the model we use to mimic the LESA dipole. We also present the solution of Eqs.~\eqref{eq:qkes} and \eqref{eq:qkes1} in the half annulus, obtained in the absence of flavor conversion. 

\subsection{{\tt FOCACCIA} simulation setup}\label{sec:setup}

We solve Eqs.~\eqref{eq:qkes} and \eqref{eq:qkes1} employing {\tt FOCACCIA}. The latter is a GPU-accelerated solver relying on the \texttt{CUDA.jl} package~\cite{Besard:2019aaa} and  allowing for efficient parallelized evaluation of the equations of motion.
We follow the method of lines (see, e.g., Ref.~\cite{Hairer:1993}) and discretize  Eqs.~\eqref{eq:qkes} and~\eqref{eq:qkes1} over the phase space to obtain a system of $N_\mathrm{tot}=8\times N_r\times N_\Theta \times N_E\times N_{\theta}\times N_{\phi}$
coupled ordinary differential equations, one for each (anti)neutrino density matrix real degree of freedom and each phase-space bin. The system state is thus described,  at each timestep, by a $6$-dimensional GPU array with double precision. Adjacent and next-to-adjacent spatial bins are coupled via the advective term (Eq.~\ref{eq:advection}), while momentum bins are coupled locally through advection and globally through the solid angle integrals in the direction-changing collision term (Eq.~\ref{eq:coll_term}) and the full momentum space integrals in the self-interaction Hamiltonian (Eq.~\ref{eq:Hnunu}). We adopt the second order central differences scheme to compute the derivatives in the spatial polar angle $\Theta$ and neutrino direction $\cos\theta$. We use the second order upwind scheme to compute radial derivatives, limiting the development of spurious numerical instabilities. In this work, we integrate the system over time using the explicit third-order variable-step Adams-Bashforth-Moulton method~\cite{Hairer:1993} and relative tolerances of $10^{-9}$.

The evaluation of the equations of motion in {\tt FOCACCIA} is performed via separate custom GPU kernels in an allocation-free pipeline. We first compute the angular integrals that appear in the direction-changing and self-interaction terms with the composite mid-point rule, concurrently for the $8$ (anti)neutrino density matrix degrees of freedom, energy, polar angle $\Theta$, and radius. Similar to Eq.~\eqref{eq:advection}, integrals involving $\cos\phi$ and $\sin\phi$ are evaluated analytically for $\cos\phi>0$ and $\cos\phi<0$, such that integrals over $\phi$ are exact in the $\phi$-symmetric limit.
The advection and collision terms are then evaluated over the full $6$-dimensional state space. As next step, the energy integrals are reduced for each density matrix degrees of freedom, polar angle, and radius with the same quadrature method, while the commutator in Eqs.~\eqref{eq:qkes} and~\eqref{eq:qkes1} is computed at last, over all phase space bins in parallel.
We verified and validated the {\tt FOCACCIA} solver against independent CPU-based solvers, reproducing the findings of, e.g., Ref.~\cite{Cornelius:2024zsb} for identical input parameters and system setup.

The evolution of the (anti)neutrino field is simulated in a half annulus (cf.~Fig.~\ref{fig:fig1}) with radial boundaries given by $r_\mathrm{min}= 20$~km and 
$r_\mathrm{max}= 45$~km to capture the evolution of neutrino flavor around the decoupling region.
The simulations presented in this paper employ a linear grid in radius, $\Theta$ and $\cos\theta$  with: $N_r=200$,
$N_\Theta=25$,  and $N_{\theta}=100$, unless stated otherwise. The energy grid employs a total of $N_E=15$ bins separated in two uniformly-spaced sub-grids with $12$ bins in the range $[0, 60]$~MeV and $3$ bins for $E\in [60, 100]$~MeV. The state of the system at each timestep is therefore described by $N_\mathrm{tot}=1.2 \times 10^8$ entries. We impose the condition that no neutrinos enter the system from the outer boundary by masking the particles with $\cos\theta<0$ in the last $6$ radial bins.
We provide additional details on the convergence of our simulations in Appendix~\ref{sec:convergence}.

Unless otherwise specified, when flavor conversion is taken into account, we set  $r_\mathrm c=22$~km and $\zeta_0=0.025$  in Eq.~\eqref{eq:Hnunu} as a compromise between the computational speed and the hierarchy between the relevant  timescales entering the problem (cf.~Sec.~\ref{sec:LSA} for a discussion on the impact of the choice of $\zeta_0$ on the growth of the flavor instabilities and Sec.~\ref{sec:quantum} for details on the related non-linear flavor evolution).  We have further verified that our choice of $r_\mathrm{c}$ in Eq.~\eqref{eq:zeta} does not affect the quasi-steady state flavor configuration. 

\subsection{LESA dipole}
In order to explore the impact of flavor conversion on LESA, we construct a toy model. We employ the hydrodynamic and thermodynamic properties of the spherically symmetric CCSN  model with mass of $18.6\,M_\odot$,  without muons, with $1.4\, M_\odot$ gravitational mass, and adopting the SFHo nuclear equation of state~\cite{SNarchive}. We focus on the outputs of the snapshot at the post-bounce time $t_\mathrm{pb}=0.5\,\mathrm{s}$ and compute the (anti)neutrino number density for each flavor solving Eqs.~\eqref{eq:qkes} and~\eqref{eq:qkes1} in the half annulus for $\mathcal{H}= \bar{\mathcal{H}}=0$, i.e., in the absence of flavor conversion.

In order to model the LESA dipolar asymmetry in the half annulus (cf.~Fig.~\ref{fig:fig1}), we  introduce 
a $\Theta$-dependent modulation  in the diagonal components of the emission term computed using the  CCSN simulation outputs: 
\begin{align}\label{eq:toy_dipole}
    \mathcal C_\mathrm{em}^{\alpha\alpha}(r, \Theta, E) = \mathcal C_\mathrm{em}^{\alpha\alpha}(r, E)\left[1 + f(A_{\alpha})\cos\Theta\right]\, ,
\end{align}
where $f(x)=\mathrm{sgn}(x){[1-|x|]}/{[1+|x|]}$.
We assume $A_{\nu_e}=-A_{\bar\nu_e}=0.8$ and $A_{\nu_x}=0.95$ to mimic the amplitude of the maximum neutrino asymmetry achieved at $t_\mathrm{pb}= 0.2$~s in  the three-dimensional, $11.2\, M_\odot$ model presented in Ref.~\cite{Tamborra:2014aua}.
This approach has the advantage of being governed by few free parameters, while  reproducing the key features of a LESA-like dipole in the neutrino field. 

The flavor-dependent number densities of the spherically symmetric model with mass of $18.6\,M_\odot$ are reproduced in  our setup with two spatial dimensions, after solving  Eqs.~\eqref{eq:qkes} and \eqref{eq:qkes1} and extracting the output for $\Theta=\pi/2$, as shown in Appendix~\ref{sec:hydro_comparison}. We compute the neutrino number density for each flavor as follows~\cite{Tamborra:2017ubu}:
\begin{equation} \label{eq:n_density}
    n_{\optbar\nu_\alpha} = \frac{1}{(2\pi)^3}\int_0^{\infty}\mathrm dEE^2\int\mathrm d\Omega\optbar\varrho_{\alpha\alpha}\, .
\end{equation}

\section{Neutrino properties in the absence of flavor conversion} \label{sec:classical}
\begin{figure*}[t]
    \centering
    \includegraphics[width=\linewidth]{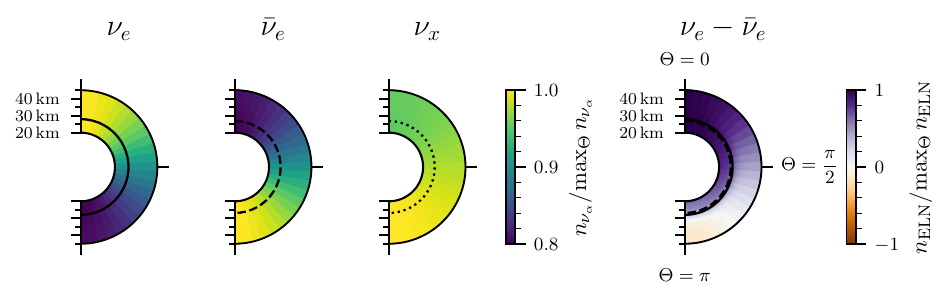}
    \caption{Number density normalized to its maximum over $\Theta$ for $\nu_e$, $\bar\nu_e$, $\nu_x$,  and the ELN ($n_\mathrm{ELN}=n_{\nu_e}-n_{\bar\nu_e}$),  from left to right respectively, in our half annulus and in the absence of flavor conversion. The black solid, dashed, and dotted lines represent the $\nu_e$, $\bar\nu_e$, and $\nu_x$ decoupling surfaces, respectively, to guide the eye. The $\nu_e$--$\bar\nu_e$ normalized dipole moment, defined as in Eq.~\eqref{eq:dipole_moment},
    reaches $d_\mathrm{ELN}=1.3$ after decoupling.
    }
    \label{fig:fig2}
\end{figure*}
We solve Eqs.~\eqref{eq:qkes} and \eqref{eq:qkes1} in the half annulus in Fig.~\ref{fig:fig1}, setting the Hamiltonian of (anti)neutrinos to zero and  assuming $\mathcal{C}_{\rm{em}}^{\alpha\alpha}$ as in Eq.~\eqref{eq:toy_dipole}.  
Figure~\ref{fig:fig2} (three leftmost panels) displays the resulting flavor-dependent  neutrino number density, normalized to its maximum over $\Theta$ to highlight its angular dependence. In agreement with the features described in Refs.~\cite{Tamborra:2014aua,Tamborra:2014hga}, the $\nu_e$ number density is the largest in the direction opposite to that of  $\bar\nu_e$  and $\nu_x$. Moreover,  the $\nu_x$ dipole amplitude is smaller than the electron-flavor one~\cite{Tamborra:2014aua,Tamborra:2014hga}. 
The solid, dashed and dotted lines in Fig.~\ref{fig:fig2} represent the energy-integrated decoupling surfaces. For a given  $\Theta$, we define the decoupling radius as the one where the  flux factor is $j_{\optbar\nu_\alpha}\equiv|\mathbf{F}_{\optbar\nu_\alpha}|/n_{\optbar\nu_\alpha}=1/3$~\cite{Wu:2017drk}; here, the flavor-dependent number flux vector is:
\begin{equation}
    \mathbf{F}_{\optbar\nu_\alpha} = \frac{1}{(2\pi)^3}\int\mathrm dEE^2\int\mathrm d\Omega\mathbf{\hat\Omega}\optbar\varrho_{\alpha\alpha}\, .
\end{equation}
The ELN density (right panel of Fig.~\ref{fig:fig2}) becomes negative outside the decoupling region and in the proximity of  the South pole.

\begin{figure}[htbp]
    \centering
    \includegraphics[width=\linewidth]{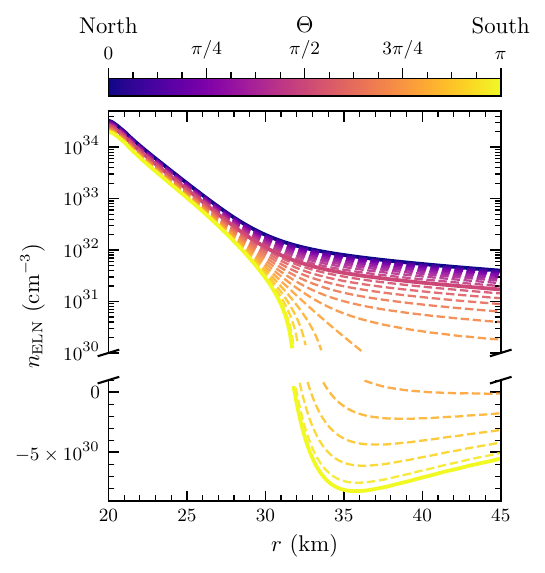}
    \vskip-4pt
    \caption{ELN density as a function of radius and for $0\leq\Theta\leq\pi$, in the absence of flavor conversion.  The North pole ($\Theta=0$), equator ($\Theta=\pi/2$) and South pole ($\Theta=\pi$) are highlighted by the blue, magenta and yellow solid lines, respectively. The dashed lines denote intermediate values of $\Theta$. The ELN is negative, starting at radii beyond neutrino decoupling in the Southern region of the half annulus.
    }
    \label{fig:fig3}
    \vskip-4pt
\end{figure}

Figure~\ref{fig:fig3} shows the ELN density, $n_\mathrm{ELN}=n_{\nu_e} - n_{\bar\nu_e}$, as a function of radius and for $0\leq\Theta\leq\pi$. Consistently with Fig.~\ref{fig:fig2},  the ELN is positive in the Northern hemisphere, whereas $\bar\nu_e$'s dominate in the Southern regions for $r\gtrsim 32$~km.

To quantify the magnitude of the dipolar asymmetry, we introduce the normalized dipole moment:
\begin{equation} \label{eq:dipole_moment}
    d_\mathrm{ELN}(r) = \frac{n_\mathrm{ELN}^{(1)}(r)}{n_\mathrm{ELN}^{(0)}(r)}\, ,
\end{equation}
where the dipole moment is
\begin{equation}
    n_\mathrm{ELN}^{(1)}(r) = \frac{3}{2}\int_0^{\pi}\mathrm d\Theta\sin\Theta\cos\Theta\,n_\mathrm{ELN}(r, \Theta)\, ,
\end{equation}
and the monopole is
\begin{equation}
    n_\mathrm{ELN}^{(0)}(r) = \frac{1}{2}\int_0^{\pi}\mathrm d\Theta\sin\Theta\,n_\mathrm{ELN}(r, \Theta)\, ,
\end{equation}
with the constant prefactors chosen to match the normalization from Refs.~\cite{Tamborra:2014aua, Tamborra:2014hga}.
Before flavor conversion,  $d_\mathrm{ELN}=1.3$ in the outermost region of the half annulus, comparable to the maximal value of $1.4$ found at $280\,\mathrm{ms}$ post-bounce in Ref.~\cite{Tamborra:2014hga}.

\begin{figure*}[htbp]
    \centering
    \includegraphics[width=0.99\textwidth]{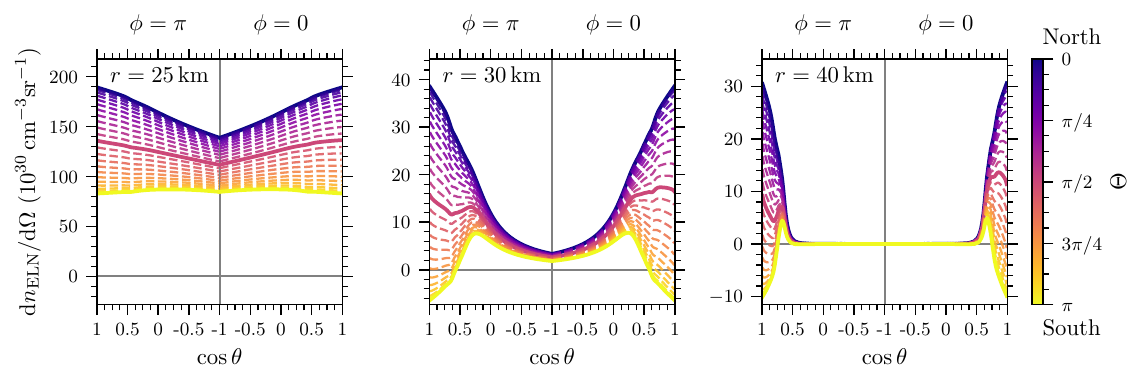}
    \vskip-4pt
    \caption{ELN angular distributions, integrated over energy and extracted at $r= 25$~km  (left), $r= 30$~km (middle), and $r= 40$~km (right), in the absence of flavor conversion. The mirrored left- and right-hand sides of each panel display the distributions for $\phi=\pi$ and $\phi=0$, respectively. The dashed lines in color  represent the angular distributions for  $0\leq\Theta\leq\pi$; the distributions extracted at the North pole ($\Theta=0$), equator ($\Theta=\pi/2$) and South pole ($\Theta=\pi$) are represented by the blue, magenta and yellow solid lines. 
    }
    \label{fig:fig4}
    \vskip-4pt
\end{figure*}
Figure~\ref{fig:fig4}  shows the ELN angular distribution ($\mathrm dn_\mathrm{ELN}/\mathrm d\Omega$) as a function of $\cos\theta$ for both $\phi =0$ and $\phi = \pi$ at three selected radii and for $0\leq\Theta\leq\pi$. We can see that ELN angular crossings occur only in the Southern regions, with $\Theta\gtrsim 0.6 \pi$. Furthermore, the $\phi =0$ and  $\phi = \pi$ directions exhibit differences in their angular distributions, e.g.~for $r \simeq 40$\,km (see right panel of Fig.~\ref{fig:fig4}). At that radius, the ELN crossings at a given $\Theta$ are deeper for $\phi=\pi$ than for $\phi=0$. Intuitively, this difference can be explained by the fact that in the Southern hemisphere ($\Theta>\pi/2$), neutrinos going in the $\phi=\pi$ direction leave a region with higher $\bar\nu_e$ content, contrary to those going in the $\phi=0$ direction (see Fig.~\ref{fig:fig1} and Fig.~\ref{fig:fig2}). These features of the angular distributions highlight multi-dimensional effects that would be missed in an azimuthally symmetric momentum space. 

\begin{figure*}[htbp]
    \centering
    \includegraphics[width=\textwidth]{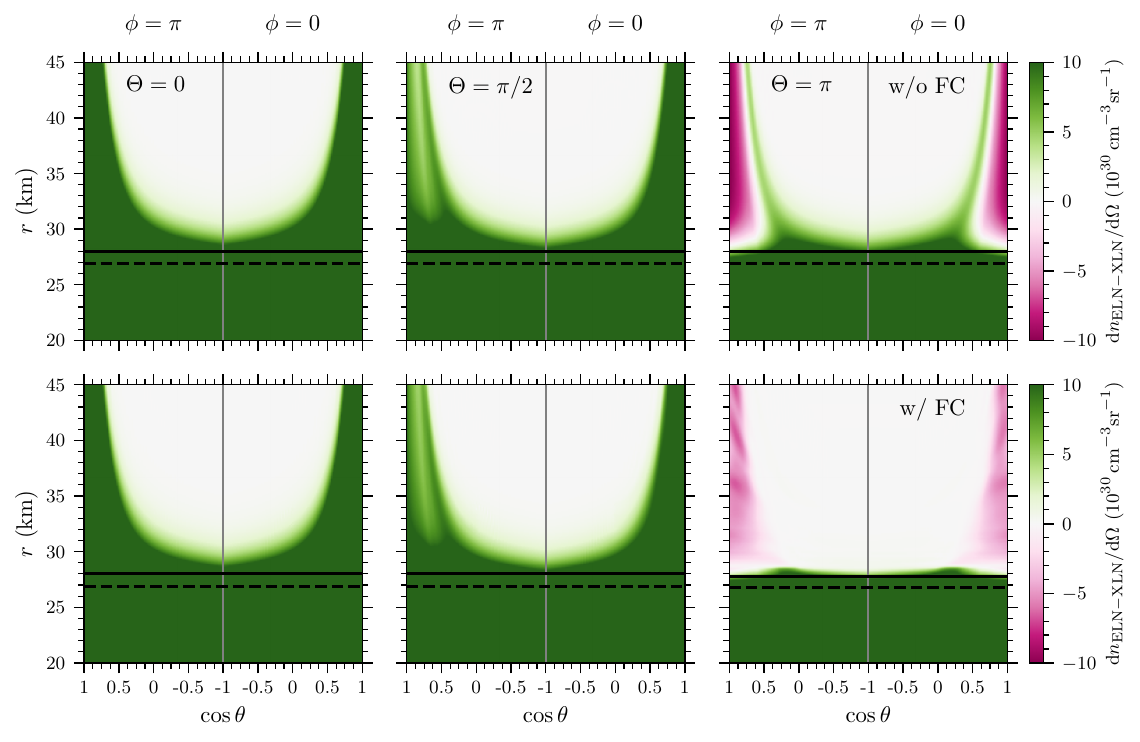}
    \vskip-4pt
    \caption{ELN--XLN angular distributions integrated over energy and in the plane spanned by $\cos\theta$ and radius for $\Theta=0$ (left), $\Theta=\pi/2$ (middle) and $\Theta=\pi$ (right) in the absence of flavor conversion  (top, w/o FC) and averaged between $t_\mathrm{sim}=275\,\mu\mathrm{s}$ and $t_\mathrm{sim}=300\,\mu\mathrm{s}$ when the quasi-steady state flavor configuration has been reached in the presence of flavor conversion (bottom, w/ FC).
    Each panel displays the distributions for $\phi=0$ (right half) and $\phi=\pi$ (left half). The solid (dashed) horizontal line marks the $\nu_e$ ($\bar\nu_e$) decoupling radius. Without flavor conversion, ELN crossings are not present at $\Theta = 0$ and $\pi/2$, but appear at $\Theta=\pi$ (with XLN crossings not being present for any $\Theta$ because $\nu_x = \bar\nu_x$). Fast flavor conversion is responsible for erasing the ELN-XLN crossings at $\Theta=\pi$.
    }
    \label{fig:fig5}
    \vskip-4pt
\end{figure*}

The top panels of Fig.~\ref{fig:fig5} represent the ELN--XLN angular distribution,
$\mathrm dn_{\mathrm{ELN}-\mathrm{XLN}}/\mathrm d\Omega$, in the plane spanned by  $\cos\theta$ and $r$, and for both $\phi=0$ and $\phi=\pi$, at the North pole (left), equator (middle) and South pole (right). Note that, since  $\nu_x$ and $\bar\nu_x$ undergo identical interactions (see Appendix~\ref{sec:collisional_rates}), the angular crossings in the  steady state configuration in the absence of flavor conversion only appear in the $e$ sector (ELN crossings).  Figure~\ref{fig:fig5} shows that ELN--XLN crossings occur for $\Theta=\pi$ at radii slightly above the $\nu_e$ decoupling radius, due to the flavor-dependent forward peakedness of the angular distributions beyond decoupling, but are not present for $\Theta = 0$ and $\pi/2$, where $\nu_e$ systematically dominates.

\begin{figure*}[htbp]
    \centering
    \includegraphics[width=0.99\textwidth]{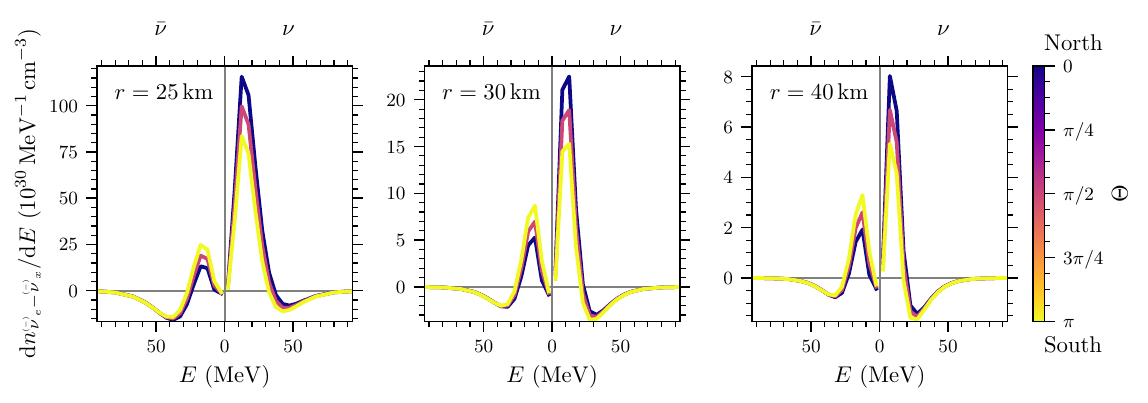}
    \vskip-4pt
    \caption{Spectral energy distribution for $\nu_e$--$\nu_x$ ($\bar\nu_e$--$\bar\nu_x$) integrated over the momentum angles on the right (left) side of each plot,  at $r= 25$~km  (left), $r= 30$~km  (middle) and $r= 40$~km (right) for $\Theta=0$ (blue), $\Theta=\pi/2$ (magenta) and $\Theta=\pi$ (yellow). Unlike ELN crossings that only appear in the Southern region of the annulus, spectral crossings are ubiquitous for both neutrinos and antineutrinos.
    }
    \label{fig:fig6}
    \vskip-4pt
\end{figure*}

Figure~\ref{fig:fig6} shows the energy spectrum, integrated over momentum angles, for $\nu_e$--$\nu_x$ ($\bar\nu_e$--$\bar\nu_x$) on the right (left) hand side of each panel,  at $r = 25$~km, $30$~km and $40$~km, from left to right. The lines of different color highlight the $\Theta$ dependence for $\Theta \in [0, \pi]$. Unlike the ELN angular crossings that are restricted to the Southern region of the annulus, spectral crossings occur for both neutrinos and antineutrinos for any $\Theta$. Consistent with the LESA dipolar asymmetry (cf.~Fig.~\ref{fig:fig2}), the difference between $\nu_e$ and $\nu_x$ is larger in the North than in the South, whereas the opposite holds for antineutrinos. 
As visible in the top panel of Fig.~\ref{fig:fig7}, representing the $e$--$x$ energy spectra in the plane spanned by the neutrino energy and radius, spectral crossings appear at all radii for $r\gtrsim 22$~km.
\section{Linear stability analysis}\label{sec:LSA}

In order to identify  the instability types responsible for triggering flavor conversion, in this  section we present  the results of the LSA performed  employing the  steady-state flavor configuration introduced in Sec.~\ref{sec:classical}.  After deriving the discretized linear eigenvalue problem, we explore its solutions and investigate the impact of the attenuation of the self-interaction Hamiltonian in Eq.~\eqref{eq:Hnunu} on the development of flavor instabilities. We also explore the role of azimuthal asymmetry in momentum space on the growth of flavor instabilities, and how flavor instabilities depend on the neutrino mass ordering. 

\subsection{Discretized eigenvalue problem}

The LSA allows us to find the eigenvalues with the largest imaginary part  by linearizing Eqs.~\eqref{eq:qkes} and \eqref{eq:qkes1}~\cite{Banerjee:2011fj}.
We discretize the system of equations as in Ref.~\cite{Shalgar:2023aca} and focus on the  homogeneous mode. Although inhomogeneous modes could drive the growth of flavor instabilities, here we are interested in identifying the types of flavor instability characterizing our system (see also  Sec.~\ref{sec:quantum}). 

Keeping only the energy and momentum direction dependence explicit, the evolution of  the off-diagonal component of the density matrix is given by 
\begin{equation} \label{eq:LSA_eq}
\begin{split}
    i\frac{\partial \varrho_{ex}(E, \mathbf{\hat\Omega})}{\partial t} &= [\mathcal H_{ee}(E, \mathbf{\hat\Omega}) - \mathcal H_{xx}(E, \mathbf{\hat\Omega})]\varrho_{ex}(E, \mathbf{\hat\Omega})\\
    &\quad- [\varrho_{ee}(E, \mathbf{\hat\Omega}) - \varrho_{xx}(E, \mathbf{\hat\Omega})]\mathcal H_{ex}(E, \mathbf{\hat\Omega})\\
    &\quad- i[\mathcal C_\mathrm{abs}^{ex}(E) + \mathcal C_\mathrm{dirch}^{(0)\,ex}(E)]\varrho_{ex}(E, \mathbf{\hat\Omega})\\
    &\quad+ \frac{i}{4\pi}\int\mathrm d\Omega'\mathcal C_\mathrm{dirch}(E, \mathbf{\hat\Omega}, \mathbf{\hat\Omega'})\varrho_{ex}(E, \mathbf{\hat\Omega}')\, .
\end{split}
\end{equation}
A  similar equation holds  for antineutrinos.
\begin{figure*}[hb!]
    \centering
    \includegraphics[width=0.95\textwidth]{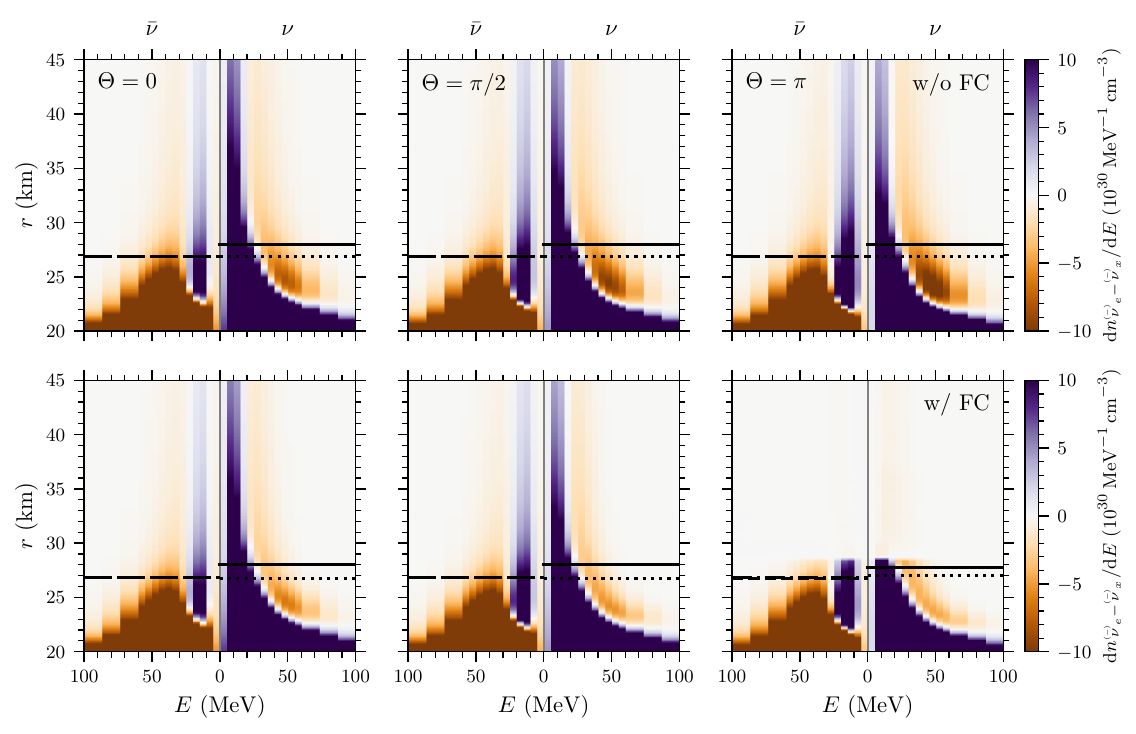}
    \vskip-4pt
    \caption{Spectra of  $\nu_e$--$\nu_x$ ($\bar\nu_e$--$\bar\nu_x$) integrated over angle on the right (left) side of each plot in the plane spanned by the neutrino energy and the radius for $\Theta=0$ (left), $\Theta=\pi/2$ (middle) and $\Theta=\pi$ (right) in the absence of flavor conversion (top, w/o FC) and averaged between $t_\mathrm{sim}=275\,\mu\mathrm{s}$ and $t_\mathrm{sim}=300\,\mu\mathrm{s}$ when a quasi-steady state configuration has been reached in the presence of flavor conversion  (bottom, w/ FC). The black solid and dashed horizontal lines represent the electron neutrino and antineutrino decoupling radii, respectively. Black dotted and dot-dashed lines refer to the heavy-lepton flavor neutrino and antineutrino decoupling radii. The dot-dashed lines overlap with the dashed lines, except in the rightmost lower panel. Spectral crossings appear across the half annulus, even before neutrino decoupling.
    }
    \label{fig:fig7}
    \vskip-4pt
\end{figure*}

We  introduce the following variables:   $\Gamma(E) = \mathcal C_\mathrm{abs}^{ex}(E) + \mathcal C_\mathrm{dirch}^{(0)\,ex}(E) $, $g(E, \mathbf{\hat\Omega}) = \sqrt{2}G_\mathrm F[\varrho_{ee}(E, \mathbf{\hat\Omega}) - \varrho_{xx}(E, \mathbf{\hat\Omega})]$, and $h(E, \mathbf{\hat\Omega}) = \mathcal H_{ee}(E, \mathbf{\hat\Omega}) - \mathcal H_{xx}(E, \mathbf{\hat\Omega}) -i\Gamma$. Moreover, we  drop the terms proportional to $\sin2\tilde\theta$ in $\mathcal H_{ex}$ that do not contribute to the growth rates.
Defining
\begin{eqnarray}
    A_0(E) &=& \int\mathrm{d}\Omega Q(E, \mathbf{\hat{\Omega}})\, ,\\
    \mathbf{A}_1(E) &=& \int\mathrm{d}\Omega\mathbf{\hat{\Omega}}Q(E, \mathbf{\hat{\Omega}})\, ,
\end{eqnarray}
together with their energy integrals
\begin{eqnarray}
    B_0 &=& \int\frac{\mathrm{d}EE^2}{(2\pi)^3} A_0(E)\, ,\\ 
    \mathbf{B}_1 &=& \int\frac{\mathrm{d}EE^2}{(2\pi)^3} \mathbf{A}_1(E)\, ,
\end{eqnarray}
and assuming $\varrho_{ex}(t, E, \mathbf{\hat\Omega}) = Q(E,\mathbf{\hat\Omega})e^{-i\gamma t}$, we obtain the following eigenvalue equation:
\begin{equation}\label{eq:LSA_linear}
\begin{split}
    \gamma Q(E, \mathbf{\hat\Omega}) &= h(E, \mathbf{\hat\Omega})Q(E, \mathbf{\hat\Omega})\\
    &\quad-g(E, \mathbf{\hat\Omega})\left[B_0 - \bar B_0 - \mathbf{\hat\Omega}\cdot(\mathbf{B}_1 - \mathbf{\bar B}_1)\right] \\
    &\quad+ i\frac{\mathcal C_\mathrm{dirch}^{(0)}(E)}{4\pi}A_0(E) + i\frac{\mathcal C_\mathrm{dirch}^{(1)}(E)}{4\pi}\mathbf{\hat\Omega}\cdot\mathbf{A}_1(E)\, ,
\end{split}
\end{equation}
along with an analogous expression for $\bar Q$.

We  solve a $2 N_E N_{\theta}N_\phi\times 2 N_E N_{\theta}N_\phi$ matrix eigenvalue problem, where the relevant integrals over energy ($E$) and momentum direction ($\hat{\mathbf{\Omega}}$) are evaluated via the composite mid-point rule. The LSA growth rate is then the eigenvalue with the largest imaginary part $\kappa=\max_i\mathrm{Im}\gamma_i$ for $1\leq i\leq 2N_EN_{\theta}N_\phi$.

\subsection{Growth rates: self-interaction Hamiltonian without attenuation}\label{sec:no_att_Hnunu}
\begin{figure*}[htbp]
    \centering
    \includegraphics[width=0.75\linewidth]{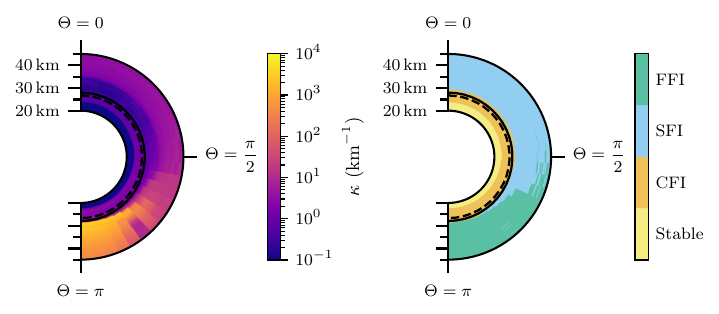}
    \vskip-4pt
    \caption{{\it Left}:  Growth rate of flavor instability in our half annulus for $\mathcal{H}_{\nu\nu}$ with $\zeta_0 =1$  (i.e., no attenuation, see Eq.~\ref{eq:Hnunu}).  {\it Right}:  Flavor instability indicatively classified by its type. In the unstable regions, we distinguish among: fast flavor instabilities (FFI, in green, if $\kappa_{\omega=0,\,\mathcal{C}=\bar{\mathcal{C}}=0}>0$), collisional flavor instabilities (CFI, in orange, if not fast and
    $\kappa_{\omega=0}>\kappa_{\mathcal{C}=\bar{\mathcal{C}}=0}$)
    and slow flavor instabilities (SFI, in blue, otherwise). The stable regions ($\kappa\leq0$) are marked in yellow. In both panels, the $\nu_e$ ($\bar\nu_e$) decoupling region is denoted by the black solid (dashed) line. 
    The growth rates are the largest in the Southern hemisphere, where fast flavor instabilities dominate, collisional instabilities appear before neutrino decoupling, whereas slow instabilities are present everywhere else at radii larger than the decoupling one.
    }
    \label{fig:fig8}
    \vskip-4pt
\end{figure*}

First, we explore the LSA outcome assuming normal mass ordering and a non-attenuated  self-interaction Hamiltonian, i.e., we use $r_\mathrm{c}=r_\mathrm{min}$, $\zeta_0=1$ in  Eq.~\eqref{eq:Hnunu}.
Figure~\ref{fig:fig8} (left panel) shows the growth rate of flavor instability in our simulated half annulus. Unstable modes appear almost everywhere, and only the innermost region of the annulus is stable. 

The right panel of Fig.~\ref{fig:fig8} indicatively classifies the flavor instability according to its dominant type: fast, slow, and collisional. This classification is performed relying on the LSA for different scenarios:
1.~if $\kappa > 0$ for $\omega =0$ and $\mathcal{C}=\bar{\mathcal{C}}=0$, we consider the flavor instability to be fast;
2.~if not fast and if the growth rate assuming $\omega =0$ and $\mathcal{C}\,,\bar{\mathcal{C}}\neq0$ is greater than the growth rate obtained for $\mathcal{C}=\bar{\mathcal{C}}=0$ and $\omega\neq0$, we consider the instability to be collisional;
3.~in all other cases, we classify the instability as slow. Due to the presence of multiple instabilities in the half annulus, such classification should not be considered as a strictly formal one, rather it is meant to guide the understanding of the physics occurring in the non-linear regime.

As expected from the shape of the angular and energy distributions presented in Sec.~\ref{sec:classical}, fast flavor instabilities are absent in the Northern hemisphere of our system, but appear in the Southern region of the annulus, with a growth rate reaching up to $3200\,\mathrm{km}^{-1}$. Slow flavor instabilities are present at any $\Theta$ thanks to the ubiquitous spectral crossings (cf.~Fig.~\ref{fig:fig7}). Interestingly, collisional instabilities are present for $23\,\mathrm{km}\lesssim r\lesssim29\,\mathrm{km}$ (i.e., within the trapped region), where the neutrino chemical potential is $\lesssim 4\,\mathrm{MeV}$. We note that the characteristic growth rate of (non-resonant, homogeneous) collisional instabilities reaches up to $2.5$\,km$^{-1}$,
which is $3$--$4$ orders of magnitude larger than the one reported in Ref.~\cite{Wang:2025vbx}  for their multi-energy setup and another CCSN model.

\subsection{Growth rates: self-interaction Hamiltonian with attenuation}

As discussed in Sec.~\ref{sec:eom}, in order to speed up the numerical solution of Eqs.~\eqref{eq:qkes} and $\eqref{eq:qkes1}$, the self-interaction Hamiltonian is attenuated. We now investigate how this choice affects the growth of flavor instabilities with respect to the non-attenuated case presented in Sec.~\ref{sec:no_att_Hnunu} (see also Sec.~\ref{sec:quantum} and Appendix~\ref{sec:1d} for the impact of the attenuation on the non-linear regime). 

\begin{figure*}[htbp]
    \centering
    \includegraphics[width=0.99\textwidth]{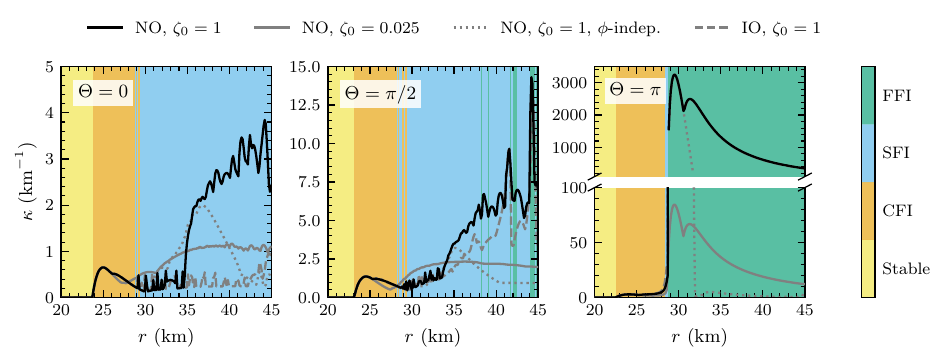}
    \vskip-4pt
    \caption{Growth rate of the flavor instability as a function of radius for $\Theta=0$ (left), $\Theta=\pi/2$ (middle) and $\Theta=\pi$ (right); note that a different $y$-axis scale is used for each panel to improve the figure readability. The different curves represent different scenarios: normal ordering and no attenuation (black solid line), normal ordering and $\zeta_0=0.025$
    (gray solid line), normal ordering, no attenuation, and azimuthal symmetry in momentum space (gray dotted line), and inverted ordering without attenuation (gray dashed line).
    The color shading highlights the instability classification as in the right panel of Fig.~\ref{fig:fig8}. Thin vertical stripes of varying color (orange-blue and blue-green) mark transition regions between different instability types in the left and middle panels.
    The attenuation factor in $\mathcal{H}_{\nu\nu}$ affects the growth rates.
    Moreover, the assumption of azimuthal symmetry in momentum space (relaxed in this work) is responsible for underestimating the growth rates close to $r_{\rm{max}}$. The growth rates of slow flavor instabilities depend on the mass ordering, with the growth rates  for inverted mass ordering being lower than the ones for normal ordering.
    }
    \label{fig:fig9}
    \vskip-4pt
\end{figure*}

Figure~\ref{fig:fig9} shows the growth rate as a function of radius in the case of normal ordering without and with attenuation of the self-interaction Hamiltonian (black vs.~gray solid lines) for $\Theta=0$ (left), $\Theta=\pi/2$ (middle) and $\Theta=\pi$ (right), respectively. As expected, the  growth rates for fast flavor instabilities are rescaled by  $\simeq \zeta_0=0.025$ for the attenuated Hamiltonian when compared to the non-attenuated case. The highest growth rate is  $83\,\mathrm{km}^{-1}$ and is achieved in the Southern region of the annulus, where fast flavor instabilities take place (cf.~right panel for $r>28\,\mathrm{km}$ and the right panel of Fig.~\ref{fig:fig8}). 

The growth rates of slow flavor instabilities are also affected by the attenuation. However, in this case, we observe a dependence on $\Theta$ and radius because of the variation of the neutrino emission properties (in particular the neutrino  average energies) across the annulus. For example, at the equator, the growth rate of slow instabilities is larger in the attenuated setup than in the non-attenuated one for $r\lesssim 34\,\mathrm{km}$, whereas  those at higher radii display an opposite trend.  
The growth rates of collisional flavor instabilities appear to be negligibly affected by the attenuation term; their growth rate is the largest for $\Theta = \pi$. 

Overall, we find that  the regions in the half annulus where   each  type of flavor instability dominates (cf.~right panel of Fig.~\ref{fig:fig8}) remain qualitatively unchanged by the attenuation term (results not shown here).
However, as pointed out in Sec.~\ref{sec:eom}, we stress that the use of an attenuation factor in $\mathcal{H}_{\nu\nu}$ is not expected to affect the quasi-steady state flavor configuration achieved in the presence of fast flavor conversion, but does affect the relative ratio among the characteristic scales entering the quantum kinetic equations when slow and collisional instabilities are dominant. We show in Sec.~\ref{sec:quantum} that the use of such simplification does not affect our conclusions.

\subsection{Dependence of the growth rate on the azimuthal angle in the momentum distribution}

We now explore the role of azimuthally-asymmetric angular distributions in the growth of flavor instabilities. In order to do so, we compare our  system setup with $\zeta_0 = 1$, in normal ordering,  with the analogous azimuthally symmetric one (i.e., the system obtained by performing an average over $\phi$ of the results presented in  Sec.~\ref{sec:classical}) and compute the eigenvalues of Eq.~\eqref{eq:LSA_linear} under the assumption of azimuthal symmetry in momentum space. The corresponding eigenvalue problem has now a size of $2N_EN_{\theta}\times2N_EN_{\theta}$.

The gray dotted curves in Fig.~\ref{fig:fig9} represent the growth rates for the system featuring azimuthal symmetry in momentum space. In the azimuthally symmetric case, flavor instabilities are suppressed in the outermost regions of the half annulus (cf.~gray dotted vs.~black curves), where the angular distributions are  forward peaked, as shown in the right panel of Fig.~\ref{fig:fig4}. Despite the presence of ELN crossings up to $r_\mathrm{max}$, fast flavor instabilities occur in a thin radial range with width of a few kilometers~\footnote{We have performed  one-dimensional simulations of Eqs.~\eqref{eq:qkes} and \eqref{eq:qkes1} at $\Theta=\pi$ accounting for flavor conversion and azimuthal symmetry in momentum space: we find that the growth rate of fast modes is restricted to the narrow peak at $28\,\mathrm{km}\lesssim r\lesssim 32\,\mathrm{km}$ appearing in Fig.~\ref{fig:fig9}  (results not shown here).}.
The azimuthal symmetry in momentum space is also responsible for suppressing the growth rate of slow instabilities beyond $r=35\,\mathrm{km}$, while the growth rate of collisional instabilities is negligibly affected. 
Our findings hint at the importance of the momentum azimuthal angle  in the development of collective flavor instabilities and corroborate earlier work that points out the possibility of spontaneous breaking of the $\phi$ symmetry due to collective flavor oscillations~\cite{Raffelt:2013rqa, Mirizzi:2013rla, Shalgar:2021oko}. 

\subsection{Dependence of the growth rate on the neutrino mass ordering}

The neutrino mass ordering is known to affect the growth of  flavor instabilities, especially slow ones, see e.g.~Refs.~\cite{Hannestad:2006nj,Fogli:2007bk,Duan:2007bt} for pioneering work in this direction within the neutrino bulb-model approximation. Figure~\ref{fig:fig9} shows the LSA results for inverted mass ordering (gray dashed line). We can see that the growth rates of slow flavor instabilities are suppressed in inverted ordering compared to the normal ordering case (cf.~gray dashed vs.~black lines and also  Fig.~\ref{fig:fig8}).

Overall, the scaling of the growth rates as a function of the mass ordering is consistent with the results of Ref.~\cite{Goimil-Garcia:2026flm} that employed a  box with periodic boundary conditions. On the other hand, Ref.~\cite{Chen:2026oym} found larger growth rates for slow instabilities in inverted mass ordering compared to normal ordering. Interestingly, in the  region of the annulus dominated by collisional instabilities, i.e.,~below the first ELN crossings at $\Theta=\pi$, slightly higher growth rates are evident in the inverted mass ordering. The impact of mass ordering on the growth rates when the bulb model approximation is relaxed depends on $\mu$, $\omega$ as well as the ratio between the lepton number and the total particle number~\cite{Goimil-Garcia:2026flm, Fiorillo:2024pns};  it deserves further dedicated work for inhomogeneous systems that is beyond the scope of this paper.

\section{Flavor evolution in multiple spatial and momentum dimensions} \label{sec:quantum}
In this section, we present the solution of  Eqs.~\eqref{eq:qkes} and \eqref{eq:qkes1} in our half annulus, in the presence of flavor conversion and for $\zeta_0 = 0.025$. After exploring the impact of flavor conversion on the  angular and spectral distributions, we investigate the impact of the attenuation of $\mathcal{H}_{\nu\nu}$ on the quasi-steady state flavor configuration.

\subsection{Quasi-steady state flavor configuration}

In order to speed up the numerical computation, we follow the approach adopted in Refs.~\cite{Shalgar:2022rjj,Shalgar:2022lvv} and solve Eqs.~\eqref{eq:qkes} and \eqref{eq:qkes1}  after computing the steady-state flavor configuration in the absence of flavor conversion in the half annulus (see Sec.~\ref{sec:classical}).
The simulation taking into account flavor conversion is evolved until $t_\mathrm{sim}=300\,\mathrm{\mu s}$,
at which point we find that a quasi-steady state has been reached.

A comparison between the numerical solution of  Eqs.~\eqref{eq:qkes} and \eqref{eq:qkes1}---that does not distinguish between homogeneous and inhomogeneous modes---and the LSA results (presented in Sec.~\ref{sec:LSA} for the homogeneous mode) reveals excellent agreement between the LSA growth rate and the linear regime of the numerical solutions (results not shown here); this suggests that the growth of non-homogeneous modes may be sub-dominant in our system.

\begin{figure*}[htbp]
    \centering
    \includegraphics[width=0.99\linewidth]{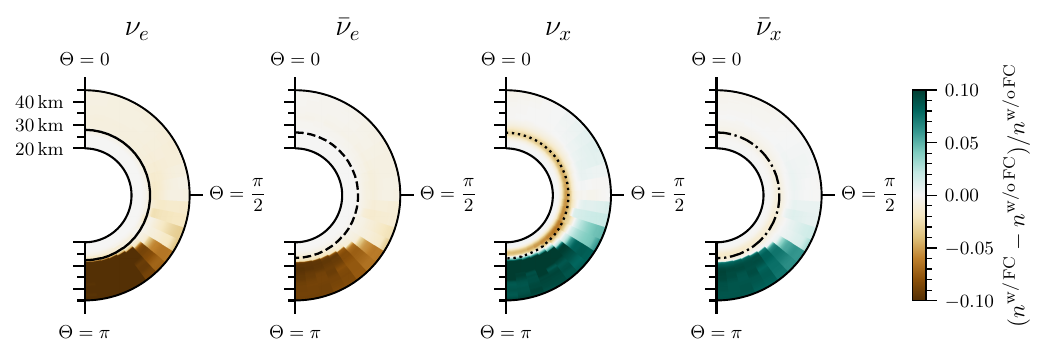}
    \vskip-4pt
    \caption{Relative difference in the neutrino number densities between the solutions with and without flavor conversion for $\nu_e$, $\bar\nu_e$, $\nu_x$, and $\bar\nu_x$, from left to right respectively, averaged between $t_\mathrm{sim}=275\,\mu\mathrm{s}$ and $t_\mathrm{sim}=300\,\mu\mathrm{s}$ when the quasi-steady state flavor configuration has been reached. The solid, dashed, dotted and dot-dashed lines mark the decoupling surfaces of $\nu_e$'s, $\bar\nu_e$'s, $\nu_x$, and $\bar\nu_x$'s, respectively.
    The impact of flavor conversion is the largest for all four neutrino species in the Southern region, where fast flavor instabilities occur, as shown in Fig.~\ref{fig:fig9}. Collisional flavor conversion affects $\nu_x$ the most around their decoupling radius.
    }
    \label{fig:fig10}
    \vskip-4pt
\end{figure*}

Figure~\ref{fig:fig10} shows the relative difference between the neutrino number densities (computed as in Eq.~\eqref{eq:n_density}) obtained in the presence of flavor conversion and when the latter is not taken into account. As hinted at by Fig.~\ref{fig:fig9} (see also Fig.~\ref{fig:fig8} for the case with $\zeta_0 = 1$), the (anti)neutrino ensemble undergoes a rich interplay between flavor conversion, collisions with the background medium, and advection. The Southern region of the half annulus is the most impacted, due to fast flavor instabilities responsible for  a net loss (gain) in number density for the $\optbar{\nu}_e$ ($\optbar{\nu}_x$) species.
The impact of  flavor conversion due to collisional instabilities is mostly visible in the proximity of the decoupling radii. It  induces a net loss of neutrino density in all species, with $\nu_x$ being the most affected. 

As visible from the animation of the time evolution of the flavor configuration in our half annulus enclosed as ancillary file in the arXiv repository and  as Supplemental Material,
flavor conversion develops during the first $80\,\mathrm{\mu s}$ of the simulation, starting at $\Theta=\pi$, where the growth rate of collisional flavor instability is the highest (see the right panel of Fig.~\ref{fig:fig9}). For the $\nu_e$ and $\bar\nu_e$ species, the associated dip in density spreads radially outwards through advection. Moreover, the high efficiency of the charged current processes in that region (see Appendix~\ref{sec:collisional_rates}) facilitates the relaxation of the $\nu_e$ and $\bar\nu_e$ number densities to equilibrium in the innermost region. These two effects combined lead to a quasi-steady state configuration that is relatively close to the one obtained without flavor conversion  for $\nu_e$'s and $\bar\nu_e$'s in the regions of the annulus dominated by collisional instabilities. For the $\nu_x$ and $\bar\nu_x$ species, the net effect of flavor  conversion  is a gain (loss) outside (inside)  their decoupling regions. The excess of $\nu_x$, caused by flavor conversion, spreads radially through advection.
The under-density inside the trapped region  relaxes to the quasi-steady state over $300\,\mathrm{\mu s}$.
The effect of collisional flavor instabilities is limited for antineutrinos compared to neutrinos by the relatively small initial difference in the $\bar\nu_e$ and $\bar\nu_x$ densities (see, e.g., Fig.~\ref{fig:fig15}). Despite the presence of regions dominated by the slow flavor instability identified in the LSA (cf. Fig.~\ref{fig:fig8}), we do not find any noticeable effects due to slow flavor conversion in the quasi-steady state configuration.

Tangential advection has a negligible impact on flavor evolution in our setup, as detailed in Appendix~\ref{sec:1d}. This is likely due to the relatively mild $\Theta$ dependence introduced by  Eq.~\eqref{eq:toy_dipole} compared to the steepness of the radial profiles. 

Neglecting both advection and collisions to better isolate the impact of slow flavor instabilities, we have verified that in a one-dimensional simulation (corresponding to the flavor configuration extracted at $\Theta=0$) with $\zeta_0=0.025$, flavor equilibration in energy is achieved in the antineutrino sector after $t_\mathrm{sim}\gtrsim800\,\mathrm{\mu s}$ in both mass orderings. However, when  advection and collisions are taken into account, the flavor waves induced by collisional instabilities develop starting at $\sim 30$--$60\,\mu$s (cf.~animation), in the proximity of $r\simeq 23\,$km to $29\,$km, corresponding to  the radii associated with  collisional instabilities in Fig.~\ref{fig:fig9}.
The (teal) flavor wave affecting $\nu_x$ and $\bar\nu_x$, after their decoupling, spreads radially outwards through advection and dominates the flavor conversion effects in the regions of slow instabilities, as shown in the animation and Fig.~\ref{fig:fig10}.
An analogous  simulation, run with $\zeta_0=1$, 
$N_r=100$ 
and without advection or collisions, shows that the flavor configuration remains far from flavor equipartition in energy after  $3\,\mathrm{ms}$, despite having reached a quasi-steady state.  Such decreased efficiency of slow flavor conversion for increased self-interaction strength is consistent with the findings of, e.g., Ref.~\cite{Goimil-Garcia:2026flm}.

\subsection{Angular and spectral distributions of neutrinos and antineutrinos}

Figure~\ref{fig:fig5} (bottom) displays the ELN-XLN angular distributions in the plane spanned by $\cos\theta$ and $r$  in the presence of flavor conversion  for selected directions of the half annulus: North (left), equator (center), and South (right). As a consequence of fast flavor conversion, the ELN-XLN crossing is erased at $\Theta=\pi$ beyond $r\gtrsim29\,\mathrm{km}$ (we note that XLN crossings are not present in the absence of flavor conversion because we assume $\nu_x = \bar\nu_x$). 

Figure~\ref{fig:fig7} (bottom) shows the spectral  energy distributions of $\nu_e$--$\nu_x$ or $\bar\nu_e$--$\bar\nu_x$ in the plane spanned by $E$ and $r$ once the quasi-steady state flavor configuration has been reached in the presence of flavor conversion. The effect of collisional instabilities in Fig.~\ref{fig:fig7} shows up between $r\simeq 23\,\mathrm{km}$ and $29\,\mathrm{km}$ in the  spectra of $\nu_e$--$\nu_x$ and $\bar\nu_e$--$\bar\nu_x$ at energies above the spectral crossings for the three selected $\Theta$'s. The spectra are, however, far from reaching equipartition in the presence of collisional instabilities, due to the efficient equilibration with the background medium caused by collisions.

The change of sign of the energy spectrum in the bottom right panel of Fig.~\ref{fig:fig7} compared to the top right panel at energies below the spectral crossing ones suggests slight flavor overconversion. Such an equipartition overshoot is consistent with the findings of  Refs.~\cite{Goimil-Garcia:2025ozm, Goimil-Garcia:2026flm} obtained adopting periodic boxes. Overall, we do not find any complete cancellation of spectral crossings due to slow flavor conversion; with slow flavor conversion leading to minimal flavor changes within our simulation time. This is different from the findings of Ref.~\cite{Chen:2026oym}, but in agreement with Ref.~\cite{Goimil-Garcia:2026flm} that showed that cancellation of spectral crossings is not a general feature of slow flavor conversion, but it depends on the shape of the spectral distributions.

\subsection{Impact of the attenuation of the self-interaction Hamiltonian} \label{sec:zeta_impact}

As discussed in Sec.~\ref{sec:eom}, we use  $\zeta_0=0.025$ in Eq.~\eqref{eq:zeta} to decrease the computational costs of our simulations. We now investigate whether  this choice may affect the quasi-steady state configuration achieved in the presence of slow and collisional instabilities. 

\begin{figure*}[htbp]
    \centering
    \includegraphics[width=0.99\linewidth]{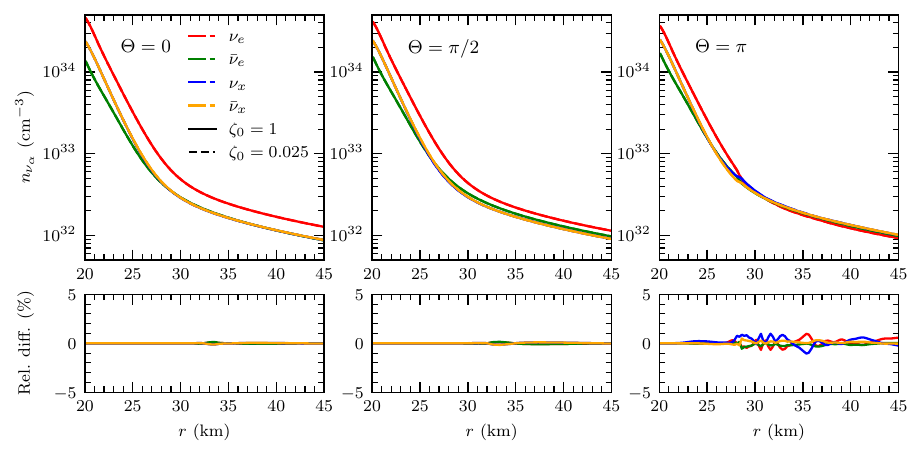}
    \vskip-4pt
    \caption{Quasi-steady state neutrino number densities (top) and relative difference (bottom) between one-dimensional runs using $N_r=100$ with $\zeta_0=1$ (solid) and $\zeta_0=0.025$ (dashed), cf.~Eq.~\eqref{eq:Hnunu}, at $\Theta=0$ (left), $\Theta=\pi/2$ (middle) and $\Theta=\pi$ (right). The results shown are averaged between $275\,\mu\mathrm{s}$ and $300\,\mu\mathrm{s}$, when a quasi-steady state flavor configuration has been achieved. The blue and orange lines overlap in the left and middle panels. The solid and dashed lines in the top panels also overlap. The relative difference between the number densities obtained from the two simulation runs is $\lesssim1\%$; hence,  the use of $\zeta_0 = 0.025$ does not have any impact on our overall conclusions.
    }
    \label{fig:fig11}
    \vskip-4pt
\end{figure*}

Figure~\ref{fig:fig11} shows the radial profiles of the number densities of $\nu_e$, $\bar\nu_e$, $\nu_x$, and $\bar\nu_x$ obtained for $\zeta_0=0.025$ and $\zeta_0=1$ from  one-dimensional simulations corresponding to the hydrodynamic backgrounds of our half annulus extracted at $\Theta=0$ (left), $\Theta=\pi/2$ (middle) and $\Theta=\pi$ (right). 
We can see that the relative difference between the two quasi-steady state configurations is $\lesssim 1\%$, even for $\Theta = \pi$ where flavor conversion has the largest impact in our half annulus (see Fig.~\ref{fig:fig10}). Moreover, the
effect of collisional instabilities remains mostly unchanged by the attenuation factor.

These findings are not generic and depend on the fact that the overall amount of flavor conversion due to slow instabilities is negligible in our half annulus.
We stress that the use of one-dimensional simulations for selected $\Theta$ is justified because, as discussed in Appendix~\ref{sec:1d}, non-radial advection has a negligible impact in our  setup.

\section{Impact of flavor conversion on LESA}\label{sec:LESA}
\begin{figure}
    \centering
    \includegraphics[width=0.9\linewidth]{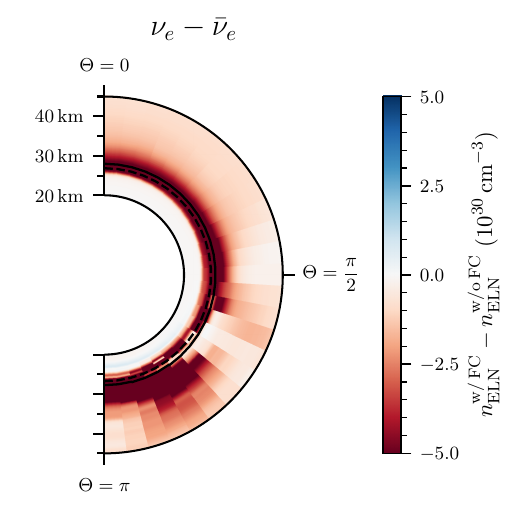}
    \vskip-4pt
    \caption{Absolute difference in the ELN density between the steady-state flavor configurations obtained with (for $\zeta_0=0.025$) and without flavor conversion. When the latter is taken into account, we  average the ELN density between $t_\mathrm{sim}=275\,\mu\mathrm{s}$ and $t_\mathrm{sim}=300\,\mu\mathrm{s}$.
    The solid (dashed) line shows the $\nu_e$ ($\bar\nu_e$) decoupling radius. The ELN density decreases as a result of flavor conversion, both due to fast  instabilities  (Southern region, for radii above  the decoupling radius) and collisional  instabilities  (in the surroundings of the decoupling surfaces for all values of $\Theta$).
    }
    \label{fig:fig12}
    \vskip-4pt
\end{figure}
We now investigate the impact of flavor conversion on the LESA dipole.
Figure~\ref{fig:fig12} shows the absolute difference in the ELN density between our numerical solutions with and without flavor conversion. We observe a loss of ELN density in the regions of fast and collisional flavor conversion (see Fig.~\ref{fig:fig8}, right panel).
As also noted in Ref.~\cite{Xiong:2024tac}, the non-vanishing difference in the $\nu_e$ and $\bar\nu_e$ collisional rates promotes the evolution of the ELN density.
Indeed, the  density deficit induced by flavor conversion is more efficiently replenished for $\nu_e$'s than for $\bar\nu_e$'s, thanks to their correspondingly higher collisional rate (see Appendix~\ref{sec:collisional_rates}).
This implies an absolute loss of $n_{\nu_e}$ smaller than the $n_{\bar\nu_e}$ one, once  the quasi-steady state flavor configuration is achieved, resulting in the observed ELN loss. 

\begin{figure}[t]
    \centering
    \includegraphics[width=\linewidth]{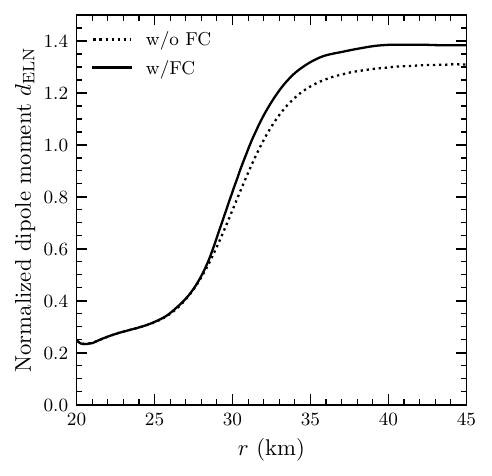}
    \vskip-4pt
    \caption{Normalized ELN dipole moment ($d_\mathrm{ELN}=n_\mathrm{ELN}^{(1)}/n_\mathrm{ELN}^{(0)}$, cf.~Eq.~\ref{eq:dipole_moment}) as a function of radius. The dotted (solid) lines represent the steady-state solution obtained in the absence (in the presence) of flavor conversion, 
    averaged between $t_\mathrm{sim}=275\,\mathrm{\mu s}$ and $t_\mathrm{sim}=300\,\mathrm{\mu s}$.
    The LESA dipole is enhanced by $5\%$ to $10\%$ as a result of flavor conversion for $r\gtrsim 30\,\mathrm{km}$.
    }
    \label{fig:fig13}
    \vskip-4pt
\end{figure}

Figure~\ref{fig:fig13} shows the normalized ELN dipole moment $d_\mathrm{ELN}$ (defined as in Eq.~\eqref{eq:dipole_moment}) as a function of radius.
Flavor conversion, which implies a loss of $\nu_e$ and $\bar\nu_e$ densities, is responsible for an increased asymmetry in $\nu_e$ and a decreased asymmetry for $\bar\nu_e$ (cf.~Fig.~\ref{fig:fig2}).
The ELN loss, being mainly concentrated in the Southern region where fast flavor conversion occurs, leads to an enhancement of $n_\mathrm{ELN}^{(1)}$ and a decrease of $n_\mathrm{ELN}^{(0)}$. Consequently, the normalized dipole moment $d_\mathrm{ELN}=n_\mathrm{ELN}^{(1)}/n_\mathrm{ELN}^{(0)}$ increases by $5\%$ to $10\%$ for $r\gtrsim 30\,\mathrm{km}$.

\section{Discussion and conclusions} \label{sec:conclusions}

Neutrinos play a crucial role in the physics of CCSNe and drive the LESA instability~\cite{Tamborra:2014aua}. The latter is responsible for a dipolar asymmetry in electron neutrino lepton number emission. In this work, for the first time, we explore the  implications of neutrino flavor conversion on LESA, by solving the neutrino quantum kinetic equations in multiple spatial and neutrino-momentum dimensions with the GPU-accelerated software {\tt FOCACCIA}. 

We consider  a half annulus located in the CCSN region of neutrino decoupling, assume spatial azimuthal symmetry and consider flavor-dependent spectral energy distributions for four neutrino species ($\nu_e$, $\bar\nu_e$, $\nu_x$ and $\bar\nu_x$, with $\nu_x = \bar\nu_x$ in the absence of flavor conversion), with  two effective  azimuthal momentum angles and a non-trivial distribution in the polar momentum angle shaped by non-forward collisions. We rely on the snapshot at $t_\mathrm{pb}=0.5\,\rm{s}$ of the spherically symmetric CCSN model with mass of $18.6\,M_\odot$~\cite{SNarchive} to compute the radial dependence of the collisional rates. We construct  the lepton number emission asymmetry by introducing an angular dependence in the neutrino emission rates responsible for creating a global dipolar feature in the neutrino number densities across the annulus. 

We find ELN crossings in the Southern region of the annulus, and $e$--$x$ crossings in the spectral distributions of both neutrinos and antineutrinos across the half annulus.
The linear stability analysis, focusing on the homogeneous mode, predicts the presence of fast, slow and collisional unstable modes in the half annulus, with  fast flavor instabilities occurring in the Southern region of the annulus, collisional instabilities appearing in the proximity of the neutrinospheres, and slow flavor instabilities being present everywhere else at radii larger than the neutrino decoupling one. Interestingly, the lack of azimuthal symmetry in  momentum space fosters the presence of fast and slow flavor instabilities in a broader spatial region with respect to the azimuthally symmetric configuration widely adopted in the literature.
This finding motivates further work towards a  full treatment of both momentum angles in global simulations of quantum neutrino transport. 

Our numerical solutions of the kinetic equations in the presence of flavor conversion show that fast  conversion is responsible for the largest variations in the flavor configuration with respect to the case without flavor conversion, followed by collisional conversion mildly affecting the (anti)neutrino field in the proximity of neutrino decoupling. Despite the existence of spectral crossings for neutrinos and antineutrinos, we find a negligible impact of slow conversion.

As a consequence of flavor conversion, the LESA dipolar asymmetry is enhanced by up to $10\%$.
Although the  mechanism driving LESA remains to be fully understood~\cite{Tamborra:2014aua,Tamborra:2014hga,Walk:2019ier,Walk:2019miz,Glas:2018vcs}, we stress that the growth of LESA is linked to an asymmetry in the proto-neutron star convection as well as a global  accretion asymmetry maintained by anisotropic neutrino heating in the gain layer. Hence, in order to reliably assess the feedback of flavor conversion on the development of LESA, one needs to  account for the physics of flavor conversion dynamically, during the pre-explosion phase of multi-dimensional CCSN simulations. 
Moreover, the global lepton asymmetry due to LESA may be smaller than what considered here, see e.g.~Refs.~\cite{Tamborra:2014aua,Walk:2019miz}. In such a scenario, ELN crossings might be less likely to form, with consequences on the quasi-steady state flavor configuration.

The CCSN model with mass of $18.6\,M_\odot$  used in this work does not account for muon production. The latter impacts the equation of state, leading to faster proto-neutron star contraction and affecting neutrino emission properties as well as the development of ELN-XLN crossings~\cite{Bollig:2017lki,Cornelius:2025nvd}. For consistency with the CCSN model, we do not account for beta processes involving muons in our half annulus; nevertheless,  XLN crossings  may develop if a more accurate treatment of muons were to be considered.

Given the  mostly-radial nature of gradients in our setup, tangential advection plays a minor role compared to radial advection. This indicates that our multi-dimensional investigation of neutrino flavor evolution can be similarly carried out by a series of one-dimensional simulations, each tracking the evolution along a specific polar angle~$\Theta$. 

In conclusion, our findings suggest a very rich flavor conversion phenomenology stemming from the multi-dimensional solution of the quantum kinetic equations  and  the presence of global asymmetries in the properties of the CCSN matter background. Such a diversity in the flavor conversion physics across  emission directions is expected to have  an anisotropic feedback on the CCSN physics that   remains to be explored.

\section*{Acknowledgements}
We thank Marie Cornelius for helpful comments on the manuscript. We also acknowledge Marie Cornelius and Shashank Shalgar for their role in early-stage code prototyping and conceptual work that helped shape the preliminary direction of this project.
This project has received support from the European Union (ERC, ANET, Project No.~101087058).
Views and opinions expressed are those of the authors only and do not necessarily reflect those of the European Union or the European Research Council. Neither the European Union nor the granting authority can be held responsible for them. The Tycho supercomputer hosted at the SCIENCE HPC center at the University of Copenhagen was used to perform the numerical simulations whose results are presented in this paper.

\appendix

\section{Collisional rates}\label{sec:collisional_rates}

In this appendix, we summarize the modeling of the reactions entering  the collision functionals in Eq.~\eqref{eq:coll_term}. These processes are charged-current $\beta$ processes, pair production and annihilation, neutrino-nucleon bremsstrahlung, and neutral current elastic scattering of neutrinos with nucleons. We compute the different reaction channels for our spherically symmetric CCSN model with mass of  $18.6\,M_\odot$ at the post-bounce time $t_{\rm pb}=0.5$~s. We also  discuss the relative importance of the different processes in the radial region  of relevance for the solution of the neutrino kinetic equations between $20$~km and $45$~km. 

The absorption opacity $\chi_\mathrm{abs}^\alpha$ for a given process and neutrino species $\nu_\alpha$ is related to the corresponding emission rate via Kirchhoff's law: $\mathcal C_\mathrm{em}^{\alpha\alpha} = \chi_\mathrm{abs}^\alpha {\varrho_{\alpha\alpha}^\mathrm{FD}}/(1-\varrho_{\alpha\alpha}^\mathrm{FD})$, with $\rho_{\alpha\alpha}^\mathrm{FD}$ denoting the equilibrium  Fermi-Dirac distribution.
Defining
\begin{equation}\label{eq:Cabs_def}
    \mathcal C^{\alpha\alpha}_\mathrm{abs} = \chi_\mathrm{abs}^\alpha/(1-\varrho_{\alpha\alpha}^\mathrm{FD})\, ,
\end{equation}
we get $\mathcal C_\mathrm{em}^{\alpha\alpha} = \varrho_{\alpha\alpha}^\mathrm{FD}\mathcal C_\mathrm{abs}^{\alpha\alpha}$. Hence,  the off-diagonal components of the absorption and direction-changing terms in the collision functionals are given by the arithmetic mean of their diagonal components, while the emission term has vanishing off-diagonal components~\cite{Shalgar:2023aca, Sigl:1993ctk, Richers:2019grc}.

The total collision coefficients appearing in Eq.~\eqref{eq:coll_term} are obtained as the sum of the contributions from the relevant reactions:
\begin{equation}
\begin{split}
    &\mathcal C_\mathrm{abs} = \mathcal C_\mathrm{abs,\,beta} + \mathcal C_\mathrm{abs,\,pair} + \mathcal C_\mathrm{abs,\,brem}\,, \\
    &\mathcal C_\mathrm{em} = \mathcal C_\mathrm{em,\,beta} + \mathcal C_\mathrm{em,\,pair} + \mathcal C_\mathrm{em,\,brem}\,, \\
    &\mathcal C_\mathrm{dirch} = \mathcal C_\mathrm{dirch,\,NC-scat}\,.
\end{split}
\end{equation}

\subsection{Beta processes}
The cross-section for the beta process, $\nu_e+n\rightarrow p + e^-$, for an electron neutrino with energy $E$ is given by~\cite{Bruenn:1985en, thompson2002}:
\begin{equation}
\begin{aligned}
        \sigma_n(E)&=\sigma_0\frac{1+3g_A^2}{4}\left(\frac{E+\Delta_{np}}{m_e}\right)^2 \sqrt{1-\left(\frac{m_e}{E+\Delta_{np}}\right)^2}\\
        \quad&\times\left[1-f_{e^-}^\mathrm{FD}(E+\Delta_{np})\right] W_M(E)\, ,
\end{aligned}
\end{equation}
where \(\sigma_0=4G_\mathrm{F}^2m_e^2/\pi\approx1.7\times10^{-44}\,\mathrm{cm}^2\), \(g_A=-1.254\), \(\Delta_{np}=m_n-m_p=1.293\,\mathrm{MeV}\), \(f_\mathrm{e^-}^\mathrm{FD}\) is the electron Fermi-Dirac distribution and  \(W_M(E)=1-1.01E/m_n\) is the weak magnetism correction~\cite{Vogel:1983hi, thompson2002}. Assuming non-degenerate nucleons, the associated absorption rate is~\cite{thompson2002}:
\begin{equation}
    \mathcal{C}_\mathrm{abs, \mathrm{beta}}^{ee}(E) = \sigma_n(E) \rho_\mathrm{B}N_\mathrm{A}\left[1-\varrho_{ee}^\mathrm{FD}(E)\right]^{-1}\,,
\end{equation}
where $\rho_\mathrm B$ is the baryon density, $N_\mathrm A$ is Avogadro's number, and  $\varrho_{ee}^\mathrm{FD}$ denotes the equilibrium  Fermi-Dirac distribution for $\nu_e$.

For electron antineutrinos absorbed via  $\bar\nu_e+p\rightarrow n+e^+$, the cross-section is~\cite{Bruenn:1985en, thompson2002}:
\begin{equation}
\begin{aligned}
        \sigma_p(E)&=\sigma_0\frac{1+3g_A^2}{4}\left(\frac{E-\Delta_{np}}{m_e}\right)^2\\
        &\quad\times\sqrt{1-\left(\frac{m_e}{E-\Delta_{np}}\right)^2}W_{\bar M}(E)
\end{aligned}
\end{equation}
for $E\geq\Delta_{np}$, where $W_{\bar M}(E)=1-7.1E/m_n$~\cite{Vogel:1983hi,thompson2002} and the Fermi-Dirac occupation number of positrons is neglected. Hence, the absorption coefficient is given by: 
\begin{equation}
    \bar{\mathcal{C}}_\mathrm{abs, \mathrm{beta}}^{ee}(E) = \sigma_p(E) \rho_\mathrm{B}N_\mathrm{A}\left[1-\bar\varrho_{ee}^\mathrm{FD}\right]^{-1}\, .
\end{equation}

\subsection{Pair processes}

We compute the emission collisional coefficients for pair processes as~\cite{Bruenn:1985en}
\begin{equation}\label{eq:pair_cem}
    \mathcal{C}^{\alpha\alpha}_\mathrm{em,\,pair}(E) = \frac{(2\pi)^3}{4\pi E^3}\frac{\mathrm{d}Q_\alpha^\mathrm{pair}}{\mathrm{d}E}\, ,
\end{equation}
where
\begin{equation}\label{eq:pair_emissivity}
    \frac{\mathrm{d}Q_\alpha^\mathrm{pair}}{\mathrm{d}E} = \frac{E^3}{8\pi^4}\int_0^\infty\mathrm d\tilde E\tilde E^2\Phi_{0,\alpha}^\mathrm{p}(E, \tilde E)\left[1-\varrho_{\bar \alpha\bar \alpha}^\mathrm{FD}(\tilde E)\right]
\end{equation}
is the corresponding emissivity~\cite{thompson2002}. The integral sums over the partner (anti)neutrino energies, for which we consider the equilibrium distribution, such that only the zero-th order term ($\Phi_{0,\alpha}^\mathrm{p}$) survives in the Legendre series expansion for the production kernel:
\begin{equation}
\begin{aligned}
    \Phi_{0,\alpha}^\mathrm{p}(E, \tilde E) = \frac{G_\mathrm F^2}{\pi}\int_0^{E+\tilde E}&\mathrm dE_{e^{-}}\left[1-f_{e^+}^\mathrm{FD}(E+\tilde E-E_{e^{-}})\right]\\
    &\times f_{e^{-}}^\mathrm{FD}(E_{e^{-}})\Phi_{0,\alpha}(E, \tilde E, E_{e^{-}})\, , 
\end{aligned}
\end{equation}
with $E_{e^{-}}$ representing the electron energy and
\begin{equation}
\begin{aligned}
    \Phi_{0,\alpha}(E, \tilde E, E_{e^{-}}) &= (C_V^\alpha+C_A^\alpha)^2J_0^{I}(E, \tilde E, E_{e^{-}})\\
    &\quad+ (C_V^\alpha-C_A^\alpha)^2J_0^{II}(E, \tilde E, E_{e^{-}})\, ;
\end{aligned}
\end{equation}
the expressions for $J_0^I$ and $J_0^{II}$ are provided in Ref.~\cite{Bruenn:1985en}, after correcting for the typo in their Eq.~(C68)~\cite{thompson2002}. The vector and axial coefficients are given by $C_V^{e}=\bar C_V^e=1+C_V^x=1/2+2\sin\theta_\mathrm W$ and $C_A^e=-\bar C_A^e=-C_A^x=1/2$, respectively, where $\theta_\mathrm W$ is the Weinberg angle~\cite{Bruenn:1985en}.

Following  Refs.~\cite{OConnor:2014sgn, Betranhandy:2020cdf}, we treat pair processes as effective absorption and emission processes, using the equilibrium distribution on the right-hand side of Eq.~\eqref{eq:pair_emissivity} and applying Kirchhoff's law to obtain $\mathcal C_\mathrm{abs}^{\alpha\alpha}$ from $\mathcal C_\mathrm{em}^{\alpha\alpha}$. This approach has the tendency to overestimate the electron-type neutrino absorption rates in the regions beyond decoupling~\cite{Betranhandy:2025amv}. However, since the  electron neutrino mean free path is dominated by beta processes (cf.~Appendix~\ref{sec:rates}), this approximation is expected to have a small effect on our results.

\subsection{Bremsstrahlung}

Similar to pair processes, we evaluate the Bremsstrahlung emission rates from the corresponding emissivity
\begin{equation}\label{eq:brem_Cem}
    \mathcal{C}^{\alpha\alpha}_\mathrm{em,\,brem}(E) = \frac{(2\pi)^3}{4\pi E^3}\frac{\mathrm{d}Q_\alpha^\mathrm{brem}}{\mathrm{d}E}\, ,
\end{equation}
where~\cite{Burrows:2000xsm}
\begin{equation}
    \frac{\mathrm dQ_\alpha^\mathrm{brem}}{\mathrm dE} \sim 0.234\frac{Q_{\nu\bar\nu}}{T/\mathrm{MeV}}\left(\frac{E}{T}\right)^{2.4}e^{-1.1E/T}\, ,
\end{equation}
with $T$ being the temperature.
Moreover, we have~\cite{Burrows:2000xsm}
\begin{equation}
\begin{aligned}
    Q_{\nu\bar\nu} &\sim2.0778\times10^{30}\upsilon\left(X_n^2+X_p^2+\frac{28}{3}X_nX_p\right)\\
    \quad&\times\left(\frac{\rho_\mathrm B}{10^{14}\,\mathrm{g\,cm}^{-3}}\right)^2\left(\frac{T_{e^-}}{\mathrm{MeV}}\right)^{5.5}\,\mathrm{erg\,cm}^{-3}\,\mathrm{s}^{-1}\,,
\end{aligned}
\end{equation}
where $\upsilon=0.5$ is a correction factor~\footnote{Note that we rely on Ref.~\cite{OConnor:2014sgn} that  considers an additional factor of $2$ with respect to  Refs.~\cite{Burrows:2000xsm, thompson2002}, correcting a typo.}.

\subsection{Neutral current scattering off nucleons}

The cross-sections for neutral current elastic scattering of neutrinos off nucleons are~\cite{Burrows:2004vq}:
\begin{equation}
    \sigma^\mathrm s_n = \frac{\sigma_0}{4}\frac{1+3g_A^2}{4}\left(\frac{E}{m_e}\right)^2 \tilde W_M^{(n)}(E)
\end{equation}
and
\begin{equation}
    \sigma^\mathrm s_p = \frac{\sigma_0}{4}\left[(C_V'-1)^2+3g_A^2(C'_A-1)^2\right]\left(\frac{E}{m_e}\right)^2 \tilde W_M^{(p)}(E)
\end{equation}
for neutrons and protons, respectively, with $C_V'=1/2+2\sin\theta_\mathrm W$ and $C'_A = 1/2$. We include the corresponding weak magnetism corrections which for $N=n,p$ are given by $\tilde W_M^{(N)}(E)=1 + \optbar{C}_{W_M}^{(N)}E/m_N$, with $C_{W_M}^{(n)}= -0.766$, $\bar C_{W_M}^{(n)}= -7.3656$, $C_{W_M}^{(p)}= -1.524$, $\bar C_{W_M}^{(p)}= -6.874$~\cite{Horowitz:2001xf, Burrows:2004vq}. As we do not distinguish the collision coefficients for $\nu_x$ and $\bar\nu_x$, we consider the average of the neutrino and antineutrino corrections for both. The corresponding differential cross-sections read~\cite{thompson2002}:
\begin{equation}
    \frac{\mathrm d\sigma_{n,p}^\mathrm s}{\mathrm d\Omega} = \frac{\sigma_{n,p}^\mathrm s}{4\pi}\left[1+\delta_{n,p}(\mathbf{\hat{\Omega}}\cdot\mathbf{\hat{\Omega}'})\right]\, ,
\end{equation}
where $\mathbf{\hat{\Omega}'}$ ($\mathbf{\hat{\Omega}}$) represents the direction of the incoming (outgoing) neutrino; the anisotropy coefficients are
\begin{equation}
    \delta_n = \frac{1-g_A^2}{1+3g_A^2} \sim-0.1
\end{equation}
and
\begin{equation}
    \delta_p = \frac{(C_V'-1)^2-g_A^2(C_A'-1)^2}{(C_V'-1)^2+3g_A^2(C_A'-1)^2} \sim-0.3\, .
\end{equation}

For all neutrino species, the isotropic part of the differential cross-section contributes to $\mathcal C_\mathrm{dirch}=\mathcal C_\mathrm{dirch}^{(0)}+\mathcal C_\mathrm{dirch}^{(1)}(\mathbf{\hat{\Omega}\cdot\mathbf{\hat{\Omega}'}})$ in Eq.~\eqref{eq:coll_term} as~\cite{thompson2002}:
\begin{equation}
    \mathcal{C}_\mathrm{dirch}^{(0)}(E) = \rho_BN_\mathrm A\left[X_n\sigma^\mathrm s_n(E) + X_p\sigma_p^\mathrm s(E)\right]\, ,
\end{equation}
whereas the anisotropic part contributes as
\begin{equation}
    \mathcal{C}_\mathrm{dirch}^{(1)}(E) = \delta_T(E)\mathcal{C}_\mathrm{dirch}^{(0)}(E)\, ,
\end{equation}
with
\begin{equation}
    \delta_T(E) = \frac{\sum_{i=n,p}\delta_in_i\sigma_i^\mathrm{s}(E)}{\sum_{i=n,p}n_i\sigma_i^\mathrm s(E)}\,,
\end{equation}
where $n_i$ represents the number density for nucleon $i$.

\subsection{Comparison among the reaction rates}\label{sec:rates}

\begin{figure*}
    \centering
    \includegraphics[width=0.75\linewidth]{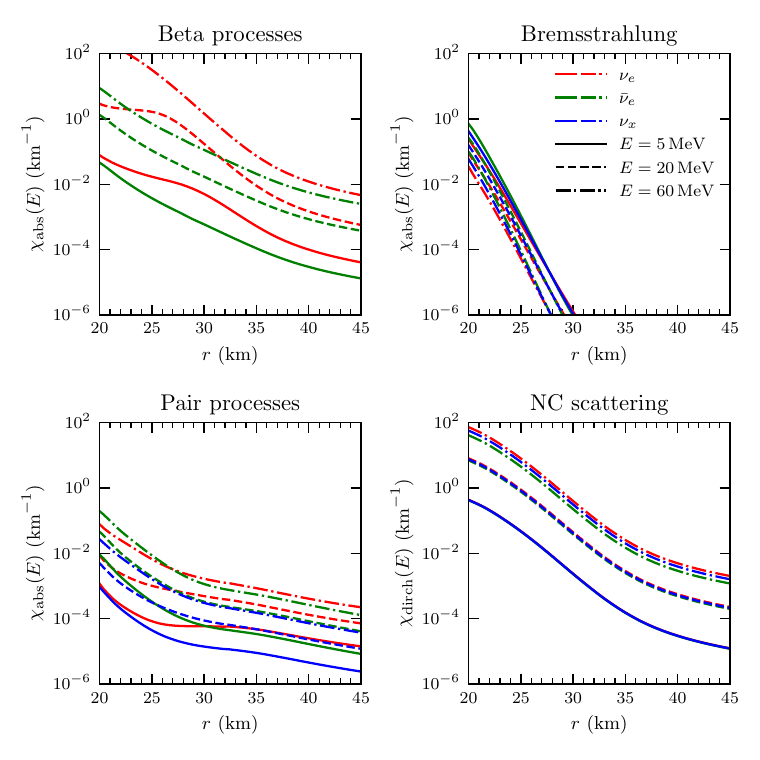}
    \caption{Absorption and scattering opacities  as  functions of radius computed for $E=5\,\mathrm{MeV}$ (solid), $E=20\,\mathrm{MeV}$ (dashed) and $E=60\,\mathrm{MeV}$ (dash-dotted) for $\nu_e$ (in red), $\bar\nu_e$ (in green) and $\nu_x$ (in blue), obtained using the hydrodynamic and thermodynamic properties of the  $18.6\,M_\odot$ model at  $t_{\rm pb}=0.5\,$s (cf.~Sec.~\ref{sec:setup}).
    Beta processes dominate the absorption of $\nu_e$ and $\bar\nu_e$, while Bremsstrahlung is negligible compared to beta and pair processes for $r\gtrsim 30\,\mathrm{km}$. The differences among the different neutrinos flavors for the neutral current scattering rates are due to the weak magnetism correction.
    }
    \label{fig:fig14}
\end{figure*}

Figure~\ref{fig:fig14} displays the absorption and scattering opacities, $\chi_\mathrm{abs}$ (cf.~Eq.~\ref{eq:Cabs_def}) and $\chi_\mathrm{scatt}\equiv \int\mathrm d\Omega'\mathcal C_\mathrm{dirch}(\mathbf{\hat{\Omega}}, \mathbf{\hat{\Omega}'})/4\pi=\mathcal C_\mathrm{dirch}^{(0)}$,
as  functions of radius for selected neutrino energies. The opacities have been computed for our CCSN model with mass of $18.6\,M_\odot$ at  $t_{\rm pb}=0.5$~s (cf.~Sec.~\ref{sec:setup})
for the $\nu_e$, $\bar\nu_e$, and $\nu_x$ species.
We can see that the beta processes dominate the absorption rates of $\nu_e$ and $\bar\nu_e$. As for $\nu_x$, elastic scattering on nucleons dominates, while pair processes become relevant  at larger radii. The differences in the neutral current scattering rates for different flavors are due to the weak magnetism correction.

\subsection{Comparison of our neutrino emission properties with the output of the benchmark supernova model} \label{sec:hydro_comparison}

\begin{figure}[htbp]
    \centering
    \includegraphics[width=0.99\linewidth]{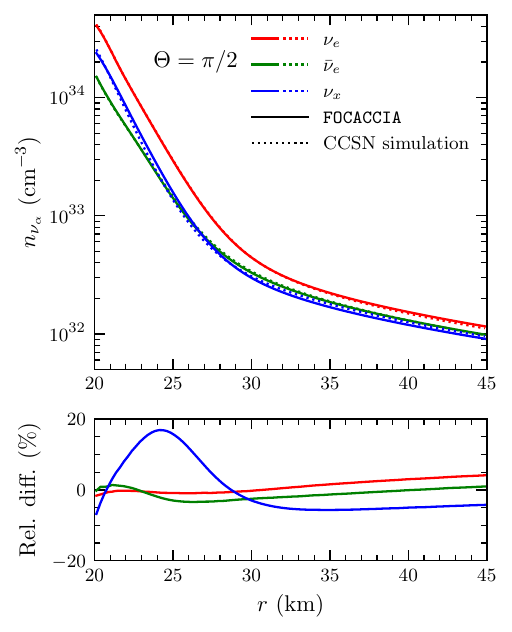}
    \vskip-4pt
    \caption{\textit{Top:} Neutrino density profiles in the absence of flavor conversion computed by solving Eqs.~\eqref{eq:qkes} and \eqref{eq:qkes1} 
    extracted at $\Theta=\pi/2$,  i.e., in the direction along which our neutrino emission properties are expected to be identical to the ones of the reference spherically symmetric CCSN simulation with mass of $18.6\, M_\odot$ at $t_{\rm{pb}}=0.5$~s (dotted lines).
    \textit{Bottom:} Relative difference in densities between our solution of the kinetic equations in the absence of flavor conversion and the reference CCSN simulation. We find sub-$5\%$ (sub-$20\%$) difference for $\optbar\nu_e$ ($\nu_x$).
    }
    \label{fig:fig15}
    \vskip-4pt
\end{figure}

Figure~\ref{fig:fig15} compares the neutrino number densities obtained by solving  Eqs.~\eqref{eq:qkes} and~\eqref{eq:qkes1} in the half annulus with {\tt FOCACCIA} in the absence of flavor conversion (solid lines) with the neutrino density profiles obtained from the  $18.6\,M_\odot$ model. The neutrino number densities are extracted at $\Theta=\pi/2$ in the half annulus,  where the $\Theta$-dependent prefactor introduced in Eq.~\eqref{eq:toy_dipole} to model the LESA dipole is equal to unity.  

As expected, our solution of the neutrino kinetic equations faithfully reproduces the output of the CCSN simulation, with a relative error $\lesssim 5\%$ ($\lesssim 20\%$) for the electron (non-electron) flavors. Such discrepancies are due to the fact that the modeling of the collisional functionals in {\tt FOCACCIA} differs from that employed in VERTEX for our benchmark CCSN simulation~\cite{Rampp:2002bq,Fiorillo:2023frv}. Moreover, the larger discrepancy visible for the non-electron flavors is due to the fact that we neglect the inelastic neutrino-electron scattering channel. Although the agreement between our results and  the CCSN output is extremely good, we stress that our goal in this paper is not to reproduce the CCSN results; we only aim to obtain realistic neutrino properties to assess the role of flavor conversion on LESA. 

\section{Numerical convergence tests}\label{sec:convergence}

We present in this Appendix the results of the convergence tests for our simulations with flavor conversion, both for what concerns the radial and angular resolutions.

\subsection{Radial resolution} \label{sec:rad_conv}
\begin{figure}[htbp]
    \centering
    \includegraphics[width=0.99\linewidth]{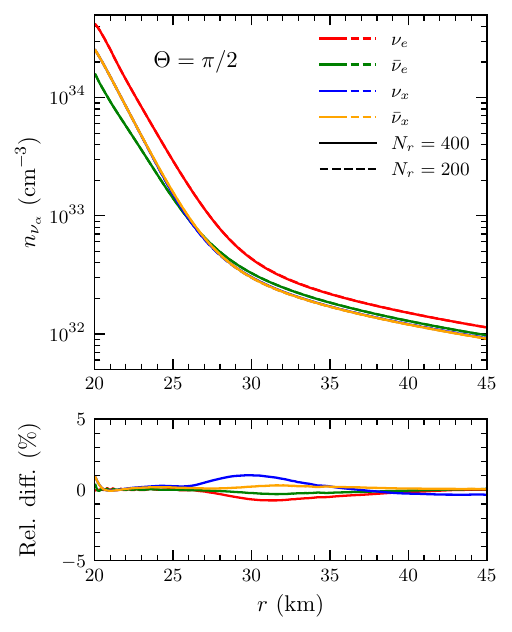}
    \vskip-4pt
    \caption{Quasi-steady state neutrino number densities (top) and relative difference (bottom) between  runs with $N_r=400$ and $N_r=200$, performed employing one spatial dimension (corresponding to $\Theta=\pi/2$ of our half annulus). The results are averaged between $275\,\mu\mathrm{s}$ and $300\,\mu\mathrm{s}$, when a quasi-steady state flavor configuration has been achieved. The relative difference between the neutrino number densities obtained from the two simulations with different radial resolutions is less than $2\%$.
    }
    \label{fig:fig16}
    \vskip-4pt
\end{figure}

Figure~\ref{fig:fig16} shows the radial profiles of the neutrino number densities obtained for a test run of {\tt FOCACCIA} in one spatial dimension. We select  $\Theta = \pi/2$ in our half annulus, include the dependence on $\phi\in\{0,\pi\}$, and employ $N_r = 200$~bins as well as $N_r = 400$ bins for $\zeta_0=0.025$.
In order to minimize transient fluctuations, we  perform a running average between $t_{\rm{sim}}= 275~\mu$s and $t_{\rm{sim}}=300~\mu$s, when the quasi-steady state flavor configuration has been reached. We can see that our default choice of $200$ radial bins for the simulations carried out in two spatial dimensions leads to $\lesssim 2\%$ relative error on the neutrino densities. 

\begin{figure*}[htbp]
    \centering
    \includegraphics[width=0.75\linewidth]{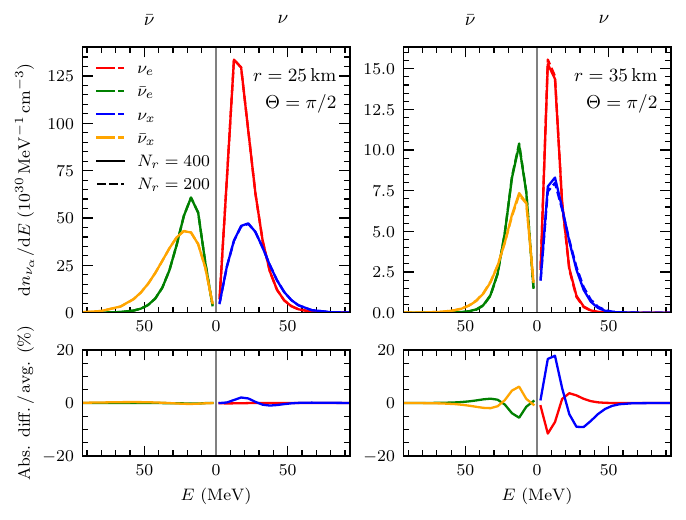}
    \vskip-4pt
    \caption{Analogue of Fig.~\ref{fig:fig16} but representing the quasi-steady state energy spectra (top) and absolute difference (bottom) between our one-dimensional runs with $N_r=400$ and $N_r=200$. The energy spectra are extracted at $r=25\,\mathrm{km}$ (left) and $r=35\,\mathrm{km}$ (right).
    The absolute difference is normalized to the spectral average $\langle\mathrm dn_{\nu_\alpha}/\mathrm dE\rangle$.
    The mirrored $x$-axis on the left-hand side of each panel represents antineutrinos, while the right-hand side represents neutrinos. 
    The larger differences  at $r=35\,\mathrm{km}$ (right panel) with opposite signs cancel each other out after energy integration, resulting in a relative error $\lesssim  4\%$  on the neutrino densities, see Fig.~\ref{fig:fig16}.
    }
    \label{fig:fig17}
    \vskip-4pt
\end{figure*}

\begin{figure*}
    \centering
    \includegraphics[width=0.75\linewidth]{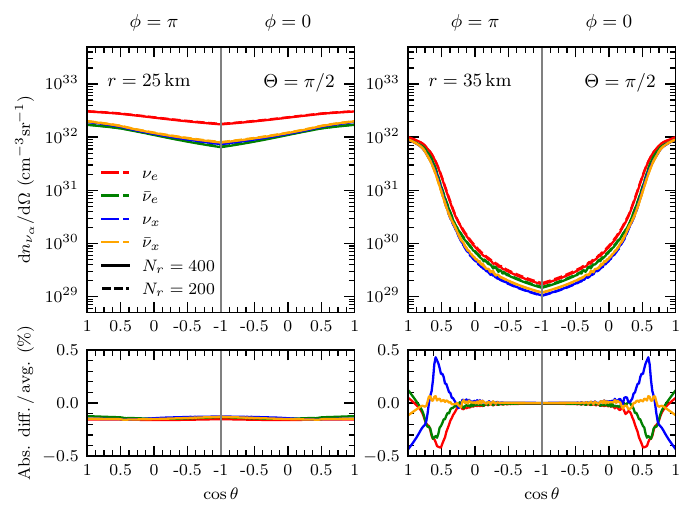}
    \vskip-4pt
    \caption{Analogue of Fig.~\ref{fig:fig16} but representing the quasi-steady state angular distributions (top) and absolute difference (bottom) between our one-dimensional runs with $N_r=400$ and $N_r=200$. The spectra are extracted at $r=25\,\mathrm{km}$ (left) and $r=35\,\mathrm{km}$ (right). The absolute difference is normalized to the average $\langle\mathrm dn_{\nu_\alpha}/\mathrm d\Omega\rangle$. The mirrored $x$-axis on the left-hand side of each panel represents $\phi=\pi$, while the right-hand side represents $\phi=0$. The absolute error is higher at $r=35\,\mathrm{km}$, close to $\cos\theta=0.5$, where it remains smaller than $0.5\%$ with respect to the average.
    }
    \label{fig:fig18}
    \vskip-4pt
\end{figure*}

Figure~\ref{fig:fig17} displays the neutrino spectra (top) obtained for  $N_r = 200$ (dashed line) and  $N_r = 400$ (solid line) for two selected radii. The bottom panels show  the corresponding absolute difference normalized to the spectral average $\langle\mathrm dn_{\nu_\alpha}/\mathrm d E\rangle$--defined as $n_{\nu_\alpha}/(E_\mathrm{max}-E_\mathrm{min})$, with $E_\mathrm{min}=0$ and $E_\mathrm{max}=100\,\mathrm{MeV}$.
The positive and negative contributions to the absolute difference  peak close to $20\%$ of the average at $r=35\,\mathrm{km}$; such an apparently large difference is due to the fact that the spectra peak significantly above $\langle\mathrm dn_{\nu_\alpha}/\mathrm d E\rangle$, however 
the relative error at the spectral peak is 
$\lesssim4\%$. The errors cancel out after energy integration, leading to the small differences in the number densities shown in  Fig.~\ref{fig:fig16}.
Figure~\ref{fig:fig18} shows the effect of the radial resolution on the angular distributions. The absolute difference with respect to the average $\langle\mathrm dn_{\nu_\alpha}/\mathrm d\Omega\rangle = n_{\nu_\alpha}/4\pi$ is $\lesssim 0.5\%$.
Therefore, our choice of $N_r = 200$~bins for the simulations in two spatial dimensions is an excellent compromise between the computational costs of the simulations and the reliability of their results.

\begin{figure}
    \centering
    \includegraphics[width=0.99\linewidth]{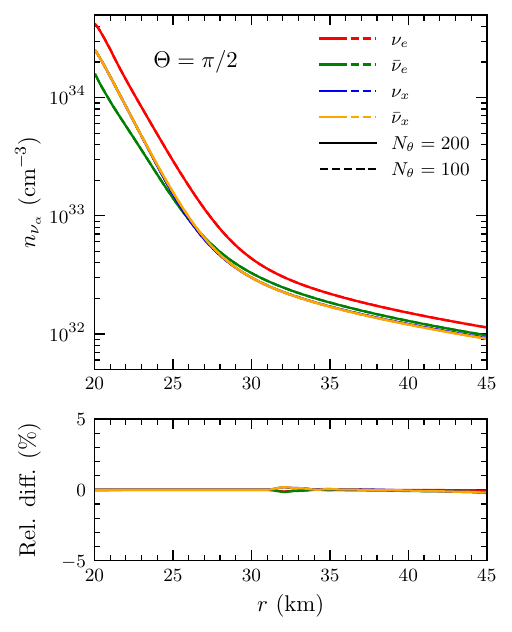}
    \vskip-4pt
    \caption{Analogue of Fig.~\ref{fig:fig16}, but representing the quasi-steady state number densities (top) and their relative difference (bottom) between our  one-dimensional runs with $N_{\theta}=200$ (solid) and $N_{\theta}=100$ (dashed). The dashed and solid lines in the upper panel overlap. The relative difference remains below $0.2\%$.
    }
    \label{fig:fig19}
    \vskip-4pt
\end{figure}
\subsection{Neutrino momentum polar angle resolution}
We compare the output of the solution of the quantum kinetic equations for $N_{\theta} = 100$ and $N_{\theta} = 200$ bins. Figure~\ref{fig:fig19} shows the comparison between the radial profiles of the flavor-dependent number densities. The difference between the two simulation runs with different resolution in $\cos\theta$ is $\lesssim 0.2\%$.

\begin{figure*}[htbp]
    \centering
    \includegraphics[width=0.75\linewidth]{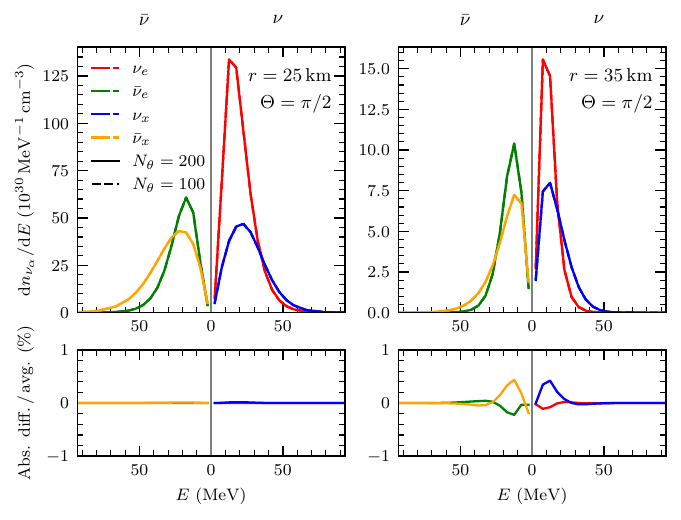}
    \vskip-4pt
    \caption{Analogue of Fig.~\ref{fig:fig17} but obtained for $N_{\theta}=200$ and $N_{\theta}=100$. The dashed and solid lines in the upper panels overlap. The absolute difference is below $1\%$ with respect to the average.
    }
    \label{fig:fig20}
    \vskip-4pt
\end{figure*}

\begin{figure*}[htbp]
    \centering
    \includegraphics[width=0.75\linewidth]{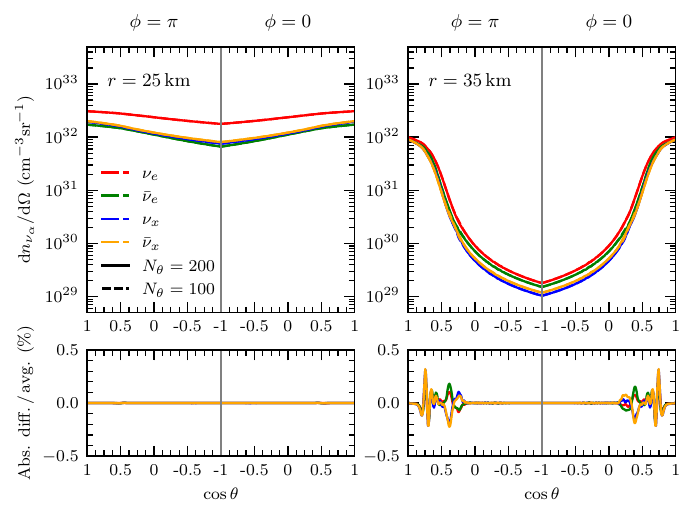}
    \vskip-4pt
    \caption{Analogue of Fig.~\ref{fig:fig18} but obtained $N_{\theta}=200$ and $N_{\theta}=100$. The absolute difference is $\lesssim 0.4\%$ with respect to the average.
    }
    \label{fig:fig21}
    \vskip-4pt
\end{figure*}

Figures~\ref{fig:fig20} and \ref{fig:fig21} show the corresponding energy and angular distributions obtained for $N_{\theta}=200$ and $N_{\theta}=100$. We can see that, in both cases, the differences between the two simulations are sub-$1\%$ of their average.
Hence, we conclude that our choice of $N_{\theta}=100$ for the simulations carried out in two spatial dimensions is an excellent compromise between the computational costs and the reliability of the simulation output.

\section{Comparison between the quasi-steady state flavor configurations obtained in one- and two-spatial dimensions} \label{sec:1d}

We discuss in this appendix how the results of attenuated ($\zeta_0=0.025$)
simulations in one spatial dimension differ from their analogous in two spatial dimensions presented in the main text.
The one-dimensional simulations were obtained by evolving a selected $\Theta$ bin from the two-dimensional  steady-state solution obtained in the absence of flavor conversion, thereby keeping the $\phi\in\{0,\pi\}$ degree of freedom. The simulation in two spatial dimensions accounts for  tangential advection, which is governed by the term proportional to $\mathrm \partial\varrho_{0,\pi}/\partial\Theta$ in Eq.~\eqref{eq:advection}. Hence, this comparison allows us to assess the impact of tangential advection on flavor conversion.

\begin{figure*}[t!]
    \centering
    \includegraphics[width=0.99\linewidth]{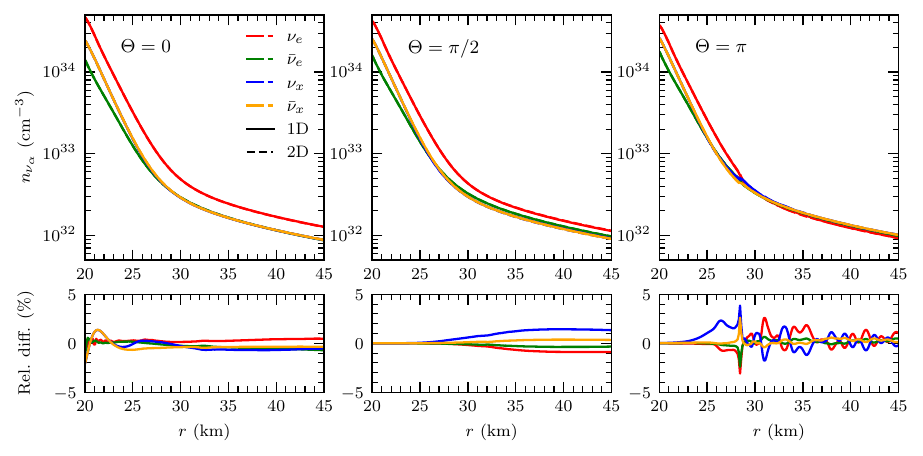}
    \vskip-4pt
    \caption{Quasi-steady state neutrino densities (top) and relative difference (bottom) between our two-dimensional run presented in the main text and one-dimensional simulations at $\Theta=0$ (left), $\Theta=\pi/2$ (middle), and $\Theta=\pi$ (right). Both simulations consider $\zeta_0=0.025$. The results are averaged between $275\,\mu\mathrm{s}$ and $300\,\mu\mathrm{s}$. In the upper panels, the dashed and solid lines overlap. Differences are lowest at
    $\Theta=0$
    and highest at $\Theta=\pi$, where they nonetheless do not exceed $4\%$. The findings suggest that tangential advection has a negligible impact on the quasi-steady state flavor configuration for our system.
    }
    \label{fig:fig22}
    \vskip-4pt
\end{figure*}
Figure~\ref{fig:fig22} shows how the neutrino densities differ between the one-dimensional simulations at $\Theta=0$ (left), $\Theta=\pi/2$ (middle) and $\Theta=\pi$ (right) and the two-dimensional one. The effect of tangential advection is $\lesssim 4\%$ across the entire simulated region and for all neutrino species. The difference is the largest at $\Theta=\pi$, where fast flavor conversion is active. 

The smallness of the differences in the results highlights that advection in the non-radial direction is negligible in our setup.
This result is not a generic one, but depends on the fact that our system exhibits a relatively mild $\Theta$ dependence (see Eq.~\ref{eq:toy_dipole}) with respect to the steepness of the radial profiles. On the other hand, Ref.~\cite{Cornelius:2024zsb} highlights the importance of  tangential advection in setups where smaller scale fluctuations in $\Theta$ are present. Our findings suggest that one may approximate a multi-dimensional system with a series of one-dimensional ones in similar cases. We have checked (results not shown here) that the combination of $7$ one-dimensional simulations, each corresponding to a selected $\Theta$, with $N_r=100$ and $\zeta_0=1$ yields a similar flavor conversion picture and, therefore, an increase of $\simeq10\%$ in the LESA dipole comparable to the two-dimensional case with $\zeta_0=0.025$ discussed in Sec.~\ref{sec:LESA}.

\bibliography{bibliography}

\begin{thebibliography}{108}%
\makeatletter
\providecommand \@ifxundefined [1]{%
 \@ifx{#1\undefined}
}%
\providecommand \@ifnum [1]{%
 \ifnum #1\expandafter \@firstoftwo
 \else \expandafter \@secondoftwo
 \fi
}%
\providecommand \@ifx [1]{%
 \ifx #1\expandafter \@firstoftwo
 \else \expandafter \@secondoftwo
 \fi
}%
\providecommand \natexlab [1]{#1}%
\providecommand \enquote  [1]{``#1''}%
\providecommand \bibnamefont  [1]{#1}%
\providecommand \bibfnamefont [1]{#1}%
\providecommand \citenamefont [1]{#1}%
\providecommand \href@noop [0]{\@secondoftwo}%
\providecommand \href [0]{\begingroup \@sanitize@url \@href}%
\providecommand \@href[1]{\@@startlink{#1}\@@href}%
\providecommand \@@href[1]{\endgroup#1\@@endlink}%
\providecommand \@sanitize@url [0]{\catcode `\\12\catcode `\$12\catcode
  `\&12\catcode `\#12\catcode `\^12\catcode `\_12\catcode `\%12\relax}%
\providecommand \@@startlink[1]{}%
\providecommand \@@endlink[0]{}%
\providecommand \url  [0]{\begingroup\@sanitize@url \@url }%
\providecommand \@url [1]{\endgroup\@href {#1}{\urlprefix }}%
\providecommand \urlprefix  [0]{URL }%
\providecommand \Eprint [0]{\href }%
\providecommand \doibase [0]{https://doi.org/}%
\providecommand \selectlanguage [0]{\@gobble}%
\providecommand \bibinfo  [0]{\@secondoftwo}%
\providecommand \bibfield  [0]{\@secondoftwo}%
\providecommand \translation [1]{[#1]}%
\providecommand \BibitemOpen [0]{}%
\providecommand \bibitemStop [0]{}%
\providecommand \bibitemNoStop [0]{.\EOS\space}%
\providecommand \EOS [0]{\spacefactor3000\relax}%
\providecommand \BibitemShut  [1]{\csname bibitem#1\endcsname}%
\let\auto@bib@innerbib\@empty
\bibitem [{\citenamefont {Mirizzi}\ \emph {et~al.}(2016)\citenamefont
  {Mirizzi}, \citenamefont {Tamborra}, \citenamefont {Janka}, \citenamefont
  {Saviano}, \citenamefont {Scholberg}, \citenamefont {Bollig}, \citenamefont
  {H{\"u}depohl},\ and\ \citenamefont {Chakraborty}}]{Mirizzi:2015eza}%
  \BibitemOpen
  \bibfield  {author} {\bibinfo {author} {\bibfnamefont {A.}~\bibnamefont
  {Mirizzi}}, \bibinfo {author} {\bibfnamefont {I.}~\bibnamefont {Tamborra}},
  \bibinfo {author} {\bibfnamefont {H.-T.}\ \bibnamefont {Janka}}, \bibinfo
  {author} {\bibfnamefont {N.}~\bibnamefont {Saviano}}, \bibinfo {author}
  {\bibfnamefont {K.}~\bibnamefont {Scholberg}}, \bibinfo {author}
  {\bibfnamefont {R.}~\bibnamefont {Bollig}}, \bibinfo {author} {\bibfnamefont
  {L.}~\bibnamefont {H{\"u}depohl}},\ and\ \bibinfo {author} {\bibfnamefont
  {S.}~\bibnamefont {Chakraborty}},\ }\bibfield  {title} {\bibinfo {title}
  {{Supernova Neutrinos: Production, Oscillations and Detection}},\ }\href
  {https://doi.org/10.1393/ncr/i2016-10120-8} {\bibfield  {journal} {\bibinfo
  {journal} {Riv. Nuovo Cim.}\ }\textbf {\bibinfo {volume} {39}},\ \bibinfo
  {pages} {1} (\bibinfo {year} {2016})},\ \Eprint
  {https://arxiv.org/abs/1508.00785} {arXiv:1508.00785 [astro-ph.HE]}
  \BibitemShut {NoStop}%
\bibitem [{\citenamefont {Raffelt}\ \emph {et~al.}(2026)\citenamefont
  {Raffelt}, \citenamefont {Janka},\ and\ \citenamefont
  {Fiorillo}}]{Raffelt:2025wty}%
  \BibitemOpen
  \bibfield  {author} {\bibinfo {author} {\bibfnamefont {G.~G.}\ \bibnamefont
  {Raffelt}}, \bibinfo {author} {\bibfnamefont {H.-T.}\ \bibnamefont {Janka}},\
  and\ \bibinfo {author} {\bibfnamefont {D.~F.~G.}\ \bibnamefont {Fiorillo}},\
  }\bibfield  {title} {\bibinfo {title} {{Neutrinos from core-collapse
  supernovae}},\ }\bibfield  {journal} {\bibinfo  {journal} {Encycl. Part.
  Phys}\ }\textbf {\bibinfo {volume} {3}},\ \href
  {https://doi.org/10.1016/B978-0-443-26598-3.00065-1}
  {10.1016/B978-0-443-26598-3.00065-1} (\bibinfo {year} {2026}),\ \Eprint
  {https://arxiv.org/abs/2509.16306} {arXiv:2509.16306 [astro-ph.HE]}
  \BibitemShut {NoStop}%
\bibitem [{\citenamefont {Tamborra}(2025)}]{Tamborra:2024fcd}%
  \BibitemOpen
  \bibfield  {author} {\bibinfo {author} {\bibfnamefont {I.}~\bibnamefont
  {Tamborra}},\ }\bibfield  {title} {\bibinfo {title} {{Neutrinos from
  explosive transients at the dawn of multi-messenger astronomy}},\ }\href
  {https://doi.org/10.1038/s42254-025-00828-2} {\bibfield  {journal} {\bibinfo
  {journal} {Nature Rev. Phys.}\ }\textbf {\bibinfo {volume} {7}},\ \bibinfo
  {pages} {285} (\bibinfo {year} {2025})},\ \Eprint
  {https://arxiv.org/abs/2412.09699} {arXiv:2412.09699 [astro-ph.HE]}
  \BibitemShut {NoStop}%
\bibitem [{\citenamefont {Janka}(2025)}]{Janka:2025tvf}%
  \BibitemOpen
  \bibfield  {author} {\bibinfo {author} {\bibfnamefont {H.-T.}\ \bibnamefont
  {Janka}},\ }\bibfield  {title} {\bibinfo {title} {{Long-Term Multidimensional
  Models of Core-Collapse Supernovae: Progress and Challenges}},\ }\href
  {https://doi.org/10.1146/annurev-nucl-121423-100945} {\bibfield  {journal}
  {\bibinfo  {journal} {Ann. Rev. Nucl. Part. Sci.}\ }\textbf {\bibinfo
  {volume} {75}},\ \bibinfo {pages} {425} (\bibinfo {year} {2025})},\ \Eprint
  {https://arxiv.org/abs/2502.14836} {arXiv:2502.14836 [astro-ph.HE]}
  \BibitemShut {NoStop}%
\bibitem [{\citenamefont {Burrows}\ and\ \citenamefont
  {Vartanyan}(2021)}]{Burrows:2020qrp}%
  \BibitemOpen
  \bibfield  {author} {\bibinfo {author} {\bibfnamefont {A.}~\bibnamefont
  {Burrows}}\ and\ \bibinfo {author} {\bibfnamefont {D.}~\bibnamefont
  {Vartanyan}},\ }\bibfield  {title} {\bibinfo {title} {{Core-Collapse
  Supernova Explosion Theory}},\ }\href
  {https://doi.org/10.1038/s41586-020-03059-w} {\bibfield  {journal} {\bibinfo
  {journal} {Nature}\ }\textbf {\bibinfo {volume} {589}},\ \bibinfo {pages}
  {29} (\bibinfo {year} {2021})},\ \Eprint {https://arxiv.org/abs/2009.14157}
  {arXiv:2009.14157 [astro-ph.SR]} \BibitemShut {NoStop}%
\bibitem [{\citenamefont {Mezzacappa}\ \emph {et~al.}(2020)\citenamefont
  {Mezzacappa}, \citenamefont {Endeve}, \citenamefont {Bronson~Messer},\ and\
  \citenamefont {Bruenn}}]{Mezzacappa:2020oyq}%
  \BibitemOpen
  \bibfield  {author} {\bibinfo {author} {\bibfnamefont {A.}~\bibnamefont
  {Mezzacappa}}, \bibinfo {author} {\bibfnamefont {E.}~\bibnamefont {Endeve}},
  \bibinfo {author} {\bibfnamefont {O.~E.}\ \bibnamefont {Bronson~Messer}},\
  and\ \bibinfo {author} {\bibfnamefont {S.~W.}\ \bibnamefont {Bruenn}},\
  }\bibfield  {title} {\bibinfo {title} {{Physical, numerical, and
  computational challenges of modeling neutrino transport in core-collapse
  supernovae}},\ }\href {https://doi.org/10.1007/s41115-020-00010-8} {\bibfield
   {journal} {\bibinfo  {journal} {Liv. Rev. Comput. Astrophys.}\ }\textbf
  {\bibinfo {volume} {6}},\ \bibinfo {pages} {4} (\bibinfo {year} {2020})},\
  \Eprint {https://arxiv.org/abs/2010.09013} {arXiv:2010.09013 [astro-ph.HE]}
  \BibitemShut {NoStop}%
\bibitem [{\citenamefont {Tamborra}\ and\ \citenamefont
  {Shalgar}(2021)}]{Tamborra:2020cul}%
  \BibitemOpen
  \bibfield  {author} {\bibinfo {author} {\bibfnamefont {I.}~\bibnamefont
  {Tamborra}}\ and\ \bibinfo {author} {\bibfnamefont {S.}~\bibnamefont
  {Shalgar}},\ }\bibfield  {title} {\bibinfo {title} {{New Developments in
  Flavor Evolution of a Dense Neutrino Gas}},\ }\href
  {https://doi.org/10.1146/annurev-nucl-102920-050505} {\bibfield  {journal}
  {\bibinfo  {journal} {Ann. Rev. Nucl. Part. Sci.}\ }\textbf {\bibinfo
  {volume} {71}},\ \bibinfo {pages} {165} (\bibinfo {year} {2021})},\ \Eprint
  {https://arxiv.org/abs/2011.01948} {arXiv:2011.01948 [astro-ph.HE]}
  \BibitemShut {NoStop}%
\bibitem [{\citenamefont {Volpe}(2024)}]{Volpe:2023met}%
  \BibitemOpen
  \bibfield  {author} {\bibinfo {author} {\bibfnamefont {M.~C.}\ \bibnamefont
  {Volpe}},\ }\bibfield  {title} {\bibinfo {title} {{Neutrinos from dense
  environments: Flavor mechanisms, theoretical approaches, observations, and
  new directions}},\ }\href {https://doi.org/10.1103/RevModPhys.96.025004}
  {\bibfield  {journal} {\bibinfo  {journal} {Rev. Mod. Phys.}\ }\textbf
  {\bibinfo {volume} {96}},\ \bibinfo {pages} {025004} (\bibinfo {year}
  {2024})},\ \Eprint {https://arxiv.org/abs/2301.11814} {arXiv:2301.11814
  [hep-ph]} \BibitemShut {NoStop}%
\bibitem [{\citenamefont {Johns}\ \emph {et~al.}(2025)\citenamefont {Johns},
  \citenamefont {Richers},\ and\ \citenamefont {Wu}}]{Johns:2025mlm}%
  \BibitemOpen
  \bibfield  {author} {\bibinfo {author} {\bibfnamefont {L.}~\bibnamefont
  {Johns}}, \bibinfo {author} {\bibfnamefont {S.}~\bibnamefont {Richers}},\
  and\ \bibinfo {author} {\bibfnamefont {M.-R.}\ \bibnamefont {Wu}},\
  }\bibfield  {title} {\bibinfo {title} {{Neutrino Oscillations in
  Core-Collapse Supernovae and Neutron Star Mergers}},\ }\href
  {https://doi.org/10.1146/annurev-nucl-121423-100853} {\bibfield  {journal}
  {\bibinfo  {journal} {Ann. Rev. Nucl. Part. Sci.}\ }\textbf {\bibinfo
  {volume} {75}},\ \bibinfo {pages} {399} (\bibinfo {year} {2025})},\ \Eprint
  {https://arxiv.org/abs/2503.05959} {arXiv:2503.05959 [astro-ph.HE]}
  \BibitemShut {NoStop}%
\bibitem [{\citenamefont {Duan}\ \emph {et~al.}(2010)\citenamefont {Duan},
  \citenamefont {Fuller},\ and\ \citenamefont {Qian}}]{Duan:2010bg}%
  \BibitemOpen
  \bibfield  {author} {\bibinfo {author} {\bibfnamefont {H.}~\bibnamefont
  {Duan}}, \bibinfo {author} {\bibfnamefont {G.~M.}\ \bibnamefont {Fuller}},\
  and\ \bibinfo {author} {\bibfnamefont {Y.-Z.}\ \bibnamefont {Qian}},\
  }\bibfield  {title} {\bibinfo {title} {{Collective Neutrino Oscillations}},\
  }\href {https://doi.org/10.1146/annurev.nucl.012809.104524} {\bibfield
  {journal} {\bibinfo  {journal} {Ann. Rev. Nucl. Part. Sci.}\ }\textbf
  {\bibinfo {volume} {60}},\ \bibinfo {pages} {569} (\bibinfo {year} {2010})},\
  \Eprint {https://arxiv.org/abs/1001.2799} {arXiv:1001.2799 [hep-ph]}
  \BibitemShut {NoStop}%
\bibitem [{\citenamefont {Fiorillo}(2026)}]{Fiorillo:2026gzy}%
  \BibitemOpen
  \bibfield  {author} {\bibinfo {author} {\bibfnamefont {D.~F.~G.}\
  \bibnamefont {Fiorillo}},\ }\bibfield  {title} {\bibinfo {title} {{Collective
  flavor conversion in dense neutrino plasmas}}\ }(\bibinfo {year} {2026})\
  \Eprint {https://arxiv.org/abs/2609.19256} {arXiv:2609.19256 [hep-ph]}
  \BibitemShut {NoStop}%
\bibitem [{\citenamefont {Johns}(2023)}]{Johns:2021qby}%
  \BibitemOpen
  \bibfield  {author} {\bibinfo {author} {\bibfnamefont {L.}~\bibnamefont
  {Johns}},\ }\bibfield  {title} {\bibinfo {title} {{Collisional Flavor
  Instabilities of Supernova Neutrinos}},\ }\href
  {https://doi.org/10.1103/PhysRevLett.130.191001} {\bibfield  {journal}
  {\bibinfo  {journal} {Phys. Rev. Lett.}\ }\textbf {\bibinfo {volume} {130}},\
  \bibinfo {pages} {191001} (\bibinfo {year} {2023})},\ \Eprint
  {https://arxiv.org/abs/2104.11369} {arXiv:2104.11369 [hep-ph]} \BibitemShut
  {NoStop}%
\bibitem [{\citenamefont {Johns}\ and\ \citenamefont
  {Xiong}(2022)}]{Johns:2022yqy}%
  \BibitemOpen
  \bibfield  {author} {\bibinfo {author} {\bibfnamefont {L.}~\bibnamefont
  {Johns}}\ and\ \bibinfo {author} {\bibfnamefont {Z.}~\bibnamefont {Xiong}},\
  }\bibfield  {title} {\bibinfo {title} {{Collisional instabilities of
  neutrinos and their interplay with fast flavor conversion in compact
  objects}},\ }\href {https://doi.org/10.1103/PhysRevD.106.103029} {\bibfield
  {journal} {\bibinfo  {journal} {Phys. Rev. D}\ }\textbf {\bibinfo {volume}
  {106}},\ \bibinfo {pages} {103029} (\bibinfo {year} {2022})},\ \Eprint
  {https://arxiv.org/abs/2208.11059} {arXiv:2208.11059 [hep-ph]} \BibitemShut
  {NoStop}%
\bibitem [{\citenamefont {Chakraborty}\ \emph
  {et~al.}(2016{\natexlab{a}})\citenamefont {Chakraborty}, \citenamefont
  {Hansen}, \citenamefont {Izaguirre},\ and\ \citenamefont
  {Raffelt}}]{Chakraborty:2016lct}%
  \BibitemOpen
  \bibfield  {author} {\bibinfo {author} {\bibfnamefont {S.}~\bibnamefont
  {Chakraborty}}, \bibinfo {author} {\bibfnamefont {R.~S.}\ \bibnamefont
  {Hansen}}, \bibinfo {author} {\bibfnamefont {I.}~\bibnamefont {Izaguirre}},\
  and\ \bibinfo {author} {\bibfnamefont {G.~G.}\ \bibnamefont {Raffelt}},\
  }\bibfield  {title} {\bibinfo {title} {{Self-induced neutrino flavor
  conversion without flavor mixing}},\ }\href
  {https://doi.org/10.1088/1475-7516/2016/03/042} {\bibfield  {journal}
  {\bibinfo  {journal} {JCAP}\ }\textbf {\bibinfo {volume} {03}},\ \bibinfo
  {pages} {042}},\ \Eprint {https://arxiv.org/abs/1602.00698} {arXiv:1602.00698
  [hep-ph]} \BibitemShut {NoStop}%
\bibitem [{\citenamefont {Chakraborty}\ \emph
  {et~al.}(2016{\natexlab{b}})\citenamefont {Chakraborty}, \citenamefont
  {Hansen}, \citenamefont {Izaguirre},\ and\ \citenamefont
  {Raffelt}}]{Chakraborty:2016yeg}%
  \BibitemOpen
  \bibfield  {author} {\bibinfo {author} {\bibfnamefont {S.}~\bibnamefont
  {Chakraborty}}, \bibinfo {author} {\bibfnamefont {R.}~\bibnamefont {Hansen}},
  \bibinfo {author} {\bibfnamefont {I.}~\bibnamefont {Izaguirre}},\ and\
  \bibinfo {author} {\bibfnamefont {G.~G.}\ \bibnamefont {Raffelt}},\
  }\bibfield  {title} {\bibinfo {title} {{Collective neutrino flavor
  conversion: Recent developments}},\ }\href
  {https://doi.org/10.1016/j.nuclphysb.2016.02.012} {\bibfield  {journal}
  {\bibinfo  {journal} {Nucl. Phys. B}\ }\textbf {\bibinfo {volume} {908}},\
  \bibinfo {pages} {366} (\bibinfo {year} {2016}{\natexlab{b}})},\ \Eprint
  {https://arxiv.org/abs/1602.02766} {arXiv:1602.02766 [hep-ph]} \BibitemShut
  {NoStop}%
\bibitem [{\citenamefont {Izaguirre}\ \emph {et~al.}(2017)\citenamefont
  {Izaguirre}, \citenamefont {Raffelt},\ and\ \citenamefont
  {Tamborra}}]{Izaguirre:2016gsx}%
  \BibitemOpen
  \bibfield  {author} {\bibinfo {author} {\bibfnamefont {I.}~\bibnamefont
  {Izaguirre}}, \bibinfo {author} {\bibfnamefont {G.~G.}\ \bibnamefont
  {Raffelt}},\ and\ \bibinfo {author} {\bibfnamefont {I.}~\bibnamefont
  {Tamborra}},\ }\bibfield  {title} {\bibinfo {title} {{Fast Pairwise
  Conversion of Supernova Neutrinos: A Dispersion-Relation Approach}},\ }\href
  {https://doi.org/10.1103/PhysRevLett.118.021101} {\bibfield  {journal}
  {\bibinfo  {journal} {Phys. Rev. Lett.}\ }\textbf {\bibinfo {volume} {118}},\
  \bibinfo {pages} {021101} (\bibinfo {year} {2017})},\ \Eprint
  {https://arxiv.org/abs/1610.01612} {arXiv:1610.01612 [hep-ph]} \BibitemShut
  {NoStop}%
\bibitem [{\citenamefont {Fiorillo}\ \emph {et~al.}(2025)\citenamefont
  {Fiorillo}, \citenamefont {Janka},\ and\ \citenamefont
  {Raffelt}}]{Fiorillo:2025gkw}%
  \BibitemOpen
  \bibfield  {author} {\bibinfo {author} {\bibfnamefont {D.~F.~G.}\
  \bibnamefont {Fiorillo}}, \bibinfo {author} {\bibfnamefont {H.-T.}\
  \bibnamefont {Janka}},\ and\ \bibinfo {author} {\bibfnamefont {G.~G.}\
  \bibnamefont {Raffelt}},\ }\bibfield  {title} {\bibinfo {title}
  {{Neutrino-Mass-Driven Instabilities as the Earliest Flavor Conversion in
  Supernovae}},\ }\href {https://doi.org/10.1103/jbmx-rbzt} {\bibfield
  {journal} {\bibinfo  {journal} {Phys. Rev. Lett.}\ }\textbf {\bibinfo
  {volume} {135}},\ \bibinfo {pages} {231003} (\bibinfo {year} {2025})},\
  \Eprint {https://arxiv.org/abs/2507.22985} {arXiv:2507.22985 [hep-ph]}
  \BibitemShut {NoStop}%
\bibitem [{\citenamefont {Fiorillo}\ and\ \citenamefont
  {Raffelt}(2026)}]{Fiorillo:2025zio}%
  \BibitemOpen
  \bibfield  {author} {\bibinfo {author} {\bibfnamefont {D.~F.~G.}\
  \bibnamefont {Fiorillo}}\ and\ \bibinfo {author} {\bibfnamefont {G.~G.}\
  \bibnamefont {Raffelt}},\ }\bibfield  {title} {\bibinfo {title} {{Dispersion
  relation of the neutrino plasma: unifying fast, slow, and collisional
  instabilities}},\ }\href {https://doi.org/10.1007/JHEP01(2026)147} {\bibfield
   {journal} {\bibinfo  {journal} {JHEP}\ }\textbf {\bibinfo {volume} {01}},\
  \bibinfo {pages} {147}},\ \Eprint {https://arxiv.org/abs/2505.20389}
  {arXiv:2505.20389 [hep-ph]} \BibitemShut {NoStop}%
\bibitem [{\citenamefont {Fiorillo}\ and\ \citenamefont
  {Raffelt}(2024{\natexlab{a}})}]{Fiorillo:2024uki}%
  \BibitemOpen
  \bibfield  {author} {\bibinfo {author} {\bibfnamefont {D.~F.~G.}\
  \bibnamefont {Fiorillo}}\ and\ \bibinfo {author} {\bibfnamefont {G.~G.}\
  \bibnamefont {Raffelt}},\ }\bibfield  {title} {\bibinfo {title} {{Theory of
  neutrino fast flavor evolution. Part II. Solutions at the edge of
  instability}},\ }\href {https://doi.org/10.1007/JHEP12(2024)205} {\bibfield
  {journal} {\bibinfo  {journal} {JHEP}\ }\textbf {\bibinfo {volume} {12}},\
  \bibinfo {pages} {205}},\ \Eprint {https://arxiv.org/abs/2409.17232}
  {arXiv:2409.17232 [hep-ph]} \BibitemShut {NoStop}%
\bibitem [{\citenamefont {Padilla-Gay}\ \emph {et~al.}(2022)\citenamefont
  {Padilla-Gay}, \citenamefont {Tamborra},\ and\ \citenamefont
  {Raffelt}}]{Padilla-Gay:2022wck}%
  \BibitemOpen
  \bibfield  {author} {\bibinfo {author} {\bibfnamefont {I.}~\bibnamefont
  {Padilla-Gay}}, \bibinfo {author} {\bibfnamefont {I.}~\bibnamefont
  {Tamborra}},\ and\ \bibinfo {author} {\bibfnamefont {G.~G.}\ \bibnamefont
  {Raffelt}},\ }\bibfield  {title} {\bibinfo {title} {{Neutrino fast flavor
  pendulum. II. Collisional damping}},\ }\href
  {https://doi.org/10.1103/PhysRevD.106.103031} {\bibfield  {journal} {\bibinfo
   {journal} {Phys. Rev. D}\ }\textbf {\bibinfo {volume} {106}},\ \bibinfo
  {pages} {103031} (\bibinfo {year} {2022})},\ \Eprint
  {https://arxiv.org/abs/2209.11235} {arXiv:2209.11235 [hep-ph]} \BibitemShut
  {NoStop}%
\bibitem [{\citenamefont {Xiong}\ \emph
  {et~al.}(2023{\natexlab{a}})\citenamefont {Xiong}, \citenamefont {Johns},
  \citenamefont {Wu},\ and\ \citenamefont {Duan}}]{Xiong:2022zqz}%
  \BibitemOpen
  \bibfield  {author} {\bibinfo {author} {\bibfnamefont {Z.}~\bibnamefont
  {Xiong}}, \bibinfo {author} {\bibfnamefont {L.}~\bibnamefont {Johns}},
  \bibinfo {author} {\bibfnamefont {M.-R.}\ \bibnamefont {Wu}},\ and\ \bibinfo
  {author} {\bibfnamefont {H.}~\bibnamefont {Duan}},\ }\bibfield  {title}
  {\bibinfo {title} {{Collisional flavor instability in dense neutrino
  gases}},\ }\href {https://doi.org/10.1103/PhysRevD.108.083002} {\bibfield
  {journal} {\bibinfo  {journal} {Phys. Rev. D}\ }\textbf {\bibinfo {volume}
  {108}},\ \bibinfo {pages} {083002} (\bibinfo {year} {2023}{\natexlab{a}})},\
  \Eprint {https://arxiv.org/abs/2212.03750} {arXiv:2212.03750 [hep-ph]}
  \BibitemShut {NoStop}%
\bibitem [{\citenamefont {Morinaga}(2022)}]{Morinaga:2021vmc}%
  \BibitemOpen
  \bibfield  {author} {\bibinfo {author} {\bibfnamefont {T.}~\bibnamefont
  {Morinaga}},\ }\bibfield  {title} {\bibinfo {title} {{Fast neutrino flavor
  instability and neutrino flavor lepton number crossings}},\ }\href
  {https://doi.org/10.1103/PhysRevD.105.L101301} {\bibfield  {journal}
  {\bibinfo  {journal} {Phys. Rev. D}\ }\textbf {\bibinfo {volume} {105}},\
  \bibinfo {pages} {L101301} (\bibinfo {year} {2022})},\ \Eprint
  {https://arxiv.org/abs/2103.15267} {arXiv:2103.15267 [hep-ph]} \BibitemShut
  {NoStop}%
\bibitem [{\citenamefont {Fiorillo}\ and\ \citenamefont
  {Raffelt}(2024{\natexlab{b}})}]{Fiorillo:2024bzm}%
  \BibitemOpen
  \bibfield  {author} {\bibinfo {author} {\bibfnamefont {D.~F.~G.}\
  \bibnamefont {Fiorillo}}\ and\ \bibinfo {author} {\bibfnamefont {G.~G.}\
  \bibnamefont {Raffelt}},\ }\bibfield  {title} {\bibinfo {title} {{Theory of
  neutrino fast flavor evolution. Part I. Linear response theory and stability
  conditions.}},\ }\href {https://doi.org/10.1007/JHEP08(2024)225} {\bibfield
  {journal} {\bibinfo  {journal} {JHEP}\ }\textbf {\bibinfo {volume} {08}},\
  \bibinfo {pages} {225}},\ \Eprint {https://arxiv.org/abs/2406.06708}
  {arXiv:2406.06708 [hep-ph]} \BibitemShut {NoStop}%
\bibitem [{\citenamefont {Dasgupta}(2022)}]{Dasgupta:2021gfs}%
  \BibitemOpen
  \bibfield  {author} {\bibinfo {author} {\bibfnamefont {B.}~\bibnamefont
  {Dasgupta}},\ }\bibfield  {title} {\bibinfo {title} {{Collective Neutrino
  Flavor Instability Requires a Crossing}},\ }\href
  {https://doi.org/10.1103/PhysRevLett.128.081102} {\bibfield  {journal}
  {\bibinfo  {journal} {Phys. Rev. Lett.}\ }\textbf {\bibinfo {volume} {128}},\
  \bibinfo {pages} {081102} (\bibinfo {year} {2022})},\ \Eprint
  {https://arxiv.org/abs/2110.00192} {arXiv:2110.00192 [hep-ph]} \BibitemShut
  {NoStop}%
\bibitem [{\citenamefont {Raffelt}\ and\ \citenamefont
  {Smirnov}(2007)}]{Raffelt:2007cb}%
  \BibitemOpen
  \bibfield  {author} {\bibinfo {author} {\bibfnamefont {G.~G.}\ \bibnamefont
  {Raffelt}}\ and\ \bibinfo {author} {\bibfnamefont {A.~Y.}\ \bibnamefont
  {Smirnov}},\ }\bibfield  {title} {\bibinfo {title} {{Self-induced spectral
  splits in supernova neutrino fluxes}},\ }\href
  {https://doi.org/10.1103/PhysRevD.76.081301} {\bibfield  {journal} {\bibinfo
  {journal} {Phys. Rev. D}\ }\textbf {\bibinfo {volume} {76}},\ \bibinfo
  {pages} {081301} (\bibinfo {year} {2007})},\ \bibinfo {note} {[Erratum:
  Phys.Rev.D 77, 029903 (2008)]},\ \Eprint {https://arxiv.org/abs/0705.1830}
  {arXiv:0705.1830 [hep-ph]} \BibitemShut {NoStop}%
\bibitem [{\citenamefont {Fiorillo}\ and\ \citenamefont
  {Raffelt}(2025{\natexlab{a}})}]{Fiorillo:2024pns}%
  \BibitemOpen
  \bibfield  {author} {\bibinfo {author} {\bibfnamefont {D.~F.~G.}\
  \bibnamefont {Fiorillo}}\ and\ \bibinfo {author} {\bibfnamefont {G.~G.}\
  \bibnamefont {Raffelt}},\ }\bibfield  {title} {\bibinfo {title} {{Theory of
  neutrino slow flavor evolution. Part I. Homogeneous medium}},\ }\href
  {https://doi.org/10.1007/JHEP04(2025)146} {\bibfield  {journal} {\bibinfo
  {journal} {JHEP}\ }\textbf {\bibinfo {volume} {04}},\ \bibinfo {pages}
  {146}},\ \Eprint {https://arxiv.org/abs/2412.02747} {arXiv:2412.02747
  [hep-ph]} \BibitemShut {NoStop}%
\bibitem [{\citenamefont {Fiorillo}\ and\ \citenamefont
  {Raffelt}(2025{\natexlab{b}})}]{Fiorillo:2025ank}%
  \BibitemOpen
  \bibfield  {author} {\bibinfo {author} {\bibfnamefont {D.~F.~G.}\
  \bibnamefont {Fiorillo}}\ and\ \bibinfo {author} {\bibfnamefont {G.~G.}\
  \bibnamefont {Raffelt}},\ }\bibfield  {title} {\bibinfo {title} {{Theory of
  neutrino slow flavor evolution. Part II. Space-time evolution of linear
  instabilities}},\ }\href {https://doi.org/10.1007/JHEP06(2025)146} {\bibfield
   {journal} {\bibinfo  {journal} {JHEP}\ }\textbf {\bibinfo {volume} {06}},\
  \bibinfo {pages} {146}},\ \Eprint {https://arxiv.org/abs/2501.16423}
  {arXiv:2501.16423 [hep-ph]} \BibitemShut {NoStop}%
\bibitem [{\citenamefont {Shalgar}\ and\ \citenamefont
  {Tamborra}(2024{\natexlab{a}})}]{Shalgar:2024gjt}%
  \BibitemOpen
  \bibfield  {author} {\bibinfo {author} {\bibfnamefont {S.}~\bibnamefont
  {Shalgar}}\ and\ \bibinfo {author} {\bibfnamefont {I.}~\bibnamefont
  {Tamborra}},\ }\bibfield  {title} {\bibinfo {title} {{Neutrino quantum
  kinetics in a core-collapse supernova}},\ }\href
  {https://doi.org/10.1088/1475-7516/2024/09/021} {\bibfield  {journal}
  {\bibinfo  {journal} {JCAP}\ }\textbf {\bibinfo {volume} {09}},\ \bibinfo
  {pages} {021}},\ \Eprint {https://arxiv.org/abs/2406.09504} {arXiv:2406.09504
  [astro-ph.HE]} \BibitemShut {NoStop}%
\bibitem [{\citenamefont {Shalgar}\ and\ \citenamefont
  {Tamborra}(2025)}]{Shalgar:2025oht}%
  \BibitemOpen
  \bibfield  {author} {\bibinfo {author} {\bibfnamefont {S.}~\bibnamefont
  {Shalgar}}\ and\ \bibinfo {author} {\bibfnamefont {I.}~\bibnamefont
  {Tamborra}},\ }\bibfield  {title} {\bibinfo {title} {{Neutrino quantum
  kinetics in three flavors}},\ }\href
  {https://doi.org/10.1088/1475-7516/2025/12/026} {\bibfield  {journal}
  {\bibinfo  {journal} {JCAP}\ }\textbf {\bibinfo {volume} {12}},\ \bibinfo
  {pages} {026}},\ \Eprint {https://arxiv.org/abs/2503.03835} {arXiv:2503.03835
  [astro-ph.HE]} \BibitemShut {NoStop}%
\bibitem [{\citenamefont {Padilla-Gay}\ \emph {et~al.}(2025)\citenamefont
  {Padilla-Gay}, \citenamefont {Chen}, \citenamefont {Abbar}, \citenamefont
  {Wu},\ and\ \citenamefont {Xiong}}]{Padilla-Gay:2025tko}%
  \BibitemOpen
  \bibfield  {author} {\bibinfo {author} {\bibfnamefont {I.}~\bibnamefont
  {Padilla-Gay}}, \bibinfo {author} {\bibfnamefont {H.-H.}\ \bibnamefont
  {Chen}}, \bibinfo {author} {\bibfnamefont {S.}~\bibnamefont {Abbar}},
  \bibinfo {author} {\bibfnamefont {M.-R.}\ \bibnamefont {Wu}},\ and\ \bibinfo
  {author} {\bibfnamefont {Z.}~\bibnamefont {Xiong}},\ }\bibfield  {title}
  {\bibinfo {title} {{Flavor equilibration of supernova neutrinos: Exploring
  the dynamics of slow modes}},\ }\href {https://doi.org/10.1103/jg14-8p4l}
  {\bibfield  {journal} {\bibinfo  {journal} {Phys. Rev. D}\ }\textbf {\bibinfo
  {volume} {112}},\ \bibinfo {pages} {043039} (\bibinfo {year} {2025})},\
  \Eprint {https://arxiv.org/abs/2505.11588} {arXiv:2505.11588 [astro-ph.HE]}
  \BibitemShut {NoStop}%
\bibitem [{\citenamefont {Shalgar}\ \emph {et~al.}(2020)\citenamefont
  {Shalgar}, \citenamefont {Padilla-Gay},\ and\ \citenamefont
  {Tamborra}}]{Shalgar:2019qwg}%
  \BibitemOpen
  \bibfield  {author} {\bibinfo {author} {\bibfnamefont {S.}~\bibnamefont
  {Shalgar}}, \bibinfo {author} {\bibfnamefont {I.}~\bibnamefont
  {Padilla-Gay}},\ and\ \bibinfo {author} {\bibfnamefont {I.}~\bibnamefont
  {Tamborra}},\ }\bibfield  {title} {\bibinfo {title} {{Neutrino propagation
  hinders fast pairwise flavor conversions}},\ }\href
  {https://doi.org/10.1088/1475-7516/2020/06/048} {\bibfield  {journal}
  {\bibinfo  {journal} {JCAP}\ }\textbf {\bibinfo {volume} {06}},\ \bibinfo
  {pages} {048}},\ \Eprint {https://arxiv.org/abs/1911.09110} {arXiv:1911.09110
  [astro-ph.HE]} \BibitemShut {NoStop}%
\bibitem [{\citenamefont {Padilla-Gay}\ \emph {et~al.}(2021)\citenamefont
  {Padilla-Gay}, \citenamefont {Shalgar},\ and\ \citenamefont
  {Tamborra}}]{Padilla-Gay:2020uxa}%
  \BibitemOpen
  \bibfield  {author} {\bibinfo {author} {\bibfnamefont {I.}~\bibnamefont
  {Padilla-Gay}}, \bibinfo {author} {\bibfnamefont {S.}~\bibnamefont
  {Shalgar}},\ and\ \bibinfo {author} {\bibfnamefont {I.}~\bibnamefont
  {Tamborra}},\ }\bibfield  {title} {\bibinfo {title} {{Multi-Dimensional
  Solution of Fast Neutrino Conversions in Binary Neutron Star Merger
  Remnants}},\ }\href {https://doi.org/10.1088/1475-7516/2021/01/017}
  {\bibfield  {journal} {\bibinfo  {journal} {JCAP}\ }\textbf {\bibinfo
  {volume} {01}},\ \bibinfo {pages} {017}},\ \Eprint
  {https://arxiv.org/abs/2009.01843} {arXiv:2009.01843 [astro-ph.HE]}
  \BibitemShut {NoStop}%
\bibitem [{\citenamefont {Martin}\ \emph {et~al.}(2021)\citenamefont {Martin},
  \citenamefont {Carlson}, \citenamefont {Cirigliano},\ and\ \citenamefont
  {Duan}}]{Martin:2021xyl}%
  \BibitemOpen
  \bibfield  {author} {\bibinfo {author} {\bibfnamefont {J.~D.}\ \bibnamefont
  {Martin}}, \bibinfo {author} {\bibfnamefont {J.}~\bibnamefont {Carlson}},
  \bibinfo {author} {\bibfnamefont {V.}~\bibnamefont {Cirigliano}},\ and\
  \bibinfo {author} {\bibfnamefont {H.}~\bibnamefont {Duan}},\ }\bibfield
  {title} {\bibinfo {title} {{Fast flavor oscillations in dense neutrino media
  with collisions}},\ }\href {https://doi.org/10.1103/PhysRevD.103.063001}
  {\bibfield  {journal} {\bibinfo  {journal} {Phys. Rev. D}\ }\textbf {\bibinfo
  {volume} {103}},\ \bibinfo {pages} {063001} (\bibinfo {year} {2021})},\
  \Eprint {https://arxiv.org/abs/2101.01278} {arXiv:2101.01278 [hep-ph]}
  \BibitemShut {NoStop}%
\bibitem [{\citenamefont {Shalgar}\ and\ \citenamefont
  {Tamborra}(2021)}]{Shalgar:2020wcx}%
  \BibitemOpen
  \bibfield  {author} {\bibinfo {author} {\bibfnamefont {S.}~\bibnamefont
  {Shalgar}}\ and\ \bibinfo {author} {\bibfnamefont {I.}~\bibnamefont
  {Tamborra}},\ }\bibfield  {title} {\bibinfo {title} {{A change of direction
  in pairwise neutrino conversion physics: The effect of collisions}},\ }\href
  {https://doi.org/10.1103/PhysRevD.103.063002} {\bibfield  {journal} {\bibinfo
   {journal} {Phys. Rev. D}\ }\textbf {\bibinfo {volume} {103}},\ \bibinfo
  {pages} {063002} (\bibinfo {year} {2021})},\ \Eprint
  {https://arxiv.org/abs/2011.00004} {arXiv:2011.00004 [astro-ph.HE]}
  \BibitemShut {NoStop}%
\bibitem [{\citenamefont {Sigl}(2022)}]{Sigl:2021tmj}%
  \BibitemOpen
  \bibfield  {author} {\bibinfo {author} {\bibfnamefont {G.}~\bibnamefont
  {Sigl}},\ }\bibfield  {title} {\bibinfo {title} {{Simulations of fast
  neutrino flavor conversions with interactions in inhomogeneous media}},\
  }\href {https://doi.org/10.1103/PhysRevD.105.043005} {\bibfield  {journal}
  {\bibinfo  {journal} {Phys. Rev. D}\ }\textbf {\bibinfo {volume} {105}},\
  \bibinfo {pages} {043005} (\bibinfo {year} {2022})},\ \Eprint
  {https://arxiv.org/abs/2109.00091} {arXiv:2109.00091 [hep-ph]} \BibitemShut
  {NoStop}%
\bibitem [{\citenamefont {Hansen}\ \emph {et~al.}(2021)\citenamefont {Hansen},
  \citenamefont {Shalgar},\ and\ \citenamefont {Tamborra}}]{Hansen:2020vgm}%
  \BibitemOpen
  \bibfield  {author} {\bibinfo {author} {\bibfnamefont {R.~S.~L.}\
  \bibnamefont {Hansen}}, \bibinfo {author} {\bibfnamefont {S.}~\bibnamefont
  {Shalgar}},\ and\ \bibinfo {author} {\bibfnamefont {I.}~\bibnamefont
  {Tamborra}},\ }\bibfield  {title} {\bibinfo {title} {{Neutrino flavor mixing
  breaks isotropy in the early universe}},\ }\href
  {https://doi.org/10.1088/1475-7516/2021/07/017} {\bibfield  {journal}
  {\bibinfo  {journal} {JCAP}\ }\textbf {\bibinfo {volume} {07}},\ \bibinfo
  {pages} {017}},\ \Eprint {https://arxiv.org/abs/2012.03948} {arXiv:2012.03948
  [astro-ph.CO]} \BibitemShut {NoStop}%
\bibitem [{\citenamefont {Sasaki}\ and\ \citenamefont
  {Takiwaki}(2022)}]{Sasaki:2021zld}%
  \BibitemOpen
  \bibfield  {author} {\bibinfo {author} {\bibfnamefont {H.}~\bibnamefont
  {Sasaki}}\ and\ \bibinfo {author} {\bibfnamefont {T.}~\bibnamefont
  {Takiwaki}},\ }\bibfield  {title} {\bibinfo {title} {{A detailed analysis of
  the dynamics of fast neutrino flavor conversions with scattering effects}},\
  }\href {https://doi.org/10.1093/ptep/ptac082} {\bibfield  {journal} {\bibinfo
   {journal} {PTEP}\ }\textbf {\bibinfo {volume} {2022}},\ \bibinfo {pages}
  {073E01} (\bibinfo {year} {2022})},\ \Eprint
  {https://arxiv.org/abs/2109.14011} {arXiv:2109.14011 [hep-ph]} \BibitemShut
  {NoStop}%
\bibitem [{\citenamefont {Shalgar}\ and\ \citenamefont
  {Tamborra}(2023{\natexlab{a}})}]{Shalgar:2022rjj}%
  \BibitemOpen
  \bibfield  {author} {\bibinfo {author} {\bibfnamefont {S.}~\bibnamefont
  {Shalgar}}\ and\ \bibinfo {author} {\bibfnamefont {I.}~\bibnamefont
  {Tamborra}},\ }\bibfield  {title} {\bibinfo {title} {{Neutrino decoupling is
  altered by flavor conversion}},\ }\href
  {https://doi.org/10.1103/PhysRevD.108.043006} {\bibfield  {journal} {\bibinfo
   {journal} {Phys. Rev. D}\ }\textbf {\bibinfo {volume} {108}},\ \bibinfo
  {pages} {043006} (\bibinfo {year} {2023}{\natexlab{a}})},\ \Eprint
  {https://arxiv.org/abs/2206.00676} {arXiv:2206.00676 [astro-ph.HE]}
  \BibitemShut {NoStop}%
\bibitem [{\citenamefont {Shalgar}\ and\ \citenamefont
  {Tamborra}(2023{\natexlab{b}})}]{Shalgar:2022lvv}%
  \BibitemOpen
  \bibfield  {author} {\bibinfo {author} {\bibfnamefont {S.}~\bibnamefont
  {Shalgar}}\ and\ \bibinfo {author} {\bibfnamefont {I.}~\bibnamefont
  {Tamborra}},\ }\bibfield  {title} {\bibinfo {title} {{Neutrino flavor
  conversion, advection, and collisions: Toward the full solution}},\ }\href
  {https://doi.org/10.1103/PhysRevD.107.063025} {\bibfield  {journal} {\bibinfo
   {journal} {Phys. Rev. D}\ }\textbf {\bibinfo {volume} {107}},\ \bibinfo
  {pages} {063025} (\bibinfo {year} {2023}{\natexlab{b}})},\ \Eprint
  {https://arxiv.org/abs/2207.04058} {arXiv:2207.04058 [astro-ph.HE]}
  \BibitemShut {NoStop}%
\bibitem [{\citenamefont {Nagakura}\ and\ \citenamefont
  {Zaizen}(2023)}]{Nagakura:2022xwe}%
  \BibitemOpen
  \bibfield  {author} {\bibinfo {author} {\bibfnamefont {H.}~\bibnamefont
  {Nagakura}}\ and\ \bibinfo {author} {\bibfnamefont {M.}~\bibnamefont
  {Zaizen}},\ }\bibfield  {title} {\bibinfo {title} {{Connecting small-scale to
  large-scale structures of fast neutrino-flavor conversion}},\ }\href
  {https://doi.org/10.1103/PhysRevD.107.063033} {\bibfield  {journal} {\bibinfo
   {journal} {Phys. Rev. D}\ }\textbf {\bibinfo {volume} {107}},\ \bibinfo
  {pages} {063033} (\bibinfo {year} {2023})},\ \Eprint
  {https://arxiv.org/abs/2211.01398} {arXiv:2211.01398 [astro-ph.HE]}
  \BibitemShut {NoStop}%
\bibitem [{\citenamefont {Xiong}\ \emph
  {et~al.}(2023{\natexlab{b}})\citenamefont {Xiong}, \citenamefont {Wu},
  \citenamefont {Abbar}, \citenamefont {Bhattacharyya}, \citenamefont
  {George},\ and\ \citenamefont {Lin}}]{Xiong:2023vcm}%
  \BibitemOpen
  \bibfield  {author} {\bibinfo {author} {\bibfnamefont {Z.}~\bibnamefont
  {Xiong}}, \bibinfo {author} {\bibfnamefont {M.-R.}\ \bibnamefont {Wu}},
  \bibinfo {author} {\bibfnamefont {S.}~\bibnamefont {Abbar}}, \bibinfo
  {author} {\bibfnamefont {S.}~\bibnamefont {Bhattacharyya}}, \bibinfo {author}
  {\bibfnamefont {M.}~\bibnamefont {George}},\ and\ \bibinfo {author}
  {\bibfnamefont {C.-Y.}\ \bibnamefont {Lin}},\ }\bibfield  {title} {\bibinfo
  {title} {{Evaluating approximate asymptotic distributions for fast neutrino
  flavor conversions in a periodic 1D box}},\ }\href
  {https://doi.org/10.1103/PhysRevD.108.063003} {\bibfield  {journal} {\bibinfo
   {journal} {Phys. Rev. D}\ }\textbf {\bibinfo {volume} {108}},\ \bibinfo
  {pages} {063003} (\bibinfo {year} {2023}{\natexlab{b}})},\ \Eprint
  {https://arxiv.org/abs/2307.11129} {arXiv:2307.11129 [astro-ph.HE]}
  \BibitemShut {NoStop}%
\bibitem [{\citenamefont {Xiong}\ \emph {et~al.}(2025)\citenamefont {Xiong},
  \citenamefont {Wu}, \citenamefont {George},\ and\ \citenamefont
  {Lin}}]{Xiong:2024pue}%
  \BibitemOpen
  \bibfield  {author} {\bibinfo {author} {\bibfnamefont {Z.}~\bibnamefont
  {Xiong}}, \bibinfo {author} {\bibfnamefont {M.-R.}\ \bibnamefont {Wu}},
  \bibinfo {author} {\bibfnamefont {M.}~\bibnamefont {George}},\ and\ \bibinfo
  {author} {\bibfnamefont {C.-Y.}\ \bibnamefont {Lin}},\ }\bibfield  {title}
  {\bibinfo {title} {{Robust Integration of Fast Flavor Conversions in
  Classical Neutrino Transport}},\ }\href
  {https://doi.org/10.1103/PhysRevLett.134.051003} {\bibfield  {journal}
  {\bibinfo  {journal} {Phys. Rev. Lett.}\ }\textbf {\bibinfo {volume} {134}},\
  \bibinfo {pages} {051003} (\bibinfo {year} {2025})},\ \Eprint
  {https://arxiv.org/abs/2403.17269} {arXiv:2403.17269 [astro-ph.HE]}
  \BibitemShut {NoStop}%
\bibitem [{\citenamefont {Zaizen}\ and\ \citenamefont
  {Nagakura}(2023{\natexlab{a}})}]{Zaizen:2023ihz}%
  \BibitemOpen
  \bibfield  {author} {\bibinfo {author} {\bibfnamefont {M.}~\bibnamefont
  {Zaizen}}\ and\ \bibinfo {author} {\bibfnamefont {H.}~\bibnamefont
  {Nagakura}},\ }\bibfield  {title} {\bibinfo {title} {{Characterizing
  quasisteady states of fast neutrino-flavor conversion by stability and
  conservation laws}},\ }\href {https://doi.org/10.1103/PhysRevD.107.123021}
  {\bibfield  {journal} {\bibinfo  {journal} {Phys. Rev. D}\ }\textbf {\bibinfo
  {volume} {107}},\ \bibinfo {pages} {123021} (\bibinfo {year}
  {2023}{\natexlab{a}})},\ \Eprint {https://arxiv.org/abs/2304.05044}
  {arXiv:2304.05044 [astro-ph.HE]} \BibitemShut {NoStop}%
\bibitem [{\citenamefont {Zaizen}\ and\ \citenamefont
  {Nagakura}(2023{\natexlab{b}})}]{Zaizen:2022cik}%
  \BibitemOpen
  \bibfield  {author} {\bibinfo {author} {\bibfnamefont {M.}~\bibnamefont
  {Zaizen}}\ and\ \bibinfo {author} {\bibfnamefont {H.}~\bibnamefont
  {Nagakura}},\ }\bibfield  {title} {\bibinfo {title} {{Simple method for
  determining asymptotic states of fast neutrino-flavor conversion}},\ }\href
  {https://doi.org/10.1103/PhysRevD.107.103022} {\bibfield  {journal} {\bibinfo
   {journal} {Phys. Rev. D}\ }\textbf {\bibinfo {volume} {107}},\ \bibinfo
  {pages} {103022} (\bibinfo {year} {2023}{\natexlab{b}})},\ \Eprint
  {https://arxiv.org/abs/2211.09343} {arXiv:2211.09343 [astro-ph.HE]}
  \BibitemShut {NoStop}%
\bibitem [{\citenamefont {Bhattacharyya}\ and\ \citenamefont
  {Dasgupta}(2022)}]{Bhattacharyya:2022eed}%
  \BibitemOpen
  \bibfield  {author} {\bibinfo {author} {\bibfnamefont {S.}~\bibnamefont
  {Bhattacharyya}}\ and\ \bibinfo {author} {\bibfnamefont {B.}~\bibnamefont
  {Dasgupta}},\ }\bibfield  {title} {\bibinfo {title} {{Elaborating the
  ultimate fate of fast collective neutrino flavor oscillations}},\ }\href
  {https://doi.org/10.1103/PhysRevD.106.103039} {\bibfield  {journal} {\bibinfo
   {journal} {Phys. Rev. D}\ }\textbf {\bibinfo {volume} {106}},\ \bibinfo
  {pages} {103039} (\bibinfo {year} {2022})},\ \Eprint
  {https://arxiv.org/abs/2205.05129} {arXiv:2205.05129 [hep-ph]} \BibitemShut
  {NoStop}%
\bibitem [{\citenamefont {Johns}(2025{\natexlab{a}})}]{Johns:2023jjt}%
  \BibitemOpen
  \bibfield  {author} {\bibinfo {author} {\bibfnamefont {L.}~\bibnamefont
  {Johns}},\ }\bibfield  {title} {\bibinfo {title} {{Thermodynamics of
  oscillating neutrinos}},\ }\href {https://doi.org/10.1103/y8qs-w8lq}
  {\bibfield  {journal} {\bibinfo  {journal} {Phys. Rev. D}\ }\textbf {\bibinfo
  {volume} {112}},\ \bibinfo {pages} {063032} (\bibinfo {year}
  {2025}{\natexlab{a}})},\ \Eprint {https://arxiv.org/abs/2306.14982}
  {arXiv:2306.14982 [hep-ph]} \BibitemShut {NoStop}%
\bibitem [{\citenamefont {Nagakura}\ \emph {et~al.}(2024)\citenamefont
  {Nagakura}, \citenamefont {Johns},\ and\ \citenamefont
  {Zaizen}}]{Nagakura:2023jfi}%
  \BibitemOpen
  \bibfield  {author} {\bibinfo {author} {\bibfnamefont {H.}~\bibnamefont
  {Nagakura}}, \bibinfo {author} {\bibfnamefont {L.}~\bibnamefont {Johns}},\
  and\ \bibinfo {author} {\bibfnamefont {M.}~\bibnamefont {Zaizen}},\
  }\bibfield  {title} {\bibinfo {title} {{Bhatnagar-Gross-Krook subgrid model
  for neutrino quantum kinetics}},\ }\href
  {https://doi.org/10.1103/PhysRevD.109.083013} {\bibfield  {journal} {\bibinfo
   {journal} {Phys. Rev. D}\ }\textbf {\bibinfo {volume} {109}},\ \bibinfo
  {pages} {083013} (\bibinfo {year} {2024})},\ \Eprint
  {https://arxiv.org/abs/2312.16285} {arXiv:2312.16285 [astro-ph.HE]}
  \BibitemShut {NoStop}%
\bibitem [{\citenamefont {Johns}(2025{\natexlab{b}})}]{Johns:2024dbe}%
  \BibitemOpen
  \bibfield  {author} {\bibinfo {author} {\bibfnamefont {L.}~\bibnamefont
  {Johns}},\ }\bibfield  {title} {\bibinfo {title} {{Subgrid modeling of
  neutrino oscillations in astrophysics}},\ }\href
  {https://doi.org/10.1103/3fr2-qttd} {\bibfield  {journal} {\bibinfo
  {journal} {Phys. Rev. D}\ }\textbf {\bibinfo {volume} {112}},\ \bibinfo
  {pages} {043024} (\bibinfo {year} {2025}{\natexlab{b}})},\ \Eprint
  {https://arxiv.org/abs/2401.15247} {arXiv:2401.15247 [astro-ph.HE]}
  \BibitemShut {NoStop}%
\bibitem [{\citenamefont {Goimil-Garc{\'\i}a}\ and\ \citenamefont
  {Tamborra}(2026)}]{Goimil-Garcia:2026flm}%
  \BibitemOpen
  \bibfield  {author} {\bibinfo {author} {\bibfnamefont {M.}~\bibnamefont
  {Goimil-Garc{\'\i}a}}\ and\ \bibinfo {author} {\bibfnamefont
  {I.}~\bibnamefont {Tamborra}},\ }\bibfield  {title} {\bibinfo {title}
  {{Quasi-steady flavor configuration of multi-energy neutrino ensembles}},\
  }\href@noop {} {\  (\bibinfo {year} {2026})},\ \Eprint
  {https://arxiv.org/abs/2608.06464} {arXiv:2608.06464 [astro-ph.HE]}
  \BibitemShut {NoStop}%
\bibitem [{\citenamefont {Goimil-Garc{\'\i}a}\ and\ \citenamefont
  {Tamborra}(2025)}]{Goimil-Garcia:2025ozm}%
  \BibitemOpen
  \bibfield  {author} {\bibinfo {author} {\bibfnamefont {M.}~\bibnamefont
  {Goimil-Garc{\'\i}a}}\ and\ \bibinfo {author} {\bibfnamefont
  {I.}~\bibnamefont {Tamborra}},\ }\bibfield  {title} {\bibinfo {title}
  {{Steady state of fast-oscillating neutrinos in an inhomogeneous medium}},\
  }\href {https://doi.org/10.1103/gdg9-rzns} {\bibfield  {journal} {\bibinfo
  {journal} {Phys. Rev. D}\ }\textbf {\bibinfo {volume} {112}},\ \bibinfo
  {pages} {103011} (\bibinfo {year} {2025})},\ \Eprint
  {https://arxiv.org/abs/2509.22805} {arXiv:2509.22805 [astro-ph.HE]}
  \BibitemShut {NoStop}%
\bibitem [{\citenamefont {Chen}\ \emph {et~al.}(2026)\citenamefont {Chen},
  \citenamefont {Padilla-Gay}, \citenamefont {Wu}, \citenamefont {Abbar},\ and\
  \citenamefont {Xiong}}]{Chen:2026oym}%
  \BibitemOpen
  \bibfield  {author} {\bibinfo {author} {\bibfnamefont {H.-H.}\ \bibnamefont
  {Chen}}, \bibinfo {author} {\bibfnamefont {I.}~\bibnamefont {Padilla-Gay}},
  \bibinfo {author} {\bibfnamefont {M.-R.}\ \bibnamefont {Wu}}, \bibinfo
  {author} {\bibfnamefont {S.}~\bibnamefont {Abbar}},\ and\ \bibinfo {author}
  {\bibfnamefont {Z.}~\bibnamefont {Xiong}},\ }\bibfield  {title} {\bibinfo
  {title} {{Energy-dependent slow collective flavor conversion of supernova
  neutrinos}},\ }\href@noop {} {\  (\bibinfo {year} {2026})},\ \Eprint
  {https://arxiv.org/abs/2608.26847} {arXiv:2608.26847 [astro-ph.HE]}
  \BibitemShut {NoStop}%
\bibitem [{\citenamefont {Ehring}\ \emph
  {et~al.}(2023{\natexlab{a}})\citenamefont {Ehring}, \citenamefont {Abbar},
  \citenamefont {Janka}, \citenamefont {Raffelt},\ and\ \citenamefont
  {Tamborra}}]{Ehring:2023abs}%
  \BibitemOpen
  \bibfield  {author} {\bibinfo {author} {\bibfnamefont {J.}~\bibnamefont
  {Ehring}}, \bibinfo {author} {\bibfnamefont {S.}~\bibnamefont {Abbar}},
  \bibinfo {author} {\bibfnamefont {H.-T.}\ \bibnamefont {Janka}}, \bibinfo
  {author} {\bibfnamefont {G.~G.}\ \bibnamefont {Raffelt}},\ and\ \bibinfo
  {author} {\bibfnamefont {I.}~\bibnamefont {Tamborra}},\ }\bibfield  {title}
  {\bibinfo {title} {{Fast Neutrino Flavor Conversions Can Help and Hinder
  Neutrino-Driven Explosions}},\ }\href
  {https://doi.org/10.1103/PhysRevLett.131.061401} {\bibfield  {journal}
  {\bibinfo  {journal} {Phys. Rev. Lett.}\ }\textbf {\bibinfo {volume} {131}},\
  \bibinfo {pages} {061401} (\bibinfo {year} {2023}{\natexlab{a}})},\ \Eprint
  {https://arxiv.org/abs/2305.11207} {arXiv:2305.11207 [astro-ph.HE]}
  \BibitemShut {NoStop}%
\bibitem [{\citenamefont {Ehring}\ \emph
  {et~al.}(2023{\natexlab{b}})\citenamefont {Ehring}, \citenamefont {Abbar},
  \citenamefont {Janka}, \citenamefont {Raffelt},\ and\ \citenamefont
  {Tamborra}}]{Ehring:2023lcd}%
  \BibitemOpen
  \bibfield  {author} {\bibinfo {author} {\bibfnamefont {J.}~\bibnamefont
  {Ehring}}, \bibinfo {author} {\bibfnamefont {S.}~\bibnamefont {Abbar}},
  \bibinfo {author} {\bibfnamefont {H.-T.}\ \bibnamefont {Janka}}, \bibinfo
  {author} {\bibfnamefont {G.~G.}\ \bibnamefont {Raffelt}},\ and\ \bibinfo
  {author} {\bibfnamefont {I.}~\bibnamefont {Tamborra}},\ }\bibfield  {title}
  {\bibinfo {title} {{Fast neutrino flavor conversion in core-collapse
  supernovae: A parametric study in 1D models}},\ }\href
  {https://doi.org/10.1103/PhysRevD.107.103034} {\bibfield  {journal} {\bibinfo
   {journal} {Phys. Rev. D}\ }\textbf {\bibinfo {volume} {107}},\ \bibinfo
  {pages} {103034} (\bibinfo {year} {2023}{\natexlab{b}})},\ \Eprint
  {https://arxiv.org/abs/2301.11938} {arXiv:2301.11938 [astro-ph.HE]}
  \BibitemShut {NoStop}%
\bibitem [{\citenamefont {Akaho}\ \emph {et~al.}(2026)\citenamefont {Akaho},
  \citenamefont {Nagakura}, \citenamefont {Iwakami}, \citenamefont {Furusawa},
  \citenamefont {Harada}, \citenamefont {Okawa}, \citenamefont {Matsufuru},
  \citenamefont {Sumiyoshi},\ and\ \citenamefont {Yamada}}]{Akaho:2026kff}%
  \BibitemOpen
  \bibfield  {author} {\bibinfo {author} {\bibfnamefont {R.}~\bibnamefont
  {Akaho}}, \bibinfo {author} {\bibfnamefont {H.}~\bibnamefont {Nagakura}},
  \bibinfo {author} {\bibfnamefont {W.}~\bibnamefont {Iwakami}}, \bibinfo
  {author} {\bibfnamefont {S.}~\bibnamefont {Furusawa}}, \bibinfo {author}
  {\bibfnamefont {A.}~\bibnamefont {Harada}}, \bibinfo {author} {\bibfnamefont
  {H.}~\bibnamefont {Okawa}}, \bibinfo {author} {\bibfnamefont
  {H.}~\bibnamefont {Matsufuru}}, \bibinfo {author} {\bibfnamefont
  {K.}~\bibnamefont {Sumiyoshi}},\ and\ \bibinfo {author} {\bibfnamefont
  {S.}~\bibnamefont {Yamada}},\ }\bibfield  {title} {\bibinfo {title}
  {{Bifurcated Impact of Neutrino Fast Flavor Conversion on Core-Collapse
  Supernovae Informed by Multiangle Neutrino Radiation Hydrodynamics}},\ }\href
  {https://doi.org/10.1103/fksy-1jtw} {\bibfield  {journal} {\bibinfo
  {journal} {Phys. Rev. Lett.}\ }\textbf {\bibinfo {volume} {136}},\ \bibinfo
  {pages} {191002} (\bibinfo {year} {2026})},\ \Eprint
  {https://arxiv.org/abs/2601.08269} {arXiv:2601.08269 [astro-ph.HE]}
  \BibitemShut {NoStop}%
\bibitem [{\citenamefont {Gogilashvili}\ and\ \citenamefont
  {Tamborra}(2026{\natexlab{a}})}]{Gogilashvili:2026kef}%
  \BibitemOpen
  \bibfield  {author} {\bibinfo {author} {\bibfnamefont {M.}~\bibnamefont
  {Gogilashvili}}\ and\ \bibinfo {author} {\bibfnamefont {I.}~\bibnamefont
  {Tamborra}},\ }\bibfield  {title} {\bibinfo {title} {{Flavor conversion
  enhances or suppresses supernova explodability independent of the progenitor
  mass}},\ }\href {https://doi.org/10.1103/j5b2-5q2p} {\bibfield  {journal}
  {\bibinfo  {journal} {Phys. Rev. D}\ }\textbf {\bibinfo {volume} {114}},\
  \bibinfo {pages} {043054} (\bibinfo {year} {2026}{\natexlab{a}})},\ \Eprint
  {https://arxiv.org/abs/2605.18972} {arXiv:2605.18972 [astro-ph.HE]}
  \BibitemShut {NoStop}%
\bibitem [{\citenamefont {Gogilashvili}\ and\ \citenamefont
  {Tamborra}(2026{\natexlab{b}})}]{Gogilashvili:2026epg}%
  \BibitemOpen
  \bibfield  {author} {\bibinfo {author} {\bibfnamefont {M.}~\bibnamefont
  {Gogilashvili}}\ and\ \bibinfo {author} {\bibfnamefont {I.}~\bibnamefont
  {Tamborra}},\ }\bibfield  {title} {\bibinfo {title} {{Neutrino flavor
  conversion shapes the rate of failed core-collapse supernovae}},\ }\href
  {https://doi.org/10.1103/pz3y-3lv5} {\bibfield  {journal} {\bibinfo
  {journal} {Phys. Rev. D}\ }\textbf {\bibinfo {volume} {114}},\ \bibinfo
  {pages} {L061304} (\bibinfo {year} {2026}{\natexlab{b}})},\ \Eprint
  {https://arxiv.org/abs/2605.16504} {arXiv:2605.16504 [astro-ph.HE]}
  \BibitemShut {NoStop}%
\bibitem [{\citenamefont {Nagakura}(2023)}]{Nagakura:2023mhr}%
  \BibitemOpen
  \bibfield  {author} {\bibinfo {author} {\bibfnamefont {H.}~\bibnamefont
  {Nagakura}},\ }\bibfield  {title} {\bibinfo {title} {{Roles of Fast
  Neutrino-Flavor Conversion on the Neutrino-Heating Mechanism of Core-Collapse
  Supernova}},\ }\href {https://doi.org/10.1103/PhysRevLett.130.211401}
  {\bibfield  {journal} {\bibinfo  {journal} {Phys. Rev. Lett.}\ }\textbf
  {\bibinfo {volume} {130}},\ \bibinfo {pages} {211401} (\bibinfo {year}
  {2023})},\ \Eprint {https://arxiv.org/abs/2301.10785} {arXiv:2301.10785
  [astro-ph.HE]} \BibitemShut {NoStop}%
\bibitem [{\citenamefont {Mori}\ \emph {et~al.}(2025)\citenamefont {Mori},
  \citenamefont {Takiwaki}, \citenamefont {Kotake},\ and\ \citenamefont
  {Horiuchi}}]{Mori:2025cke}%
  \BibitemOpen
  \bibfield  {author} {\bibinfo {author} {\bibfnamefont {K.}~\bibnamefont
  {Mori}}, \bibinfo {author} {\bibfnamefont {T.}~\bibnamefont {Takiwaki}},
  \bibinfo {author} {\bibfnamefont {K.}~\bibnamefont {Kotake}},\ and\ \bibinfo
  {author} {\bibfnamefont {S.}~\bibnamefont {Horiuchi}},\ }\bibfield  {title}
  {\bibinfo {title} {{Three-dimensional core-collapse supernova models with
  phenomenological treatment of neutrino flavor conversions}},\ }\href
  {https://doi.org/10.1093/pasj/psaf007} {\bibfield  {journal} {\bibinfo
  {journal} {Publ. Astron. Soc. Jap.}\ }\textbf {\bibinfo {volume} {77}},\
  \bibinfo {pages} {L9} (\bibinfo {year} {2025})},\ \Eprint
  {https://arxiv.org/abs/2501.15256} {arXiv:2501.15256 [astro-ph.HE]}
  \BibitemShut {NoStop}%
\bibitem [{\citenamefont {Cornelius}\ \emph
  {et~al.}(2024{\natexlab{a}})\citenamefont {Cornelius}, \citenamefont
  {Shalgar},\ and\ \citenamefont {Tamborra}}]{Cornelius:2023eop}%
  \BibitemOpen
  \bibfield  {author} {\bibinfo {author} {\bibfnamefont {M.}~\bibnamefont
  {Cornelius}}, \bibinfo {author} {\bibfnamefont {S.}~\bibnamefont {Shalgar}},\
  and\ \bibinfo {author} {\bibfnamefont {I.}~\bibnamefont {Tamborra}},\
  }\bibfield  {title} {\bibinfo {title} {{Perturbing fast neutrino flavor
  conversion}},\ }\href {https://doi.org/10.1088/1475-7516/2024/02/038}
  {\bibfield  {journal} {\bibinfo  {journal} {JCAP}\ }\textbf {\bibinfo
  {volume} {02}},\ \bibinfo {pages} {038}},\ \Eprint
  {https://arxiv.org/abs/2312.03839} {arXiv:2312.03839 [astro-ph.HE]}
  \BibitemShut {NoStop}%
\bibitem [{\citenamefont {Cornelius}\ \emph
  {et~al.}(2024{\natexlab{b}})\citenamefont {Cornelius}, \citenamefont
  {Shalgar},\ and\ \citenamefont {Tamborra}}]{Cornelius:2024zsb}%
  \BibitemOpen
  \bibfield  {author} {\bibinfo {author} {\bibfnamefont {M.}~\bibnamefont
  {Cornelius}}, \bibinfo {author} {\bibfnamefont {S.}~\bibnamefont {Shalgar}},\
  and\ \bibinfo {author} {\bibfnamefont {I.}~\bibnamefont {Tamborra}},\
  }\bibfield  {title} {\bibinfo {title} {{Neutrino quantum kinetics in two
  spatial dimensions}},\ }\href {https://doi.org/10.1088/1475-7516/2024/11/060}
  {\bibfield  {journal} {\bibinfo  {journal} {JCAP}\ }\textbf {\bibinfo
  {volume} {11}},\ \bibinfo {pages} {060}},\ \Eprint
  {https://arxiv.org/abs/2407.04769} {arXiv:2407.04769 [astro-ph.HE]}
  \BibitemShut {NoStop}%
\bibitem [{\citenamefont {Richers}\ \emph {et~al.}(2021)\citenamefont
  {Richers}, \citenamefont {Willcox},\ and\ \citenamefont
  {Ford}}]{Richers:2021xtf}%
  \BibitemOpen
  \bibfield  {author} {\bibinfo {author} {\bibfnamefont {S.}~\bibnamefont
  {Richers}}, \bibinfo {author} {\bibfnamefont {D.}~\bibnamefont {Willcox}},\
  and\ \bibinfo {author} {\bibfnamefont {N.}~\bibnamefont {Ford}},\ }\bibfield
  {title} {\bibinfo {title} {{Neutrino fast flavor instability in three
  dimensions}},\ }\href {https://doi.org/10.1103/PhysRevD.104.103023}
  {\bibfield  {journal} {\bibinfo  {journal} {Phys. Rev. D}\ }\textbf {\bibinfo
  {volume} {104}},\ \bibinfo {pages} {103023} (\bibinfo {year} {2021})},\
  \Eprint {https://arxiv.org/abs/2109.08631} {arXiv:2109.08631 [astro-ph.HE]}
  \BibitemShut {NoStop}%
\bibitem [{\citenamefont {Shalgar}\ and\ \citenamefont
  {Tamborra}(2022)}]{Shalgar:2021oko}%
  \BibitemOpen
  \bibfield  {author} {\bibinfo {author} {\bibfnamefont {S.}~\bibnamefont
  {Shalgar}}\ and\ \bibinfo {author} {\bibfnamefont {I.}~\bibnamefont
  {Tamborra}},\ }\bibfield  {title} {\bibinfo {title} {{Symmetry breaking
  induced by pairwise conversion of neutrinos in compact sources}},\ }\href
  {https://doi.org/10.1103/PhysRevD.105.043018} {\bibfield  {journal} {\bibinfo
   {journal} {Phys. Rev. D}\ }\textbf {\bibinfo {volume} {105}},\ \bibinfo
  {pages} {043018} (\bibinfo {year} {2022})},\ \Eprint
  {https://arxiv.org/abs/2106.15622} {arXiv:2106.15622 [hep-ph]} \BibitemShut
  {NoStop}%
\bibitem [{\citenamefont {Janka}\ \emph {et~al.}(2016)\citenamefont {Janka},
  \citenamefont {Melson},\ and\ \citenamefont {Summa}}]{Janka:2016fox}%
  \BibitemOpen
  \bibfield  {author} {\bibinfo {author} {\bibfnamefont {H.-T.}\ \bibnamefont
  {Janka}}, \bibinfo {author} {\bibfnamefont {T.}~\bibnamefont {Melson}},\ and\
  \bibinfo {author} {\bibfnamefont {A.}~\bibnamefont {Summa}},\ }\bibfield
  {title} {\bibinfo {title} {{Physics of Core-Collapse Supernovae in Three
  Dimensions: a Sneak Preview}},\ }\href
  {https://doi.org/10.1146/annurev-nucl-102115-044747} {\bibfield  {journal}
  {\bibinfo  {journal} {Ann. Rev. Nucl. Part. Sci.}\ }\textbf {\bibinfo
  {volume} {66}},\ \bibinfo {pages} {341} (\bibinfo {year} {2016})},\ \Eprint
  {https://arxiv.org/abs/1602.05576} {arXiv:1602.05576 [astro-ph.SR]}
  \BibitemShut {NoStop}%
\bibitem [{\citenamefont {Burrows}\ \emph {et~al.}(2007)\citenamefont
  {Burrows}, \citenamefont {Dessart}, \citenamefont {Ott},\ and\ \citenamefont
  {Livne}}]{Burrows:2006ci}%
  \BibitemOpen
  \bibfield  {author} {\bibinfo {author} {\bibfnamefont {A.}~\bibnamefont
  {Burrows}}, \bibinfo {author} {\bibfnamefont {L.}~\bibnamefont {Dessart}},
  \bibinfo {author} {\bibfnamefont {C.~D.}\ \bibnamefont {Ott}},\ and\ \bibinfo
  {author} {\bibfnamefont {E.}~\bibnamefont {Livne}},\ }\bibfield  {title}
  {\bibinfo {title} {{Multi-Dimensional Explorations in Supernova Theory}},\
  }\href {https://doi.org/10.1016/j.physrep.2007.02.001} {\bibfield  {journal}
  {\bibinfo  {journal} {Phys. Rept.}\ }\textbf {\bibinfo {volume} {442}},\
  \bibinfo {pages} {23} (\bibinfo {year} {2007})},\ \Eprint
  {https://arxiv.org/abs/astro-ph/0612460} {arXiv:astro-ph/0612460}
  \BibitemShut {NoStop}%
\bibitem [{\citenamefont {Tamborra}\ \emph
  {et~al.}(2014{\natexlab{a}})\citenamefont {Tamborra}, \citenamefont {Hanke},
  \citenamefont {Janka}, \citenamefont {M{\"u}ller}, \citenamefont {Raffelt},\
  and\ \citenamefont {Marek}}]{Tamborra:2014aua}%
  \BibitemOpen
  \bibfield  {author} {\bibinfo {author} {\bibfnamefont {I.}~\bibnamefont
  {Tamborra}}, \bibinfo {author} {\bibfnamefont {F.}~\bibnamefont {Hanke}},
  \bibinfo {author} {\bibfnamefont {H.-T.}\ \bibnamefont {Janka}}, \bibinfo
  {author} {\bibfnamefont {B.}~\bibnamefont {M{\"u}ller}}, \bibinfo {author}
  {\bibfnamefont {G.~G.}\ \bibnamefont {Raffelt}},\ and\ \bibinfo {author}
  {\bibfnamefont {A.}~\bibnamefont {Marek}},\ }\bibfield  {title} {\bibinfo
  {title} {{Self-sustained asymmetry of lepton-number emission: A new
  phenomenon during the supernova shock-accretion phase in three dimensions}},\
  }\href {https://doi.org/10.1088/0004-637X/792/2/96} {\bibfield  {journal}
  {\bibinfo  {journal} {Astrophys. J.}\ }\textbf {\bibinfo {volume} {792}},\
  \bibinfo {pages} {96} (\bibinfo {year} {2014}{\natexlab{a}})},\ \Eprint
  {https://arxiv.org/abs/1402.5418} {arXiv:1402.5418 [astro-ph.SR]}
  \BibitemShut {NoStop}%
\bibitem [{\citenamefont {Janka}(2017)}]{Janka:2016nak}%
  \BibitemOpen
  \bibfield  {author} {\bibinfo {author} {\bibfnamefont {H.-T.}\ \bibnamefont
  {Janka}},\ }\bibfield  {title} {\bibinfo {title} {{Neutron star kicks by the
  gravitational tug-boat mechanism in asymmetric supernova explosions:
  progenitor and explosion dependence}},\ }\href
  {https://doi.org/10.3847/1538-4357/aa618e} {\bibfield  {journal} {\bibinfo
  {journal} {Astrophys. J.}\ }\textbf {\bibinfo {volume} {837}},\ \bibinfo
  {pages} {84} (\bibinfo {year} {2017})},\ \Eprint
  {https://arxiv.org/abs/1611.07562} {arXiv:1611.07562 [astro-ph.HE]}
  \BibitemShut {NoStop}%
\bibitem [{\citenamefont {Glas}\ \emph {et~al.}(2018)\citenamefont {Glas},
  \citenamefont {Janka}, \citenamefont {Melson}, \citenamefont {Stockinger},\
  and\ \citenamefont {Just}}]{Glas:2018vcs}%
  \BibitemOpen
  \bibfield  {author} {\bibinfo {author} {\bibfnamefont {R.}~\bibnamefont
  {Glas}}, \bibinfo {author} {\bibfnamefont {H.-T.}\ \bibnamefont {Janka}},
  \bibinfo {author} {\bibfnamefont {T.}~\bibnamefont {Melson}}, \bibinfo
  {author} {\bibfnamefont {G.}~\bibnamefont {Stockinger}},\ and\ \bibinfo
  {author} {\bibfnamefont {O.}~\bibnamefont {Just}},\ }\bibfield  {title}
  {\bibinfo {title} {{Effects of LESA in Three-Dimensional Supernova
  Simulations with Multi-Dimensional and Ray-by-Ray-plus Neutrino Transport}}\
  }\href {https://doi.org/10.3847/1538-4357/ab275c} {10.3847/1538-4357/ab275c}
  (\bibinfo {year} {2018}),\ \Eprint {https://arxiv.org/abs/1809.10150}
  {arXiv:1809.10150 [astro-ph.HE]} \BibitemShut {NoStop}%
\bibitem [{\citenamefont {Janka}\ and\ \citenamefont
  {Kresse}(2024)}]{Janka:2024xbp}%
  \BibitemOpen
  \bibfield  {author} {\bibinfo {author} {\bibfnamefont {H.-T.}\ \bibnamefont
  {Janka}}\ and\ \bibinfo {author} {\bibfnamefont {D.}~\bibnamefont {Kresse}},\
  }\bibfield  {title} {\bibinfo {title} {{Interplay between neutrino kicks and
  hydrodynamic kicks of neutron stars and black holes}},\ }\href
  {https://doi.org/10.1007/s10509-024-04343-1} {\bibfield  {journal} {\bibinfo
  {journal} {Astrophys. Space Sci.}\ }\textbf {\bibinfo {volume} {369}},\
  \bibinfo {pages} {80} (\bibinfo {year} {2024})},\ \Eprint
  {https://arxiv.org/abs/2401.13817} {arXiv:2401.13817 [astro-ph.HE]}
  \BibitemShut {NoStop}%
\bibitem [{\citenamefont {Stockinger}\ \emph {et~al.}(2020)\citenamefont
  {Stockinger} \emph {et~al.}}]{Stockinger:2020hse}%
  \BibitemOpen
  \bibfield  {author} {\bibinfo {author} {\bibfnamefont {G.}~\bibnamefont
  {Stockinger}} \emph {et~al.},\ }\bibfield  {title} {\bibinfo {title}
  {{Three-dimensional Models of Core-collapse Supernovae From Low-mass
  Progenitors With Implications for Crab}},\ }\href
  {https://doi.org/10.1093/mnras/staa1691} {\bibfield  {journal} {\bibinfo
  {journal} {Mon. Not. Roy. Astron. Soc.}\ }\textbf {\bibinfo {volume} {496}},\
  \bibinfo {pages} {2039} (\bibinfo {year} {2020})},\ \Eprint
  {https://arxiv.org/abs/2005.02420} {arXiv:2005.02420 [astro-ph.HE]}
  \BibitemShut {NoStop}%
\bibitem [{\citenamefont {Chakraborty}\ \emph {et~al.}(2015)\citenamefont
  {Chakraborty}, \citenamefont {Raffelt}, \citenamefont {Janka},\ and\
  \citenamefont {M{\"u}ller}}]{Chakraborty:2014lsa}%
  \BibitemOpen
  \bibfield  {author} {\bibinfo {author} {\bibfnamefont {S.}~\bibnamefont
  {Chakraborty}}, \bibinfo {author} {\bibfnamefont {G.~G.}\ \bibnamefont
  {Raffelt}}, \bibinfo {author} {\bibfnamefont {H.-T.}\ \bibnamefont {Janka}},\
  and\ \bibinfo {author} {\bibfnamefont {B.}~\bibnamefont {M{\"u}ller}},\
  }\bibfield  {title} {\bibinfo {title} {{Supernova deleptonization asymmetry:
  Impact on self-induced flavor conversion}},\ }\href
  {https://doi.org/10.1103/PhysRevD.92.105002} {\bibfield  {journal} {\bibinfo
  {journal} {Phys. Rev. D}\ }\textbf {\bibinfo {volume} {92}},\ \bibinfo
  {pages} {105002} (\bibinfo {year} {2015})},\ \Eprint
  {https://arxiv.org/abs/1412.0670} {arXiv:1412.0670 [hep-ph]} \BibitemShut
  {NoStop}%
\bibitem [{\citenamefont {Abbar}\ and\ \citenamefont
  {Duan}(2015)}]{Abbar:2015fwa}%
  \BibitemOpen
  \bibfield  {author} {\bibinfo {author} {\bibfnamefont {S.}~\bibnamefont
  {Abbar}}\ and\ \bibinfo {author} {\bibfnamefont {H.}~\bibnamefont {Duan}},\
  }\bibfield  {title} {\bibinfo {title} {{Neutrino flavor instabilities in a
  time-dependent supernova model}},\ }\href
  {https://doi.org/10.1016/j.physletb.2015.10.019} {\bibfield  {journal}
  {\bibinfo  {journal} {Phys. Lett. B}\ }\textbf {\bibinfo {volume} {751}},\
  \bibinfo {pages} {43} (\bibinfo {year} {2015})},\ \Eprint
  {https://arxiv.org/abs/1509.01538} {arXiv:1509.01538 [astro-ph.HE]}
  \BibitemShut {NoStop}%
\bibitem [{\citenamefont {Dasgupta}\ and\ \citenamefont
  {Mirizzi}(2015)}]{Dasgupta:2015iia}%
  \BibitemOpen
  \bibfield  {author} {\bibinfo {author} {\bibfnamefont {B.}~\bibnamefont
  {Dasgupta}}\ and\ \bibinfo {author} {\bibfnamefont {A.}~\bibnamefont
  {Mirizzi}},\ }\bibfield  {title} {\bibinfo {title} {{Temporal Instability
  Enables Neutrino Flavor Conversions Deep Inside Supernovae}},\ }\href
  {https://doi.org/10.1103/PhysRevD.92.125030} {\bibfield  {journal} {\bibinfo
  {journal} {Phys. Rev. D}\ }\textbf {\bibinfo {volume} {92}},\ \bibinfo
  {pages} {125030} (\bibinfo {year} {2015})},\ \Eprint
  {https://arxiv.org/abs/1509.03171} {arXiv:1509.03171 [hep-ph]} \BibitemShut
  {NoStop}%
\bibitem [{\citenamefont {Dasgupta}\ \emph {et~al.}(2017)\citenamefont
  {Dasgupta}, \citenamefont {Mirizzi},\ and\ \citenamefont
  {Sen}}]{Dasgupta:2016dbv}%
  \BibitemOpen
  \bibfield  {author} {\bibinfo {author} {\bibfnamefont {B.}~\bibnamefont
  {Dasgupta}}, \bibinfo {author} {\bibfnamefont {A.}~\bibnamefont {Mirizzi}},\
  and\ \bibinfo {author} {\bibfnamefont {M.}~\bibnamefont {Sen}},\ }\bibfield
  {title} {\bibinfo {title} {{Fast neutrino flavor conversions near the
  supernova core with realistic flavor-dependent angular distributions}},\
  }\href {https://doi.org/10.1088/1475-7516/2017/02/019} {\bibfield  {journal}
  {\bibinfo  {journal} {JCAP}\ }\textbf {\bibinfo {volume} {02}},\ \bibinfo
  {pages} {019}},\ \Eprint {https://arxiv.org/abs/1609.00528} {arXiv:1609.00528
  [hep-ph]} \BibitemShut {NoStop}%
\bibitem [{\citenamefont {Abbar}\ and\ \citenamefont
  {Duan}(2018)}]{Abbar:2017pkh}%
  \BibitemOpen
  \bibfield  {author} {\bibinfo {author} {\bibfnamefont {S.}~\bibnamefont
  {Abbar}}\ and\ \bibinfo {author} {\bibfnamefont {H.}~\bibnamefont {Duan}},\
  }\bibfield  {title} {\bibinfo {title} {{Fast neutrino flavor conversion:
  roles of dense matter and spectrum crossing}},\ }\href
  {https://doi.org/10.1103/PhysRevD.98.043014} {\bibfield  {journal} {\bibinfo
  {journal} {Phys. Rev. D}\ }\textbf {\bibinfo {volume} {98}},\ \bibinfo
  {pages} {043014} (\bibinfo {year} {2018})},\ \Eprint
  {https://arxiv.org/abs/1712.07013} {arXiv:1712.07013 [hep-ph]} \BibitemShut
  {NoStop}%
\bibitem [{\citenamefont {Sigl}\ and\ \citenamefont
  {Raffelt}(1993)}]{Sigl:1993ctk}%
  \BibitemOpen
  \bibfield  {author} {\bibinfo {author} {\bibfnamefont {G.}~\bibnamefont
  {Sigl}}\ and\ \bibinfo {author} {\bibfnamefont {G.~G.}\ \bibnamefont
  {Raffelt}},\ }\bibfield  {title} {\bibinfo {title} {{General kinetic
  description of relativistic mixed neutrinos}},\ }\href
  {https://doi.org/10.1016/0550-3213(93)90175-O} {\bibfield  {journal}
  {\bibinfo  {journal} {Nucl. Phys. B}\ }\textbf {\bibinfo {volume} {406}},\
  \bibinfo {pages} {423} (\bibinfo {year} {1993})}\BibitemShut {NoStop}%
\bibitem [{\citenamefont {Nagakura}(2022)}]{Nagakura:2022qko}%
  \BibitemOpen
  \bibfield  {author} {\bibinfo {author} {\bibfnamefont {H.}~\bibnamefont
  {Nagakura}},\ }\bibfield  {title} {\bibinfo {title} {{General-relativistic
  quantum-kinetics neutrino transport}},\ }\href
  {https://doi.org/10.1103/PhysRevD.106.063011} {\bibfield  {journal} {\bibinfo
   {journal} {Phys. Rev. D}\ }\textbf {\bibinfo {volume} {106}},\ \bibinfo
  {pages} {063011} (\bibinfo {year} {2022})},\ \Eprint
  {https://arxiv.org/abs/2206.04098} {arXiv:2206.04098 [astro-ph.HE]}
  \BibitemShut {NoStop}%
\bibitem [{\citenamefont {Hannestad}\ \emph {et~al.}(2006)\citenamefont
  {Hannestad}, \citenamefont {Raffelt}, \citenamefont {Sigl},\ and\
  \citenamefont {Wong}}]{Hannestad:2006nj}%
  \BibitemOpen
  \bibfield  {author} {\bibinfo {author} {\bibfnamefont {S.}~\bibnamefont
  {Hannestad}}, \bibinfo {author} {\bibfnamefont {G.~G.}\ \bibnamefont
  {Raffelt}}, \bibinfo {author} {\bibfnamefont {G.}~\bibnamefont {Sigl}},\ and\
  \bibinfo {author} {\bibfnamefont {Y.~Y.~Y.}\ \bibnamefont {Wong}},\
  }\bibfield  {title} {\bibinfo {title} {{Self-induced conversion in dense
  neutrino gases: Pendulum in flavour space}},\ }\href
  {https://doi.org/10.1103/PhysRevD.74.105010} {\bibfield  {journal} {\bibinfo
  {journal} {Phys. Rev. D}\ }\textbf {\bibinfo {volume} {74}},\ \bibinfo
  {pages} {105010} (\bibinfo {year} {2006})},\ \bibinfo {note} {[Erratum:
  Phys.Rev.D 76, 029901 (2007)]},\ \Eprint
  {https://arxiv.org/abs/astro-ph/0608695} {arXiv:astro-ph/0608695}
  \BibitemShut {NoStop}%
\bibitem [{\citenamefont {Thompson}(2002)}]{thompson2002}%
  \BibitemOpen
  \bibfield  {author} {\bibinfo {author} {\bibfnamefont {T.~A.}\ \bibnamefont
  {Thompson}},\ }\href@noop {} {\bibinfo {title} {{Topics in the theory of
  core-collapse supernovae, Ph.D.~thesis, U.~of Arizona}}} (\bibinfo {year}
  {2002})\BibitemShut {NoStop}%
\bibitem [{\citenamefont {O'Connor}(2015)}]{OConnor:2014sgn}%
  \BibitemOpen
  \bibfield  {author} {\bibinfo {author} {\bibfnamefont {E.}~\bibnamefont
  {O'Connor}},\ }\bibfield  {title} {\bibinfo {title} {{An Open-Source Neutrino
  Radiation Hydrodynamics Code for Core-Collapse Supernovae}},\ }\href
  {https://doi.org/10.1088/0067-0049/219/2/24} {\bibfield  {journal} {\bibinfo
  {journal} {Astrophys. J. Suppl.}\ }\textbf {\bibinfo {volume} {219}},\
  \bibinfo {pages} {24} (\bibinfo {year} {2015})},\ \Eprint
  {https://arxiv.org/abs/1411.7058} {arXiv:1411.7058 [astro-ph.HE]}
  \BibitemShut {NoStop}%
\bibitem [{\citenamefont {Nagakura}\ and\ \citenamefont
  {Zaizen}(2022)}]{Nagakura:2022kic}%
  \BibitemOpen
  \bibfield  {author} {\bibinfo {author} {\bibfnamefont {H.}~\bibnamefont
  {Nagakura}}\ and\ \bibinfo {author} {\bibfnamefont {M.}~\bibnamefont
  {Zaizen}},\ }\bibfield  {title} {\bibinfo {title} {{Time-Dependent and
  Quasisteady Features of Fast Neutrino-Flavor Conversion}},\ }\href
  {https://doi.org/10.1103/PhysRevLett.129.261101} {\bibfield  {journal}
  {\bibinfo  {journal} {Phys. Rev. Lett.}\ }\textbf {\bibinfo {volume} {129}},\
  \bibinfo {pages} {261101} (\bibinfo {year} {2022})},\ \Eprint
  {https://arxiv.org/abs/2206.04097} {arXiv:2206.04097 [astro-ph.HE]}
  \BibitemShut {NoStop}%
\bibitem [{\citenamefont {Besard}\ \emph {et~al.}(2019)\citenamefont {Besard},
  \citenamefont {Foket},\ and\ \citenamefont {De~Sutter}}]{Besard:2019aaa}%
  \BibitemOpen
  \bibfield  {author} {\bibinfo {author} {\bibfnamefont {T.}~\bibnamefont
  {Besard}}, \bibinfo {author} {\bibfnamefont {C.}~\bibnamefont {Foket}},\ and\
  \bibinfo {author} {\bibfnamefont {B.}~\bibnamefont {De~Sutter}},\ }\bibfield
  {title} {\bibinfo {title} {Effective extensible programming: Unleashing julia
  on gpus},\ }\href {https://doi.org/10.1109/TPDS.2018.2872064} {\bibfield
  {journal} {\bibinfo  {journal} {IEEE Trans. Parallel Distrib. Syst.}\
  }\textbf {\bibinfo {volume} {30}},\ \bibinfo {pages} {827–841} (\bibinfo
  {year} {2019})}\BibitemShut {NoStop}%
\bibitem [{Hai()}]{Hairer:1993}%
  \BibitemOpen
  \bibfield  {title} {\bibinfo {title} {Multistep {{Methods}} and {{General
  Linear Methods}}},\ }in\ \href {https://doi.org/10.1007/978-3-540-78862-1_3}
  {\emph {\bibinfo {booktitle} {Solving {{Ordinary Differential Equations I}}:
  {{Nonstiff Problems}}}}},\ \bibinfo {editor} {edited by\ \bibinfo {editor}
  {\bibfnamefont {E.}~\bibnamefont {Hairer}}, \bibinfo {editor} {\bibfnamefont
  {G.}~\bibnamefont {Wanner}},\ and\ \bibinfo {editor} {\bibfnamefont {S.~P.}\
  \bibnamefont {Nørsett}}}\ (\bibinfo  {publisher} {Springer})\ pp.\ \bibinfo
  {pages} {355--474}\BibitemShut {NoStop}%
\bibitem [{\citenamefont {{MPA Supernova Archive}}()}]{SNarchive}%
  \BibitemOpen
  \bibfield  {author} {\bibinfo {author} {\bibnamefont {{MPA Supernova
  Archive}}},\ }\href@noop {} {}\bibinfo {howpublished}
  {\url{https://wwwmpa.mpa-garching.mpg.de/ccsnarchive/data/Bollig2016/}}\BibitemShut
  {NoStop}%
\bibitem [{\citenamefont {Tamborra}\ \emph {et~al.}(2017)\citenamefont
  {Tamborra}, \citenamefont {H{\"u}edepohl}, \citenamefont {Raffelt},\ and\
  \citenamefont {Janka}}]{Tamborra:2017ubu}%
  \BibitemOpen
  \bibfield  {author} {\bibinfo {author} {\bibfnamefont {I.}~\bibnamefont
  {Tamborra}}, \bibinfo {author} {\bibfnamefont {L.}~\bibnamefont
  {H{\"u}edepohl}}, \bibinfo {author} {\bibfnamefont {G.~G.}\ \bibnamefont
  {Raffelt}},\ and\ \bibinfo {author} {\bibfnamefont {H.-T.}\ \bibnamefont
  {Janka}},\ }\bibfield  {title} {\bibinfo {title} {{Flavor-dependent neutrino
  angular distribution in core-collapse supernovae}},\ }\href
  {https://doi.org/10.3847/1538-4357/aa6a18} {\bibfield  {journal} {\bibinfo
  {journal} {Astrophys. J.}\ }\textbf {\bibinfo {volume} {839}},\ \bibinfo
  {pages} {132} (\bibinfo {year} {2017})},\ \Eprint
  {https://arxiv.org/abs/1702.00060} {arXiv:1702.00060 [astro-ph.HE]}
  \BibitemShut {NoStop}%
\bibitem [{\citenamefont {Tamborra}\ \emph
  {et~al.}(2014{\natexlab{b}})\citenamefont {Tamborra}, \citenamefont
  {Raffelt}, \citenamefont {Hanke}, \citenamefont {Janka},\ and\ \citenamefont
  {M{\"u}ller}}]{Tamborra:2014hga}%
  \BibitemOpen
  \bibfield  {author} {\bibinfo {author} {\bibfnamefont {I.}~\bibnamefont
  {Tamborra}}, \bibinfo {author} {\bibfnamefont {G.~G.}\ \bibnamefont
  {Raffelt}}, \bibinfo {author} {\bibfnamefont {F.}~\bibnamefont {Hanke}},
  \bibinfo {author} {\bibfnamefont {H.-T.}\ \bibnamefont {Janka}},\ and\
  \bibinfo {author} {\bibfnamefont {B.}~\bibnamefont {M{\"u}ller}},\ }\bibfield
   {title} {\bibinfo {title} {{Neutrino emission characteristics and detection
  opportunities based on three-dimensional supernova simulations}},\ }\href
  {https://doi.org/10.1103/PhysRevD.90.045032} {\bibfield  {journal} {\bibinfo
  {journal} {Phys. Rev. D}\ }\textbf {\bibinfo {volume} {90}},\ \bibinfo
  {pages} {045032} (\bibinfo {year} {2014}{\natexlab{b}})},\ \Eprint
  {https://arxiv.org/abs/1406.0006} {arXiv:1406.0006 [astro-ph.SR]}
  \BibitemShut {NoStop}%
\bibitem [{\citenamefont {Wu}\ \emph {et~al.}(2017)\citenamefont {Wu},
  \citenamefont {Tamborra}, \citenamefont {Just},\ and\ \citenamefont
  {Janka}}]{Wu:2017drk}%
  \BibitemOpen
  \bibfield  {author} {\bibinfo {author} {\bibfnamefont {M.-R.}\ \bibnamefont
  {Wu}}, \bibinfo {author} {\bibfnamefont {I.}~\bibnamefont {Tamborra}},
  \bibinfo {author} {\bibfnamefont {O.}~\bibnamefont {Just}},\ and\ \bibinfo
  {author} {\bibfnamefont {H.-T.}\ \bibnamefont {Janka}},\ }\bibfield  {title}
  {\bibinfo {title} {{Imprints of neutrino-pair flavor conversions on
  nucleosynthesis in ejecta from neutron-star merger remnants}},\ }\href
  {https://doi.org/10.1103/PhysRevD.96.123015} {\bibfield  {journal} {\bibinfo
  {journal} {Phys. Rev. D}\ }\textbf {\bibinfo {volume} {96}},\ \bibinfo
  {pages} {123015} (\bibinfo {year} {2017})},\ \Eprint
  {https://arxiv.org/abs/1711.00477} {arXiv:1711.00477 [astro-ph.HE]}
  \BibitemShut {NoStop}%
\bibitem [{\citenamefont {Banerjee}\ \emph {et~al.}(2011)\citenamefont
  {Banerjee}, \citenamefont {Dighe},\ and\ \citenamefont
  {Raffelt}}]{Banerjee:2011fj}%
  \BibitemOpen
  \bibfield  {author} {\bibinfo {author} {\bibfnamefont {A.}~\bibnamefont
  {Banerjee}}, \bibinfo {author} {\bibfnamefont {A.}~\bibnamefont {Dighe}},\
  and\ \bibinfo {author} {\bibfnamefont {G.~G.}\ \bibnamefont {Raffelt}},\
  }\bibfield  {title} {\bibinfo {title} {{Linearized flavor-stability analysis
  of dense neutrino streams}},\ }\href
  {https://doi.org/10.1103/PhysRevD.84.053013} {\bibfield  {journal} {\bibinfo
  {journal} {Phys. Rev. D}\ }\textbf {\bibinfo {volume} {84}},\ \bibinfo
  {pages} {053013} (\bibinfo {year} {2011})},\ \Eprint
  {https://arxiv.org/abs/1107.2308} {arXiv:1107.2308 [hep-ph]} \BibitemShut
  {NoStop}%
\bibitem [{\citenamefont {Shalgar}\ and\ \citenamefont
  {Tamborra}(2024{\natexlab{b}})}]{Shalgar:2023aca}%
  \BibitemOpen
  \bibfield  {author} {\bibinfo {author} {\bibfnamefont {S.}~\bibnamefont
  {Shalgar}}\ and\ \bibinfo {author} {\bibfnamefont {I.}~\bibnamefont
  {Tamborra}},\ }\bibfield  {title} {\bibinfo {title} {{Do neutrinos become
  flavor unstable due to collisions with matter in the supernova decoupling
  region?}},\ }\href {https://doi.org/10.1103/PhysRevD.109.103011} {\bibfield
  {journal} {\bibinfo  {journal} {Phys. Rev. D}\ }\textbf {\bibinfo {volume}
  {109}},\ \bibinfo {pages} {103011} (\bibinfo {year} {2024}{\natexlab{b}})},\
  \Eprint {https://arxiv.org/abs/2307.10366} {arXiv:2307.10366 [astro-ph.HE]}
  \BibitemShut {NoStop}%
\bibitem [{\citenamefont {Wang}\ \emph {et~al.}(2025)\citenamefont {Wang},
  \citenamefont {Nagakura}, \citenamefont {Johns},\ and\ \citenamefont
  {Burrows}}]{Wang:2025vbx}%
  \BibitemOpen
  \bibfield  {author} {\bibinfo {author} {\bibfnamefont {T.}~\bibnamefont
  {Wang}}, \bibinfo {author} {\bibfnamefont {H.}~\bibnamefont {Nagakura}},
  \bibinfo {author} {\bibfnamefont {L.}~\bibnamefont {Johns}},\ and\ \bibinfo
  {author} {\bibfnamefont {A.}~\bibnamefont {Burrows}},\ }\bibfield  {title}
  {\bibinfo {title} {{Effect of the collisional flavor instability on
  core-collapse supernova models}},\ }\href {https://doi.org/10.1103/z3qh-nj18}
  {\bibfield  {journal} {\bibinfo  {journal} {Phys. Rev. D}\ }\textbf {\bibinfo
  {volume} {112}},\ \bibinfo {pages} {063039} (\bibinfo {year} {2025})},\
  \Eprint {https://arxiv.org/abs/2507.01100} {arXiv:2507.01100 [astro-ph.HE]}
  \BibitemShut {NoStop}%
\bibitem [{\citenamefont {Raffelt}\ \emph {et~al.}(2013)\citenamefont
  {Raffelt}, \citenamefont {Sarikas},\ and\ \citenamefont
  {de~Sousa~Seixas}}]{Raffelt:2013rqa}%
  \BibitemOpen
  \bibfield  {author} {\bibinfo {author} {\bibfnamefont {G.~G.}\ \bibnamefont
  {Raffelt}}, \bibinfo {author} {\bibfnamefont {S.}~\bibnamefont {Sarikas}},\
  and\ \bibinfo {author} {\bibfnamefont {D.}~\bibnamefont {de~Sousa~Seixas}},\
  }\bibfield  {title} {\bibinfo {title} {{Axial Symmetry Breaking in
  Self-Induced Flavor Conversion of Supernova Neutrino Fluxes}},\ }\href
  {https://doi.org/10.1103/PhysRevLett.111.091101} {\bibfield  {journal}
  {\bibinfo  {journal} {Phys. Rev. Lett.}\ }\textbf {\bibinfo {volume} {111}},\
  \bibinfo {pages} {091101} (\bibinfo {year} {2013})},\ \bibinfo {note}
  {[Erratum: Phys.Rev.Lett. 113, 239903 (2014)]},\ \Eprint
  {https://arxiv.org/abs/1305.7140} {arXiv:1305.7140 [hep-ph]} \BibitemShut
  {NoStop}%
\bibitem [{\citenamefont {Mirizzi}(2013)}]{Mirizzi:2013rla}%
  \BibitemOpen
  \bibfield  {author} {\bibinfo {author} {\bibfnamefont {A.}~\bibnamefont
  {Mirizzi}},\ }\bibfield  {title} {\bibinfo {title} {{Multi-azimuthal-angle
  effects in self-induced supernova neutrino flavor conversions without axial
  symmetry}},\ }\href {https://doi.org/10.1103/PhysRevD.88.073004} {\bibfield
  {journal} {\bibinfo  {journal} {Phys. Rev. D}\ }\textbf {\bibinfo {volume}
  {88}},\ \bibinfo {pages} {073004} (\bibinfo {year} {2013})},\ \Eprint
  {https://arxiv.org/abs/1308.1402} {arXiv:1308.1402 [hep-ph]} \BibitemShut
  {NoStop}%
\bibitem [{\citenamefont {Fogli}\ \emph {et~al.}(2007)\citenamefont {Fogli},
  \citenamefont {Lisi}, \citenamefont {Marrone},\ and\ \citenamefont
  {Mirizzi}}]{Fogli:2007bk}%
  \BibitemOpen
  \bibfield  {author} {\bibinfo {author} {\bibfnamefont {G.~L.}\ \bibnamefont
  {Fogli}}, \bibinfo {author} {\bibfnamefont {E.}~\bibnamefont {Lisi}},
  \bibinfo {author} {\bibfnamefont {A.}~\bibnamefont {Marrone}},\ and\ \bibinfo
  {author} {\bibfnamefont {A.}~\bibnamefont {Mirizzi}},\ }\bibfield  {title}
  {\bibinfo {title} {{Collective neutrino flavor transitions in supernovae and
  the role of trajectory averaging}},\ }\href
  {https://doi.org/10.1088/1475-7516/2007/12/010} {\bibfield  {journal}
  {\bibinfo  {journal} {JCAP}\ }\textbf {\bibinfo {volume} {12}},\ \bibinfo
  {pages} {010}},\ \Eprint {https://arxiv.org/abs/0707.1998} {arXiv:0707.1998
  [hep-ph]} \BibitemShut {NoStop}%
\bibitem [{\citenamefont {Duan}\ \emph {et~al.}(2007)\citenamefont {Duan},
  \citenamefont {Fuller}, \citenamefont {Carlson},\ and\ \citenamefont
  {Qian}}]{Duan:2007bt}%
  \BibitemOpen
  \bibfield  {author} {\bibinfo {author} {\bibfnamefont {H.}~\bibnamefont
  {Duan}}, \bibinfo {author} {\bibfnamefont {G.~M.}\ \bibnamefont {Fuller}},
  \bibinfo {author} {\bibfnamefont {J.}~\bibnamefont {Carlson}},\ and\ \bibinfo
  {author} {\bibfnamefont {Y.-Z.}\ \bibnamefont {Qian}},\ }\bibfield  {title}
  {\bibinfo {title} {{Neutrino Mass Hierarchy and Stepwise Spectral Swapping of
  Supernova Neutrino Flavors}},\ }\href
  {https://doi.org/10.1103/PhysRevLett.99.241802} {\bibfield  {journal}
  {\bibinfo  {journal} {Phys. Rev. Lett.}\ }\textbf {\bibinfo {volume} {99}},\
  \bibinfo {pages} {241802} (\bibinfo {year} {2007})},\ \Eprint
  {https://arxiv.org/abs/0707.0290} {arXiv:0707.0290 [astro-ph]} \BibitemShut
  {NoStop}%
\bibitem [{\citenamefont {Xiong}\ \emph {et~al.}(2024)\citenamefont {Xiong},
  \citenamefont {Wu}, \citenamefont {George}, \citenamefont {Lin},
  \citenamefont {Largani}, \citenamefont {Fischer},\ and\ \citenamefont
  {Mart{\'\i}nez-Pinedo}}]{Xiong:2024tac}%
  \BibitemOpen
  \bibfield  {author} {\bibinfo {author} {\bibfnamefont {Z.}~\bibnamefont
  {Xiong}}, \bibinfo {author} {\bibfnamefont {M.-R.}\ \bibnamefont {Wu}},
  \bibinfo {author} {\bibfnamefont {M.}~\bibnamefont {George}}, \bibinfo
  {author} {\bibfnamefont {C.-Y.}\ \bibnamefont {Lin}}, \bibinfo {author}
  {\bibfnamefont {N.~K.}\ \bibnamefont {Largani}}, \bibinfo {author}
  {\bibfnamefont {T.}~\bibnamefont {Fischer}},\ and\ \bibinfo {author}
  {\bibfnamefont {G.}~\bibnamefont {Mart{\'\i}nez-Pinedo}},\ }\bibfield
  {title} {\bibinfo {title} {{Fast neutrino flavor conversions in a supernova:
  Emergence, evolution, and effects}},\ }\href
  {https://doi.org/10.1103/PhysRevD.109.123008} {\bibfield  {journal} {\bibinfo
   {journal} {Phys. Rev. D}\ }\textbf {\bibinfo {volume} {109}},\ \bibinfo
  {pages} {123008} (\bibinfo {year} {2024})},\ \Eprint
  {https://arxiv.org/abs/2402.19252} {arXiv:2402.19252 [astro-ph.HE]}
  \BibitemShut {NoStop}%
\bibitem [{\citenamefont {Walk}\ \emph {et~al.}(2019)\citenamefont {Walk},
  \citenamefont {Tamborra}, \citenamefont {Janka},\ and\ \citenamefont
  {Summa}}]{Walk:2019ier}%
  \BibitemOpen
  \bibfield  {author} {\bibinfo {author} {\bibfnamefont {L.}~\bibnamefont
  {Walk}}, \bibinfo {author} {\bibfnamefont {I.}~\bibnamefont {Tamborra}},
  \bibinfo {author} {\bibfnamefont {H.-T.}\ \bibnamefont {Janka}},\ and\
  \bibinfo {author} {\bibfnamefont {A.}~\bibnamefont {Summa}},\ }\bibfield
  {title} {\bibinfo {title} {{Effects of the standing accretion-shock
  instability and the lepton-emission self-sustained asymmetry in the neutrino
  emission of rotating supernovae}},\ }\href
  {https://doi.org/10.1103/PhysRevD.100.063018} {\bibfield  {journal} {\bibinfo
   {journal} {Phys. Rev. D}\ }\textbf {\bibinfo {volume} {100}},\ \bibinfo
  {pages} {063018} (\bibinfo {year} {2019})},\ \Eprint
  {https://arxiv.org/abs/1901.06235} {arXiv:1901.06235 [astro-ph.HE]}
  \BibitemShut {NoStop}%
\bibitem [{\citenamefont {Walk}\ \emph {et~al.}(2020)\citenamefont {Walk},
  \citenamefont {Tamborra}, \citenamefont {Janka}, \citenamefont {Summa},\ and\
  \citenamefont {Kresse}}]{Walk:2019miz}%
  \BibitemOpen
  \bibfield  {author} {\bibinfo {author} {\bibfnamefont {L.}~\bibnamefont
  {Walk}}, \bibinfo {author} {\bibfnamefont {I.}~\bibnamefont {Tamborra}},
  \bibinfo {author} {\bibfnamefont {H.-T.}\ \bibnamefont {Janka}}, \bibinfo
  {author} {\bibfnamefont {A.}~\bibnamefont {Summa}},\ and\ \bibinfo {author}
  {\bibfnamefont {D.}~\bibnamefont {Kresse}},\ }\bibfield  {title} {\bibinfo
  {title} {{Neutrino emission characteristics of black hole formation in
  three-dimensional simulations of stellar collapse}},\ }\href
  {https://doi.org/10.1103/PhysRevD.101.123013} {\bibfield  {journal} {\bibinfo
   {journal} {Phys. Rev. D}\ }\textbf {\bibinfo {volume} {101}},\ \bibinfo
  {pages} {123013} (\bibinfo {year} {2020})},\ \Eprint
  {https://arxiv.org/abs/1910.12971} {arXiv:1910.12971 [astro-ph.HE]}
  \BibitemShut {NoStop}%
\bibitem [{\citenamefont {Bollig}\ \emph {et~al.}(2017)\citenamefont {Bollig},
  \citenamefont {Janka}, \citenamefont {Lohs}, \citenamefont {Martinez-Pinedo},
  \citenamefont {Horowitz},\ and\ \citenamefont {Melson}}]{Bollig:2017lki}%
  \BibitemOpen
  \bibfield  {author} {\bibinfo {author} {\bibfnamefont {R.}~\bibnamefont
  {Bollig}}, \bibinfo {author} {\bibfnamefont {H.-T.}\ \bibnamefont {Janka}},
  \bibinfo {author} {\bibfnamefont {A.}~\bibnamefont {Lohs}}, \bibinfo {author}
  {\bibfnamefont {G.}~\bibnamefont {Martinez-Pinedo}}, \bibinfo {author}
  {\bibfnamefont {C.~J.}\ \bibnamefont {Horowitz}},\ and\ \bibinfo {author}
  {\bibfnamefont {T.}~\bibnamefont {Melson}},\ }\bibfield  {title} {\bibinfo
  {title} {{Muon Creation in Supernova Matter Facilitates Neutrino-driven
  Explosions}},\ }\href {https://doi.org/10.1103/PhysRevLett.119.242702}
  {\bibfield  {journal} {\bibinfo  {journal} {Phys. Rev. Lett.}\ }\textbf
  {\bibinfo {volume} {119}},\ \bibinfo {pages} {242702} (\bibinfo {year}
  {2017})},\ \Eprint {https://arxiv.org/abs/1706.04630} {arXiv:1706.04630
  [astro-ph.HE]} \BibitemShut {NoStop}%
\bibitem [{\citenamefont {Cornelius}\ \emph {et~al.}(2025)\citenamefont
  {Cornelius}, \citenamefont {Tamborra}, \citenamefont {Heinlein},
  \citenamefont {Shalgar},\ and\ \citenamefont {Janka}}]{Cornelius:2025nvd}%
  \BibitemOpen
  \bibfield  {author} {\bibinfo {author} {\bibfnamefont {M.}~\bibnamefont
  {Cornelius}}, \bibinfo {author} {\bibfnamefont {I.}~\bibnamefont {Tamborra}},
  \bibinfo {author} {\bibfnamefont {M.}~\bibnamefont {Heinlein}}, \bibinfo
  {author} {\bibfnamefont {S.}~\bibnamefont {Shalgar}},\ and\ \bibinfo {author}
  {\bibfnamefont {H.-T.}\ \bibnamefont {Janka}},\ }\bibfield  {title} {\bibinfo
  {title} {{Electron-neutrino lepton number crossings: Variations with the
  supernova core physics}},\ }\href {https://doi.org/10.1103/n2h5-3qwz}
  {\bibfield  {journal} {\bibinfo  {journal} {Phys. Rev. D}\ }\textbf {\bibinfo
  {volume} {112}},\ \bibinfo {pages} {063006} (\bibinfo {year} {2025})},\
  \Eprint {https://arxiv.org/abs/2507.13429} {arXiv:2507.13429 [astro-ph.HE]}
  \BibitemShut {NoStop}%
\bibitem [{\citenamefont {Richers}\ \emph {et~al.}(2019)\citenamefont
  {Richers}, \citenamefont {McLaughlin}, \citenamefont {Kneller},\ and\
  \citenamefont {Vlasenko}}]{Richers:2019grc}%
  \BibitemOpen
  \bibfield  {author} {\bibinfo {author} {\bibfnamefont {S.~A.}\ \bibnamefont
  {Richers}}, \bibinfo {author} {\bibfnamefont {G.~C.}\ \bibnamefont
  {McLaughlin}}, \bibinfo {author} {\bibfnamefont {J.~P.}\ \bibnamefont
  {Kneller}},\ and\ \bibinfo {author} {\bibfnamefont {A.}~\bibnamefont
  {Vlasenko}},\ }\bibfield  {title} {\bibinfo {title} {{Neutrino Quantum
  Kinetics in Compact Objects}},\ }\href
  {https://doi.org/10.1103/PhysRevD.99.123014} {\bibfield  {journal} {\bibinfo
  {journal} {Phys. Rev. D}\ }\textbf {\bibinfo {volume} {99}},\ \bibinfo
  {pages} {123014} (\bibinfo {year} {2019})},\ \bibinfo {note} {[Erratum:
  Phys.Rev.D 109, 129902 (2024)]},\ \Eprint {https://arxiv.org/abs/1903.00022}
  {arXiv:1903.00022 [astro-ph.HE]} \BibitemShut {NoStop}%
\bibitem [{\citenamefont {Bruenn}(1985)}]{Bruenn:1985en}%
  \BibitemOpen
  \bibfield  {author} {\bibinfo {author} {\bibfnamefont {S.~W.}\ \bibnamefont
  {Bruenn}},\ }\bibfield  {title} {\bibinfo {title} {{Stellar core collapse:
  Numerical model and infall epoch}},\ }\href {https://doi.org/10.1086/191056}
  {\bibfield  {journal} {\bibinfo  {journal} {Astrophys. J. Suppl.}\ }\textbf
  {\bibinfo {volume} {58}},\ \bibinfo {pages} {771} (\bibinfo {year}
  {1985})}\BibitemShut {NoStop}%
\bibitem [{\citenamefont {Vogel}(1984)}]{Vogel:1983hi}%
  \BibitemOpen
  \bibfield  {author} {\bibinfo {author} {\bibfnamefont {P.}~\bibnamefont
  {Vogel}},\ }\bibfield  {title} {\bibinfo {title} {{Analysis of the
  Anti-neutrino Capture on Protons}},\ }\href
  {https://doi.org/10.1103/PhysRevD.29.1918} {\bibfield  {journal} {\bibinfo
  {journal} {Phys. Rev. D}\ }\textbf {\bibinfo {volume} {29}},\ \bibinfo
  {pages} {1918} (\bibinfo {year} {1984})}\BibitemShut {NoStop}%
\bibitem [{\citenamefont {Betranhandy}\ and\ \citenamefont
  {O'Connor}(2020)}]{Betranhandy:2020cdf}%
  \BibitemOpen
  \bibfield  {author} {\bibinfo {author} {\bibfnamefont {A.}~\bibnamefont
  {Betranhandy}}\ and\ \bibinfo {author} {\bibfnamefont {E.}~\bibnamefont
  {O'Connor}},\ }\bibfield  {title} {\bibinfo {title} {{Impact of neutrino
  pair-production rates in Core-Collapse Supernovae}},\ }\href
  {https://doi.org/10.1103/PhysRevD.102.123015} {\bibfield  {journal} {\bibinfo
   {journal} {Phys. Rev. D}\ }\textbf {\bibinfo {volume} {102}},\ \bibinfo
  {pages} {123015} (\bibinfo {year} {2020})},\ \Eprint
  {https://arxiv.org/abs/2010.02261} {arXiv:2010.02261 [astro-ph.HE]}
  \BibitemShut {NoStop}%
\bibitem [{\citenamefont {Betranhandy}\ and\ \citenamefont
  {O'Connor}(2025)}]{Betranhandy:2025amv}%
  \BibitemOpen
  \bibfield  {author} {\bibinfo {author} {\bibfnamefont {A.}~\bibnamefont
  {Betranhandy}}\ and\ \bibinfo {author} {\bibfnamefont {E.}~\bibnamefont
  {O'Connor}},\ }\bibfield  {title} {\bibinfo {title} {{New approximation for
  heavy-lepton neutrino pair processes in simulations of core-collapse
  supernovae}},\ }\href {https://doi.org/10.1103/fv13-5jtg} {\bibfield
  {journal} {\bibinfo  {journal} {Phys. Rev. D}\ }\textbf {\bibinfo {volume}
  {111}},\ \bibinfo {pages} {103049} (\bibinfo {year} {2025})},\ \Eprint
  {https://arxiv.org/abs/2501.17141} {arXiv:2501.17141 [astro-ph.HE]}
  \BibitemShut {NoStop}%
\bibitem [{\citenamefont {Burrows}\ \emph {et~al.}(2000)\citenamefont
  {Burrows}, \citenamefont {Young}, \citenamefont {Young}, \citenamefont
  {Pinto}, \citenamefont {Eastman}, \citenamefont {Thompson},\ and\
  \citenamefont {Thompson}}]{Burrows:2000xsm}%
  \BibitemOpen
  \bibfield  {author} {\bibinfo {author} {\bibfnamefont {A.}~\bibnamefont
  {Burrows}}, \bibinfo {author} {\bibfnamefont {T.}~\bibnamefont {Young}},
  \bibinfo {author} {\bibfnamefont {T.}~\bibnamefont {Young}}, \bibinfo
  {author} {\bibfnamefont {P.}~\bibnamefont {Pinto}}, \bibinfo {author}
  {\bibfnamefont {R.}~\bibnamefont {Eastman}}, \bibinfo {author} {\bibfnamefont
  {T.}~\bibnamefont {Thompson}},\ and\ \bibinfo {author} {\bibfnamefont
  {T.~A.}\ \bibnamefont {Thompson}},\ }\bibfield  {title} {\bibinfo {title} {{A
  new algorithm for supernova neutrino transport and some applications}},\
  }\href {https://doi.org/10.1086/309244} {\bibfield  {journal} {\bibinfo
  {journal} {Astrophys. J.}\ }\textbf {\bibinfo {volume} {539}},\ \bibinfo
  {pages} {865} (\bibinfo {year} {2000})},\ \Eprint
  {https://arxiv.org/abs/astro-ph/9905132} {arXiv:astro-ph/9905132}
  \BibitemShut {NoStop}%
\bibitem [{\citenamefont {Burrows}\ \emph {et~al.}(2006)\citenamefont
  {Burrows}, \citenamefont {Reddy},\ and\ \citenamefont
  {Thompson}}]{Burrows:2004vq}%
  \BibitemOpen
  \bibfield  {author} {\bibinfo {author} {\bibfnamefont {A.}~\bibnamefont
  {Burrows}}, \bibinfo {author} {\bibfnamefont {S.}~\bibnamefont {Reddy}},\
  and\ \bibinfo {author} {\bibfnamefont {T.~A.}\ \bibnamefont {Thompson}},\
  }\bibfield  {title} {\bibinfo {title} {{Neutrino opacities in nuclear
  matter}},\ }\href {https://doi.org/10.1016/j.nuclphysa.2004.06.012}
  {\bibfield  {journal} {\bibinfo  {journal} {Nucl. Phys. A}\ }\textbf
  {\bibinfo {volume} {777}},\ \bibinfo {pages} {356} (\bibinfo {year}
  {2006})},\ \Eprint {https://arxiv.org/abs/astro-ph/0404432}
  {arXiv:astro-ph/0404432} \BibitemShut {NoStop}%
\bibitem [{\citenamefont {Horowitz}(2002)}]{Horowitz:2001xf}%
  \BibitemOpen
  \bibfield  {author} {\bibinfo {author} {\bibfnamefont {C.~J.}\ \bibnamefont
  {Horowitz}},\ }\bibfield  {title} {\bibinfo {title} {{Weak magnetism for
  anti-neutrinos in supernovae}},\ }\href
  {https://doi.org/10.1103/PhysRevD.65.043001} {\bibfield  {journal} {\bibinfo
  {journal} {Phys. Rev. D}\ }\textbf {\bibinfo {volume} {65}},\ \bibinfo
  {pages} {043001} (\bibinfo {year} {2002})},\ \Eprint
  {https://arxiv.org/abs/astro-ph/0109209} {arXiv:astro-ph/0109209}
  \BibitemShut {NoStop}%
\bibitem [{\citenamefont {Rampp}\ and\ \citenamefont
  {Janka}(2002)}]{Rampp:2002bq}%
  \BibitemOpen
  \bibfield  {author} {\bibinfo {author} {\bibfnamefont {M.}~\bibnamefont
  {Rampp}}\ and\ \bibinfo {author} {\bibfnamefont {H.-T.}\ \bibnamefont
  {Janka}},\ }\bibfield  {title} {\bibinfo {title} {{Radiation hydrodynamics
  with neutrinos: Variable Eddington factor method for core collapse supernova
  simulations}},\ }\href {https://doi.org/10.1051/0004-6361:20021398}
  {\bibfield  {journal} {\bibinfo  {journal} {Astron. Astrophys.}\ }\textbf
  {\bibinfo {volume} {396}},\ \bibinfo {pages} {361} (\bibinfo {year}
  {2002})},\ \Eprint {https://arxiv.org/abs/astro-ph/0203101}
  {arXiv:astro-ph/0203101} \BibitemShut {NoStop}%
\bibitem [{\citenamefont {Fiorillo}\ \emph {et~al.}(2023)\citenamefont
  {Fiorillo}, \citenamefont {Heinlein}, \citenamefont {Janka}, \citenamefont
  {Raffelt}, \citenamefont {Vitagliano},\ and\ \citenamefont
  {Bollig}}]{Fiorillo:2023frv}%
  \BibitemOpen
  \bibfield  {author} {\bibinfo {author} {\bibfnamefont {D.~F.~G.}\
  \bibnamefont {Fiorillo}}, \bibinfo {author} {\bibfnamefont {M.}~\bibnamefont
  {Heinlein}}, \bibinfo {author} {\bibfnamefont {H.-T.}\ \bibnamefont {Janka}},
  \bibinfo {author} {\bibfnamefont {G.~G.}\ \bibnamefont {Raffelt}}, \bibinfo
  {author} {\bibfnamefont {E.}~\bibnamefont {Vitagliano}},\ and\ \bibinfo
  {author} {\bibfnamefont {R.}~\bibnamefont {Bollig}},\ }\bibfield  {title}
  {\bibinfo {title} {{Supernova simulations confront SN 1987A neutrinos}},\
  }\href {https://doi.org/10.1103/PhysRevD.108.083040} {\bibfield  {journal}
  {\bibinfo  {journal} {Phys. Rev. D}\ }\textbf {\bibinfo {volume} {108}},\
  \bibinfo {pages} {083040} (\bibinfo {year} {2023})},\ \Eprint
  {https://arxiv.org/abs/2308.01403} {arXiv:2308.01403 [astro-ph.HE]}
  \BibitemShut {NoStop}%
\end{thebibliography}%

\end{document}